%% file: main.tex
\documentclass[11pt,preprintnumbers,superscriptaddress,amsmath,amssymb,nofootinbib]{revtex4}
\usepackage{graphicx} 
\usepackage{dcolumn}
\usepackage{bm}
\usepackage{amssymb}
\usepackage{amsmath}
\usepackage{mathtools}
\usepackage{epsfig}    
\usepackage{color}
\usepackage{slashed}
\usepackage{subfiles}
\usepackage{ifthen}
\usepackage{float}
\usepackage{physics}
\allowdisplaybreaks[1]
\newcommand{\red}[1]{{\color[rgb]{0,0,0} #1}}

\newcommand{\BR}{\text{BR}}

\newcommand{\cba}{c_{\beta-\alpha}}
\newcommand{\sba}{s_{\beta-\alpha}}
\newcommand{\tb}{\tan\beta}
\newcommand{\mHp}{m_{H^\pm}}
\newcommand{\fp}{f^\prime}
\newcommand{\qp}{q^\prime}
\newcommand{\mfp}{m_{f^\prime}}

\newcommand{\Hpm}{H^\pm}
\newcommand{\Hp}{H^+}
\newcounter{mainflag}

\begin{document}

\title{Radiative corrections to decays of charged Higgs bosons \\
in two Higgs doublet models}


\preprint{OU-HET 1100}
\preprint{TU 1128}

\author{Masashi Aiko}
\email{m-aikou@het.phys.sci.osaka-u.ac.jp}
\affiliation{Department of Physics, Osaka University, Toyonaka, Osaka 560-0043, Japan}

\author{Shinya Kanemura}
\email{kanemu@het.phys.sci.osaka-u.ac.jp}
\affiliation{Department of Physics, Osaka University, Toyonaka, Osaka 560-0043, Japan}

\author{Kodai Sakurai}
\email{kodai.sakurai.e3@tohoku.ac.jp}
\affiliation{Department of Physics, Tohoku University, Sendai, Miyagi 980-8578, Japan}

\begin{abstract}
We calculate the next-to-leading order (NLO) electroweak (EW) corrections to decay rates of charged Higgs bosons for various decay modes in the four types of two Higgs doublet models (THDMs) with the softly broken discrete $Z_{2}$ symmetry. 
Decay branching ratios of charged Higgs bosons are evaluated including NLO EW corrections, as well as  QCD corrections up to next-to-next-to-leading order (NNLO).   
We comprehensively study impacts of the NLO EW corrections to the branching ratios in nearly alignment scenarios where the couplings constants of the Higgs boson with the mass of 125 GeV are close to those predicted in the standard model.  
Furthermore, in the nearly alignment scenario, we discuss whether or not the four types of THDMs can be distinguished via the decays of charged Higgs bosons. 
We find that characteristic predictions of charged Higgs branching ratios can be obtained for all types of the THDMs, by which each type of the THDMs are separated, and information on the internal parameters of the THDMs can be extracted from the magnitudes of the various decay branching ratios. 
\end{abstract}

\setcounter{mainflag}{1} 

\maketitle
\newpage
\tableofcontents

\newpage
\input{1_Introduction}

\input{2_Model}
%
\input{3_Decay_rate}
%
\input{4_NLO_EW}
%
\input{5_numerical_result}

\section{Conclusions}\label{sec:conclusions}
We have calculated NLO EW corrections to the decay rates for the charged Higgs boson decays into a pair of fermions, the $W^{+}$ boson and the neutral Higgs bosons $H^{+}\to W^{+} \phi$ $(\phi=h,\;H,\;A)$. 
The loop-induced decay rates $H^{+}\to W^{+}V$ ($V=Z,\;\gamma$) have been calculated at LO.
We have shown how one can obtain analytical formulae for the decay rates of $H^{-}$ in Appendix~\ref{sec:ApB3}. 
We have applied the improved on-shell scheme~\cite{Kanemura:2017wtm} to the computations of the NLO EW corrections.
In addition, we have removed IR divergences arising from virtual photon loop corrections by adding contributions from real photon emissions.

In order to study the phenomenological impacts in the case that the charged Higgs bosons are discovered, we have investigated whether or not four types of THDMs can be distinguished in nearly alignment scenarios by decay branching ratios of the charged Higgs bosons.
We have found that characteristic predictions of $\BR(\Hp\to W^{+}h)$ for Type-I and those of $\BR(\Hp\to \tau^{+}\nu)$ for Type-X can be obtained even if the significant deviation from the SM prediction in the $hZZ$ coupling is not detected. 
In addition, Type-II and Type-Y can be separated by looking at the magnitudes of $\BR(\Hp\to c\bar{b})$ and $\BR(\Hp\to \tau^{+}\nu)$.
Furthermore, we have investigated the impact of NLO EW corrections to the branching ratios in nearly alignment scenarios.
We have found that the NLO EW corrections to the branching ratios similarly behave without depending on the types of THDMs in the low $\tb$ region. 
Relatively large corrections to $\Hp\to f\bar{f}^{\prime}$ can arise in the high $\tb$ region even if we consider the heavy charged Higgs bosons with the mass of 1000~GeV. 
We have also found that the NLO EW corrections to $\Hp\to W^{+}h$ can be sizable due to the non-decoupling effect of the additional Higgs bosons in the case that $\cba$ is close to zero.
In the alignment limit, i.e., $\cba=0$, the tree-level $\Hp W^{-}h$ coupling vanishes, so that the decay rate of $\Hp\to W^{+}h$ disappears at NLO.
Even in such a case, this decay mode is induced through the squared one-loop amplitude. 
We have revealed that $\BR(\Hp\to W^{+}h)$ in the alignment limit can reach $0.1\%$ when the mass of the charged Higgs bosons is at the EW scale. 

\begin{acknowledgments}
We would like to thank Mariko Kikuchi and Kei Yagyu for their useful comments. 
This work is supported in part by the Grant-in-Aid on Innovative Areas, the Ministry of Education, Culture, 
Sports, Science and Technology, No.~16H06492 and No.~20H00160 [S.K.],  
K.S. is supported by JSPS KAKENHI Grant No.~20H01894. 
M. A. was supported in part by the Sasakawa Scientific Research Grant from The Japan Science Society. 
\end{acknowledgments}

\appendix
\input{Apendix}

\bibliographystyle{junsrt}
\bibliography{temp}

\end{document}

%% file: 1_Introduction.tex
%

\section{Introduction}
After the discovery of a Higgs boson with the mass of 125 GeV at LHC in 2012~\cite{Aad:2012tfa, Chatrchyan:2012ufa}, 
it has turned out that the Higgs sector of the standard model (SM) is consistent with the LHC data under the experimental and theoretical uncertainties. 
However, although the Higgs boson was found the whole structure of the Higgs sector remains unknown and the origin of electroweak symmetry breaking is still a mystery. 
While the Higgs sector of the SM is assumed to be composed of an isospin doublet scalar field without any theoretical principle, there can be a possibility for non-minimal Higgs sectors. 
Such extended Higgs sectors are often introduced in new physics models to explain phenomena that cannot be explained in the SM, e.g., neutrino masses, dark matter and baryon asymmetry of the Universe.
In addition, they also necessarily appear in some of the new physics models to solve the hierarchy problem and the strong CP problem. 
Therefore, it is significant to consider the physics of extended Higgs sectors as a probe of new physics.

Extended Higgs sectors predict one or more additional scalar fields.  
Hence, direct searches of new particles at collider experiments are a powerful way to test the extended Higgs sectors.
On the other hand, additional Higgs bosons can be indirectly explored through precision measurements of properties of the discovered Higgs boson, because effects of the additional Higgs bosons can appear in the observables for the discovered Higgs boson through mixing of the Higgs states and/or loop corrections. 
One of the clear signatures of non-minimal Higgs sectors would be existence of charged Higgs bosons since they are predicted in extended Higgs models with multi Higgs doublet fields. 
Among various extended Higgs models, two Higgs doublet models (THDMs) are representative models that contain a pair of charged Higgs bosons, in addition to two CP-even Higgs bosons and a CP-odd Higgs boson. 

In the following, we briefly review the previous studies on the productions of the charged Higgs bosons at both hadron and lepton colliders, especially in \red{THDMs with softly-broken $Z_{2}$ symmetry.} (see e.g. \cite{Akeroyd:2016ymd, Djouadi:2005gj} as a comprehensive review).
\red{Due to the symmetry, there are four types of Yukawa interactions as described in Sec.~I\ref{sec:model}. Production modes of charged Higgs bosons with Yukawa interactions depend on the type of them. }
At both hadron and lepton colliders, charged Higgs bosons with a mass below the top-quark mass are produced via the decay of top quarks \cite{Chang:1978ke, Bigi:1986jk, Czarnecki:1992ig, Czarnecki:1992zm}.
Various hadronic production channels of the light charged Higgs bosons are comprehensively studied in Ref.~\cite{Aoki:2011wd}.
If charged Higgs bosons are heavier than the top quark, they can be produced mainly via single production processces at hadron colliders such as $pp \to H^{\pm}tb$ \cite{Degrande:2015vpa, Degrande:2016hyf}, $pp \to H^{\pm}W^{\mp}$ \cite{Dicus:1989vf, BarrientosBendezu:1998gd, BarrientosBendezu:1999vd, Moretti:1998xq, Brein:2000cv, Asakawa:2005nx, Eriksson:2006yt, Hashemi:2010ce} and $pp \to H^{\pm}h/H/A$ \cite{Kanemura:2001hz, Akeroyd:2003bt, Akeroyd:2003jp, Cao:2003tr, Belyaev:2006rf, Arhrib:2021xmc}.
In addition to these channels, there are single production process via $W^{\pm}Z$ fusion $pp\to W^{\pm*}Z^{*}X\to H^{\pm}X$ \cite{Asakawa:2005gv}, pair production process $pp\to H^{+}H^{-}$ \cite{Willenbrock:1986ry, Jiang:1997cg, Krause:1997rc, BarrientosBendezu:1999gp, Brein:1999sy}, same-sign pair production process $pp\to H^{+}H^{+}jj$ \cite{Aiko:2019mww, Arhrib:2019ywg} and so on.
At lepton colliders, charged Higgs bosons can be produced mainly via the pair production process $e^{+}e^{-}\to H^{+}H^{-}$ \cite{Grau:1981ig, Deshpande:1983xu, Komamiya:1988rs}~\red{\footnote{\red{ In addition, one-loop corrections to pair production of charged Higgs bosons via photon-photon collisions, i.e., $\gamma\gamma\to H^{+}H^{-}$ are studied in Refs.~\cite{Ma:1996nq,Demirci:2020zgt}.}}}.
The associated production with $W^{\pm}$ boson $e^{+}e^{-}\to H^{\pm}W^{\mp}$ has been studied in Refs.~\cite{Kanemura:1999tg, Arhrib:1999rg}.
There are also the associated productions with bottom quarks $e^{+}e^{-}\to b\bar{b}W^{\pm}H^{\mp}$ and $e^{+}e^{-}\to b\bar{b}H^{+}H^{-}$ \cite{Moretti:1997nh}.
Various channels of single productions of the charged Higgs bosons at lepton colliders are comprehensively studied in Ref.~\cite{Kanemura:2000cw}.

In the indirect probe of extended Higgs sectors, accurate calculations of the Higgs boson couplings are important since the effect of higher-order corrections can be comparable with the precise measurements in the future collider experiments, such as 
the HL-LHC~\cite{ApollinariG.:2017ojx}, the International Linear Collider (ILC)~\cite{Baer:2013cma,Fujii:2017vwa,Asai:2017pwp,Fujii:2019zll}, 
the Future Circular Collider (FCC-ee)~\cite{Gomez-Ceballos:2013zzn} and the Circular Electron Positron Collider (CEPC)~\cite{CEPC-SPPCStudyGroup:2015csa}. 
In Refs.~\cite{Aoki:2009ha,Kanemura:2014bqa}, it has been pointed out that various extended Higgs models can be discriminated by comparing patterns of deviations from the SM in Higgs boson couplings at tree level analysis. 
This study then has been extended including one-loop corrections~\cite{Kanemura:2004mg,Kanemura:2014dja,Kanemura:2015mxa,Kanemura:2015fra,Kanemura:2016lkz,Kanemura:2016sos,Kanemura:2017wtm}. 
In the context of THDMs, many studies for electroweak (EW) corrections to the Higgs boson couplings and/or decays have been performed~\cite{Arhrib:2003ph,Arhrib:2016snv,Kanemura:2004mg,Kanemura:2014dja,Kanemura:2015mxa,Kanemura:2017wtm,Kanemura:2018yai,Kanemura:2019kjg,Gu:2017ckc,Chen:2018shg,Han:2020lta,LopezVal:2010vk,Castilla-Valdez:2015sng,Xie:2018yiv, Altenkamp:2017ldc,Altenkamp:2017kxk,Altenkamp:2018bcs,Kanemura:2016sos,Arhrib:2015hoa,Krause:2019qwe}. 
Several numerical computation tools, e.g., {\tt H-COUP}~\cite{Kanemura:2017gbi,Kanemura:2019slf}, {\tt 2HDECAY}~\cite{Krause:2018wmo} and {\tt Prophecy4f}~\cite{Denner:2019fcr}, have been published. 

The current measurement of the discovered Higgs boson at \red{ATLAS}~\cite{Aad:2019mbh} and CMS \cite{CMS:2020gsy} shows that the properties of the discovered Higgs boson are close to those in the SM.
THDMs can fit with such situations by taking the alignment limit, where the coupling constants of the Higgs boson with SM particles approach the predictions in the SM.
Although direct searches of additional Higgs bosons at \red{ATLAS}~\cite{Aad:2020zxo, Aad:2019zwb, Aaboud:2018mjh, Aaboud:2018knk, Aaboud:2017gsl, Aad:2020fpj, Aad:2020tps, Aad:2021xzu, Aad:2020zxo} and CMS \cite{Khachatryan:2016qkc, Sirunyan:2018taj, Sirunyan:2019wph, Sirunyan:2017isc, Sirunyan:2019pqw, Sirunyan:2019xjg, Sirunyan:2020hwv, Sirunyan:2019hkq} give lower bounds of the masses, the decoupling properties have not been determined. 
Thus, two possibilities still remain, a case with a decoupling limit where additional Higgs bosons are sufficiently heavy or a case without decoupling limit where they exist at the EW scale. 
In short, alignment with or without decoupling scenario is favored, considering current situations from the LHC experiments. 
In Ref.~\cite{Aiko:2020ksl}, we have studied how these scenarios can be probed by the synergy of direct searches at (HL-) LHC and indirect searches at future lepton colliders. 
In particular, we found that rather wide parameter regions can be probed by additional Higgs boson decay channels such as $A\to Zh$~\cite{Barger:1991ed,Djouadi:1995gv} and $H\to hh$~\cite{Djouadi:1995gv} (also, see studies on next-to-leading order (NLO) corrections to the former process \cite{Chankowski:1992es,Krause:2019qwe} and the latter process \cite{Barger:1991ed,Chankowski:1992es,Osland:1998hv,Philippov:2006th,Williams:2007dc,Williams:2011bu,Krause:2016xku}) in nearly alignment regions. 
This is a contrast to the exact alignment scenario since these decay modes do not open in this scenario. 
In other words, in the nearly alignment regions predictions of additional Higgs boson decays can be drastically changed, depending on other model parameters. 
Considering such remarkable behaviors at leading order (LO), 
radiative corrections to additional Higgs bosons decays might be significant in the nearly alignment cases. 

In this paper, we dedicate ourselves to investigating the impacts of one-loop EW corrections to various charged Higgs boson decays in THDMs with softly-broken $Z_{2}$ symmetry. 
As earlier works for calculations of higher order corrections to charged Higgs boson decays, NLO EW corrections to the charged Higgs boson decay into the $W^\pm$ boson and the CP-odd Higgs boson $\Hpm \to W^{\pm}A$~\cite{Akeroyd:1998uw, Akeroyd:2000xa}, the decay into the $W^\pm$ boson and the CP-even Higgs bosons $\Hpm \to W^{\pm}h/H$~\cite{Santos:1996hs, Krause:2016oke} and the loop-induced decays at LO, $\Hpm\to W^{\pm}V$ ($V=Z,\gamma$)~\cite{Mendez:1990epa,CapdequiPeyranere:1990qk,Raychaudhuri:1992tm,Raychaudhuri:1994dc,Kanemura:1997ej,DiazCruz:2001tn,HernandezSanchez:2004tq,Hernandez-Sanchez:2004yid,Arhrib:2006wd,Arhrib:2007rm,Gastmans:2011ks,Abbas:2018pfp} have been studied. 
For charged Higgs boson decays into a pair of fermions, the NLO QCD corrections have been studied in THDMs~\cite{Li:1990ag,Mendez:1990jr}. 
In the context of the minimal supersymmetric standard model (MSSM), the NLO QCD corrections~\cite{Djouadi:1994gf,Jimenez:1995wf,Bartl:1995tx,Herrero:2001yg,Djouadi:2005gj} and the EW corrections~\cite{Diaz:1993kn} for these decays have been calculated. 
In addition, EW corrections to the decay into the $W^{\pm}$ boson and the CP-even Higgs boson, $\Hpm \to W^{\pm} h$ have been studied in the MSSM~\cite{Yang:2000sk}. 

We independently calculate NLO EW corrections to decay rates for the charged Higgs boson decays into a pair of fermions, $W^{\pm}$ boson and neutral Higgs bosons. 
Loop-induced decays, $\Hpm\to W^{\pm}V$, are calculated at LO in this paper. 
We present the explicit formulae for the decay rates with NLO EW corrections as well as QCD corrections.
The former is systematically written by the renormalized vertex functions for the charged Higgs bosons, and the analytical formulae are given in Appendix~\ref{sec:ApB}. 
All analytical formulae presented in this paper will be implemented in a new version of our developing program {\tt H-COUP~v3}~\cite{HCOUPv3}. 
With analytical formulae of the decay rates, we investigate theoretical behaviors for the EW corrections in detail to clarify what kind of contributions can be significant. 
We then discuss possibilities that four types of THDMs can be distinguished with the branching ratios of the charged Higgs bosons in the alignment scenarios. 
We also comprehensively investigate impacts of NLO EW corrections to the branching ratios of charged Higgs bosons in the alignment scenarios. 
In addition, we discuss effects of squared one-loop amplitudes to the decay into the $W^{\pm}$ boson and the lighter CP-even Higgs boson, $\Hpm \to W^{\pm} h$. 
They correspond to corrections at  next-to-next-to-leading order (NNLO) and can be significant in the alignment scenarios because the tree-level amplitude is suppressed in such scenarios. 

What is new in this paper is the following. 
First, we derive analytical formulae for the NLO EW corrections to various decay rates of charged Higgs bosons, using the improved on-shell renormalization scheme~\cite{Kanemura:2017wtm}, which has been applied in {\tt H-COUP}~\cite{Kanemura:2017gbi,Kanemura:2019slf}. 
Our calculations are performed independently of the above earlier works. 
Second, taking into account theoretical bounds and current experimental bounds on THDMs, we investigate impacts of the EW corrections to the branching ratios of charged Higgs bosons, while possible magnitudes of the EW corrections to decays of additional neutral Higgs bosons have been investigated in Refs.~\cite{Krause:2016xku, Krause:2019qwe,NeutralKKY}. 
Finally, we discuss how one can discriminate four types of THDMs via charged Higgs bosons decays in the nearly alignment regions. 
 
This paper is organized as follows. In Sec.~\ref{sec:model}, Lagrangian of THDMs are introduced, and the constraints for model parameter space are discussed. 
In Sec.~\ref{sec:decayrate}, we give formulae for the decay rates of the charged Higgs bosons after we define the vertex functions of the charged Higgs bosons. 
In Sec.~\ref{sec:IV}, we examine the theoretical behaviors of NLO EW corrections to the decay rates and model parameter dependence on the branching ratios with NLO corrections. 
Sec.~\ref{sec:V} is devoted to discussions for discrimination of four types of THDMs, the impact of NLO EW corrections to the branching ratios and the effect of squared one-loop amplitudes on $\Hpm \to W^{\pm} h$. 
Conclusions are given in \ref{sec:conclusions}. 
In Appendices, we give analytic expressions for scalar couplings, self-energies and vertex functions of the charged Higgs bosons, and decay rates with real photon emission.



%% file: 2_Model.tex

\section{THDMs with softly broken $Z_{2}$ symmetry}\label{sec:model}
We here define Lagrangian for THDMs with the softly broken $Z_{2}$ symmetry in order to specify input parameters and to fix notation. 
Relevant interactions for charged Higgs boson decays are presented. 
We also briefly review the theoretical and experimental constraints on the models. 

\subsection{Lagrangian}
In THDMs, there are two ${\rm SU(2)_{L}}$ doublet Higgs fields $\Phi_{1}$ and $\Phi_{2}$ with the hypercharge $Y=1/2$. 
The component fields are parametrized as
\begin{align}
\Phi_{i}=
\begin{pmatrix}
\omega^{+}_{i}\\
\frac{1}{\sqrt{2}}(v_{i}+h_{i}+i z_{i})
\end{pmatrix}
\quad(i=1,2),
\end{align}
where $v_{1}$ and $v_{2}$ denote the vacuum expectation values (VEVs) for $\Phi_{1}$ and $\Phi_{2}$, which give the electroweak VEV $v^{}=\sqrt{v_{1}^{2}+v_{2}^{2}}$. 
In order to avoid flavor changing neutral currents (FCNCs) at tree level, a softly-broken discrete $Z_{2}$ symmetry is imposed, for which the Higgs fields are transformed as $\Phi_{1}\to +\Phi_{1}$ and $\Phi_{2}\to -\Phi_{2}$~\cite{Glashow:1976nt, Paschos:1976ay}. 
The Higgs potential under the softly-broken $Z_{2}$ symmetry is given by 
\begin{align}
V & = m_1^2|\Phi_1|^2+m_2^2|\Phi_2|^2 - (m_3^2\Phi_1^\dagger \Phi_2 +\text{h.c.})\notag\\
& +\frac{\lambda_1}{2}|\Phi_1|^4+\frac{\lambda_2}{2}|\Phi_2|^4+\lambda_3|\Phi_1|^2|\Phi_2|^2+\lambda_4|\Phi_1^\dagger\Phi_2|^2
+\left[\frac{\lambda_5}{2}(\Phi_1^\dagger\Phi_2)^2+\text{h.c.}\right], \label{eq:pot-thdm}
\end{align}
where $m_{3}^{2}$ corresponds to the soft-breaking parameter of the $Z_{2}$ symmetry. 
We consider a CP-conserving potential, so that $m_{3}^{2}$ and $\lambda_{5}$ are taken to be real.  

Mass eigenstates for the charged Higgs and CP-odd Higgs sectors are obtained by the orthogonal transformation with the mixing angle $\beta$, which are defined by $\tan\beta=v_{1}/v_{2}$, as 
\begin{align}
\left(\begin{array}{c}
\omega^{+}_1\\
\omega^{+}_2
\end{array}\right)=R(\beta)
\left(\begin{array}{c}
G^{+}\\
H^{+}
\end{array}\right)
,\;
\left(\begin{array}{c}
z_1\\
z_2
\end{array}\right)=R(\beta)
\left(\begin{array}{c}
G^{0}\\
A
\end{array}\right)
,\quad
\mbox{with} ~
R(\theta)=
\begin{pmatrix}
c_{\theta} & -s_{\theta} \\
s_{\theta} & c_{\theta} \\
\end{pmatrix},
\end{align}
where we have introduced shorthanded notation, $s_{\theta}=\sin\theta$ and $c_{\theta}=\cos\theta$. 
The Nambu-Goldstone bosons $G^{\pm}$ and $G^{0}$ are accompanied by physical states, the charged Higgs bosons $\Hpm$ and the CP-odd Higgs boson $A$. 
The masses for $\Hpm$ and $A$ are given by
\begin{align}
m_{H^\pm}^2 = M^2 - \frac{v^2}{2}(\lambda_4+\lambda_5), \quad\quad
m_A^2    = M^2 - v^2\lambda_5, 
\end{align}
where the soft-breaking scale $M$ is given by $M=\sqrt{m_{3}^{2}/(s_{\beta}c_{\beta})}$. 
On the other hand, the CP-even Higgs sector is diagonalized by the rotation matrix with the mixing angle $\alpha$, 
 \begin{align}
\left(\begin{array}{c}
h_1\\
h_2
\end{array}\right)=R(\alpha)
\left(\begin{array}{c}
H\\
h
\end{array}\right). \label{eq:mixing:even}
\end{align}
Throughout this paper, we identify $h$ and $H$ as the observed Higgs boson with the mass 125 {\rm GeV} and an additional CP-even Higgs boson, respectively. 
Mass formulae for these states and the mixing angle $\alpha$ can be expressed by 
\begin{align}
m_H^2   & = M_{11}^2c^2_{\beta-\alpha} + M_{22}^2s^2_{\beta-\alpha} - M_{12}^2s_{2(\beta-\alpha)},\\ 
m_h^2   & = M_{11}^2s^2_{\beta-\alpha} + M_{22}^2c^2_{\beta-\alpha} + M_{12}^2s_{2(\beta-\alpha)},\\
&\tan 2(\beta-\alpha)= -\frac{2M_{12}^2}{M_{11}^2-M_{22}^2}, \label{eq:b-a}
\end{align}
using the mass matrix elements $M_{ij}$ in the basis of $(h_{1},h_{2}) R(\beta)$,
\begin{align}
M_{11}^2&=v^2(\lambda_1c^4_\beta+\lambda_2s^4_\beta +2\lambda_{345}s^2_{\beta}c^2_{\beta}),  \\
M_{22}^2&=M^2 + \frac{v^2}{4}s^2_{2\beta}(\lambda_1+\lambda_2-2\lambda_{345}), \label{m22} \\
M_{12}^2&=\frac{v^2}{2} s_{2\beta}( -\lambda_1c^2_\beta + \lambda_2s^2_\beta + \lambda_{345}c_{2\beta}), 
\end{align}
with $\lambda_{345}\equiv \lambda_3+\lambda_4+\lambda_5$.

Original eight parameters in the Higgs potential can be replaced by the physical parameters. 
While the mass $m_{h}$ of the Higgs boson and the electroweak VEV are fixed to be experimental values $m_{h}=125.09\ {\rm GeV}$ and $v= (\sqrt{2}G_{F})^{-1/2}\simeq 246\ {\rm GeV}$, we choose the following six parameters as inputs:
\begin{align}
m_H^{},~~m_A^{},~~m_{H^\pm}^{},~~M^2,~~\tan\beta,~~s_{\beta-\alpha}, \label{eq:thdminput}
\end{align} 
where regions of the mixing angles are taken to be $0\le\beta\le\pi/2$ and $0\le\beta-\alpha\le\pi$. 

\begin{table}[t] 
\begin{center}
\begin{tabular}{l||ccccc|ccc}\hline\hline
&$Q_L$&$L_L$&$u_R$&$d_R$&$e_R$&$\zeta_u$ &$\zeta_d$&$\zeta_e$ \\\hline\hline
Type-I &$+$&$+$&
$-$&$-$&$-$&$\cot\beta$&$\cot\beta$&$\cot\beta$ \\\hline
Type-II&$+$&$+$&
$-$
&$+$&$+$& $\cot\beta$&$-\tan\beta$&$-\tan\beta$ \\\hline
Type-X (lepton specific)&$+$&$+$&
$-$
&$-$&$+$&$\cot\beta$&$\cot\beta$&$-\tan\beta$ \\\hline
Type-Y (flipped) &$+$&$+$&
$-$
&$+$&$-$& $\cot\beta$&$-\tan\beta$&$\cot\beta$ \\\hline\hline
\end{tabular}
\caption{$Z_2$ charge assignments and $\zeta_f$ ($f=u,d,e$) factors~\cite{Aoki:2009ha} in four types of THDMs. 
\label{tab:z2}
}
\end{center}
\end{table}

Four types of THDMs are characterized by the shape of Yukawa interactions~\cite{Barger:1989fj,Aoki:2009ha}. 
Under the $Z_{2}$ symmetry it is generally written in terms of $\Phi_{1}$ and $\Phi_{2}$ by 
\begin{align}\label{eq:Yukawa}
{\mathcal L}_Y =
&-Y_{u}{\bar Q}_L\tilde{\Phi}_uu_R^{}
-Y_{d}{\bar Q}_L\Phi_dd_R^{}
-Y_{e}{\bar L}_L\Phi_e e_R^{}+\text{h.c.},
\end{align}
where $\tilde{\Phi}^{}_{u}= i\sigma_2 \Phi^{*}_{u}$, and $\Phi_{u,d,e}$ denote $\Phi_1$ or $\Phi_2$. 
While the ${\rm SU(2)_{L}}$ quark doublet $Q_{L}$ and the ${\rm SU(2)_{L}}$ lepton doublet $L_{L}$ are assigned by $Z_{2}$ even, there are four possible $Z_{2}$ charge assignment for right-handed quarks and leptons~\cite{Barger:1989fj,Aoki:2009ha}, as given in Table.~\ref{tab:z2}. 
We call them Type-I, Type-II, Type-X and Type-Y, respectively~\cite{Aoki:2009ha}. 
From the original Lagrangian Eq.~\eqref{eq:Yukawa}, one can extract Yukawa interactions for the charged Higgs bosons as
\begin{align}
{\cal L}_Y^{\Hpm} &=
+ \frac{\sqrt{2}}{v}\left[ \overline{u}
(m_{u} V_{ud} \zeta_u P_L - V_{ud} m_{d} \zeta_d P_R)dH^+
- \zeta_e\overline{\nu}m_{e}  P_R e H^+
+\text{h.c.} \right], 
\end{align}
with Cabibbo-Kobayashi-Maskawa (CKM) matrix elements $V_{ud}$,
where the projection operators $P_{L/R}=(1\mp \gamma_{5})/2$, and $\zeta_f$ ($f=u,d,e$) factors are presented in Table.~\ref{tab:z2}.

Kinetic terms for the Higgs doublet fields are given by
\begin{align}
\mathcal{L}_{\rm kin}=|D_{\mu}\Phi_{1}|^{2}+|D_{\mu}\Phi_{2}|^{2}
\end{align}
with the covariant derivative $D_{\mu}=\partial_{\mu} -ig I^{a}W_{\mu}^{a}-ig^{\prime}YB_{\mu}$, where $I^{a}$ $(a=1,2,3)$ and $Y$ are the ${\rm SU(2)_{L}}$ generators and the hypercharge for a field, respectively. 
Derivative couplings for the charged Higgs bosons with neutral Higgs bosons and a $W^\pm$ boson are obtained as
\begin{align}
\mathcal{L}_{\rm kin}^{\Hpm}=g_{\phi H^{\pm} W^{\mp}} \big\{(\partial^{\mu} \phi) H^{\pm} -\phi(\partial^{\mu} H^{\pm})\big\} W^{\mp}_{\mu},~~ (\phi=h,H,A). 
\end{align}
The corresponding couplings are given by
\begin{align}\label{eq:HpmWS}
g_{h H^\pm W^{\mp} }^{}&= \mp i \frac{m_W}{v}c_{\beta-\alpha}, \ \ \ 
g_{H H^\pm W^{\mp} }^{}= \pm i \frac{m_W}{v}s_{\beta-\alpha},\ \ \ 
g_{A H^\pm W^{\mp} }^{} =- \frac{m_W}{v}.
\end{align} 

\subsection{Constraints }\label{sec:constraint}
We here discuss theoretical constraints and experimental constraints on THDMs. 

\noindent
$\bullet$ Theoretical constraints

As a theoretical constraint, perturbative unitarity~\cite{Kanemura:1993hm, Akeroyd:2000wc, Ginzburg:2005dt, Kanemura:2015ska} and vacuum stability bounds at tree level~\cite{Deshpande:1977rw, Klimenko:1984qx, Sher:1988mj, Nie:1998yn, Kanemura:1999xf} are taken into account in our study.  
The perturbative unitarity gives constraints on s-wave amplitudes for two-body to two-body scattering processes of scalar boson states in the high energy limit, that is, the eigenvalues of the s-wave amplitudes $a_{i}$ should satisfy the conditions; $|a_{i}|<1/2$. 
For the vacuum stability bounds, it is required that the Higgs potential is bounded from below in any direction at large values of Higgs fields.
They are expressed in terms of the scalar couplings $\lambda_{1-5}$. 
Hence, these theoretical constraints bound all the model inputs parameters in Eq.~\eqref{eq:thdminput}. 
In particular, in the case of $s_{\beta-\alpha}\simeq1$, these theoretical bounds are sensitive to the difference between $M^{2}$ and squared masses of the additional Higgs bosons, as seen from concrete expressions of $\lambda_{1-5}$ in Appendix.~A in Ref.~\cite{Aiko:2020ksl}. 

\noindent
$\bullet$ Electroweak precision tests

Constraints from electroweak precision tests are imposed via the S and T parameters, which were proposed by Peskin and Takeuchi~\cite{Peskin:1990zt,Peskin:1991sw}. 
We use the experimental data for these parameters given in Ref.~\cite{Baak:2012kk}, i.e., $S=0.05\pm 0.09$ and $T=0.08\pm 0.07$ with the $U$ parameter being fixed as $U=0$. 
The new physics contributions to these electroweak precision parameters in THDMs are defined by $\Delta S=S_{\rm THDM}-S_{\rm SM}$ and $\Delta T= T_{\rm THDM}-T_{\rm SM}$. 
The analytical formulae are given in Refs.~\cite{Bertolini:1985ia,Grimus:2008nb,Kanemura:2011sj,Kanemura:2015mxa}. 
We impose the constraints from the S and T parameters by using {\tt H-COUP}, in which the numerical computations of $\Delta S$ and $\Delta T$ are implemented. 

\noindent
$\bullet$ Measurement of the discovered Higgs boson

One of the important bounds from the current data of collider experiments is measurements of signal strength for the Higgs boson with a mss of 125 {\rm GeV}. 
The current data from the ATLAS~\cite{Aad:2019mbh} and CMS~\cite{CMS:2020gsy} indicate that the Higgs boson couplings with weak gauge bosons and fermions should be close to the predictions in the SM. 
This situation can be realized by taking so-called the alignment limit, i.e., $\sba=1$, in which the Higgs boson couplings coincide with those of the SM at tree level, i.e., $g_{hVV}^{\rm THDM}= g^{\rm SM}_{hVV}$ and $g_{hff}^{\rm THDM}=g^{\rm SM}_{hff}$. 
Also, nearly alignment regions $\sba\simeq 1$ still can fit the current bounds at the LHC Run II, choosing a suitable value of $\tb$. 

\noindent
$\bullet$ Direct searches of additional Higgs bosons

In Ref.~\cite{Aiko:2020ksl}, constraints from direct searches of additional Higgs bosons at the LHC 13 TeV have been investigated in the alignment limit and nearly alignment regions for all the types of THDMs, in which
upper bounds for production cross sections times branching ratios with 36 ${\rm fb^{-1}}$ data are used. 
By the charged Higgs boson search with $H^{+}\to t\bar{b}$, the lower bounds of $\mHp \gtrsim 500~(950)~{\rm GeV}$ are given at $\tan\beta =1~(0.5)$ for the all types of THDMs without depending on the values of $\sba$. 
The channel $A\to \tau^{+}\tau^{-}$ gives the lower bounds on $m_{A}$ for Type-II and Type-X in the alignment limit, e.g., $m_{A} \gtrsim 350~(370)~{\rm GeV}$ for $\tan\beta\lesssim2.3~(8)$ in Type-II (X). 
In nearly alignment regions, much tighter bounds on $m_{A}$ are given from the channel $A\to Zh$, e.g., $m_{A}\gtrsim 940~(1100)~{\rm GeV}$ at $s_{\beta-\alpha}=0.995$ and $\tan\beta=1$ for Type-I (II) and Type-X (Y).

\noindent
$\bullet$ Flavor constraints

Charged Higgs bosons are related to the $B$ meson flavor violating decay $B\to X_{s}\gamma$~\cite{Misiak:2017bgg,Misiak:2020vlo}, which gives severe constraints especially for Type-II and Type-Y. 
The excluded regions for these models are $m_{H^{\pm}}\lesssim$ 800{\rm GeV} in $\tan\beta>2$~\cite{Misiak:2017bgg,Misiak:2020vlo}.
For Type-I and Type-X constraints by this channel are weaker than Type-II and Type-Y. The excluded regions are only given in lower $\tan\beta$ regions with $\tan\beta<2$, e.g, $m_{H^{\pm}}\lesssim 400~(180)~{\rm GeV}$ at $\tan\beta=1~(1.5)$~\cite{Misiak:2017bgg}. 
On the other hand, for Type-II, high $\tan\beta$ regions are constrained by the $B$ meson rare leptonic decay $B_{s}\to \mu^{+}\mu^{-}$~\cite{Cheng:2015yfu}. At $\mHp=800~{\rm GeV}$, The region with $\tan\beta\gtrsim 10$ is excluded~\cite{Haller:2018nnx}. 
Also, comprehensive studies for constraint from various flavor observables such as $B$ meson decays, $D$ meson decays and $B_{0}-\bar{B}_{0}$ mixing, are performed in Refs.~\cite{Haller:2018nnx,Enomoto:2015wbn}. 



%% file: 3_Decay_rate.tex

\section{Decay rates with higher order corrections }\label{sec:decayrate}
In this section, we explain calculations of the decay rates with NLO EW corrections for two-body decay of the charged Higgs bosons, i.e., $\Hpm\to f\bar{f^{\prime}} $ and $\Hpm\to W^{\pm}\phi$ $(\phi=h,H,A)$. For the decay into a pair of quarks $ \Hpm\to q\bar{q}^{\prime} $, QCD corrections up to NNLO are presented. 
Formulae for the loop-induced decay processes $\Hpm\to W^{\pm}V$ $(V=Z,\gamma)$ are given at LO. 

For the computations of the NLO EW corrections to the charged Higgs boson decays, we adopt the improved on-shell renormalization scheme~\cite{Kanemura:2017wtm}. 
While we do not give all descriptions for the renormalization scheme in this paper, we here highlight the main points, for details see Ref.~\cite{Kanemura:2017wtm}. 
In the Higgs sector, there are six free parameters given in Eq.~\eqref{eq:thdminput}. 
Together with the wave function renormalization constants, the masses of the additional Higgs bosons and the mixing angles are renormalized by the on-shell conditions for the Higgs bosons in the mass eigenstates. 
However, the gauge dependence appears in renormalization of mixing angles~\cite{Yamada:2001px}, which are resolved by applying the pinch technique~\cite{Krause:2016oke, Kanemura:2017wtm}. 
The remaining parameter in the Higgs sector, $M^{2}$, is renormalized by the minimal subtraction~\cite{Kanemura:2004mg}. 
On the other hand, renormalization of the gauge sector and the fermion sector are performed by using on-shell conditions~\cite{Hollik:1988ii}. 
In our calculation for $ \Hpm\to q\bar{q}^{\prime} $, we do not take into account contributions arising by quark mixing at one-loop level, which is always suppressed by the CKM matrix element. 
Hence renormalization of the CKM matrix does not have to be carried out. 

While the UV divergences are correctly removed by virtue of the mentioned above renormalization, IR divergences appear in one-loop diagrams containing a virtual photon, which cancel with those of real photon emission diagrams. 
We regularize them by introducing a small photon mass $\mu$, and numerically confirm that IR divergences are canceled out when virtual corrections and real emissions are summed up. 

\subsection{Form factors for vertex functions of charged Higgs bosons}
Before we give formulae for decay rates with higher corrections, we first define renormalized vertex functions of the charged Higgs bosons. All NLO EW corrections are expressed in terms of the form factors of the vertex functions.
\subsubsection{$H^{\pm} ff^\prime$ vertex}\label{sec:Hpffpvetex}
\begin{figure}[t]
 \centering
 \includegraphics[scale=0.45]{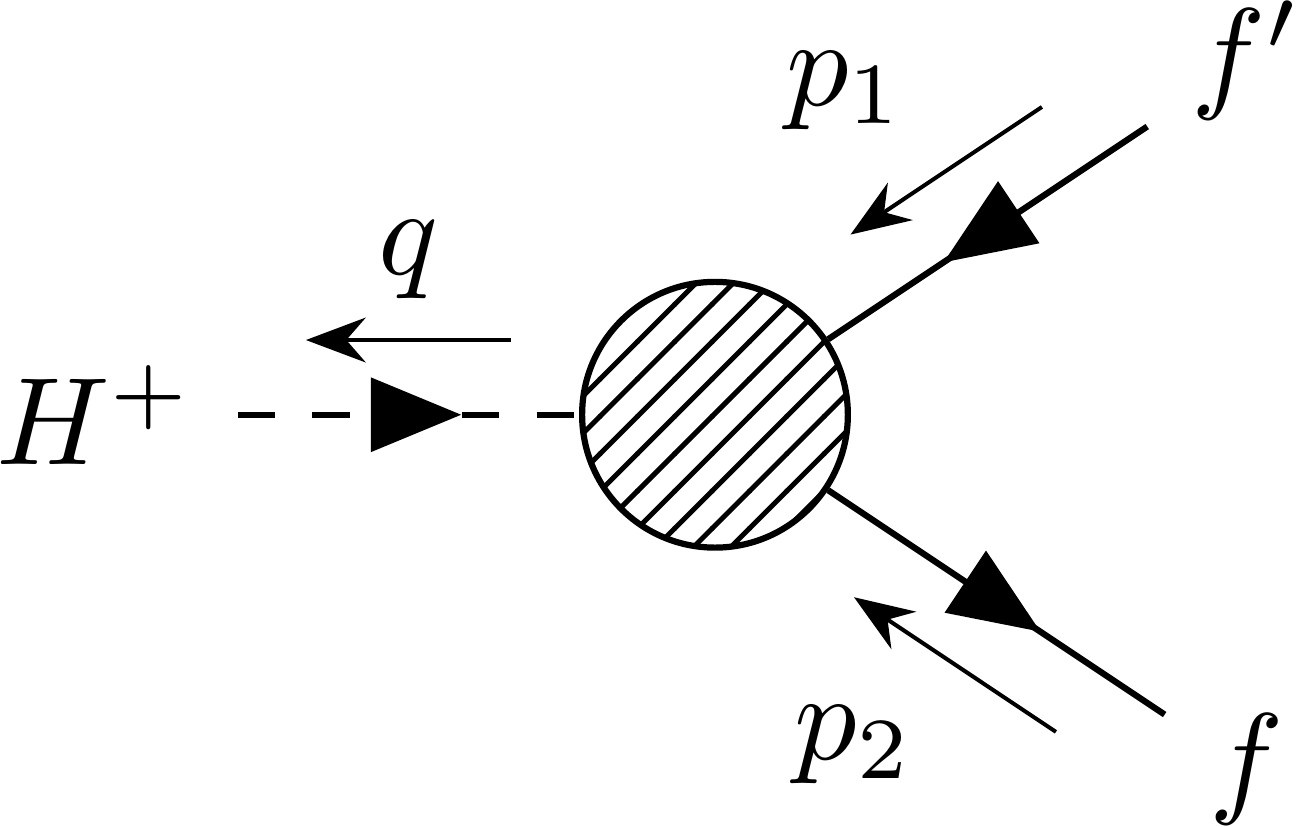}
 \caption{Momentum assignment for the renormalized $H^{+}f\bar{f}'$ vertex.}
 \label{Mom_Hpm_to_ffp}
\end{figure}

In the computations of the charged Higgs boson decays into two fermions $H^\pm \to f\bar{f^\prime}$, the renormalized $\Hpm ff^\prime$ vertex functions are needed. 
In general, the vertex functions can be expressed as \cite{Kanemura:2014dja}
\begin{align}
\hat{\Gamma}_{H^\pm ff^\prime}&=\hat{\Gamma}_{H^\pm ff^\prime}^{\rm S}+\gamma_5\hat{\Gamma}_{H^\pm ff^\prime}^{\rm P}+\slashed p_1\hat{\Gamma}_{H^\pm ff^\prime}^{\rm V1}+\slashed p_2\hat{\Gamma}_{H^\pm ff^\prime}^{\rm V2} \notag \\
&+\slashed p_1\gamma_5\hat{\Gamma}_{H^\pm ff^\prime}^{\rm A1}+\slashed p_2\gamma_5\hat{\Gamma}_{H^\pm ff^\prime}^{\rm A2}+\red{\slashed p_1\slashed p_2\hat{\Gamma}_{H^\pm ff^\prime}^{\rm T}}+\slashed p_1\slashed p_2\gamma_5\hat{\Gamma}_{H^\pm ff^\prime}^{\rm PT}, 
\end{align}
where $p_1$ and $p_2$ denote incoming momenta of a fermion $f^{\prime}$ and the SU(2) partner $f$, and $q$ is the outgoing momentum of the charged Higgs bosons (see Fig.~\ref{Mom_Hpm_to_ffp}). 
The renormalized form factors can be commonly written by the tree-level part and the one-loop part as
\begin{align}
\hat{\Gamma}^X_{\Hpm ff^\prime}={\Gamma}^{X,{\rm tree}}_{\Hpm ff^\prime}+{\Gamma}^{X,{\rm loop}}_{\Hpm ff^\prime},\ \ \ \mbox{(X=S,\ P,\ V1,\ V2,\ A1,\ A2,\ T,\ PT)},
\end{align}
where the one-loop parts are further divided into contributions from 1PI diagrams and counterterms, i.e., 
$\red{{\Gamma}^{X,{\rm loop}}_{\Hpm ff^\prime}}={\Gamma}^{X,{\rm 1PI}}_{\Hpm ff^\prime}+\delta {\Gamma}^{X}_{\Hpm ff^\prime}$. 
The 1PI diagrams contributions ${\Gamma}^{X,{\rm 1PI}}_{\Hpm ff^\prime}$ are given in Appendix~\ref{sec:ApB}. 

The tree-level couplings for $H^\pm ff^\prime$ vertices are given by,
\begin{align}
&{\Gamma}^{S,{\rm tree}}_{H^\pm ff^\prime}=\red{\pm}\frac{V_{f\fp}}{\sqrt{2}v}(m_f\zeta_f-m_{f^\prime}\zeta_{f^\prime}), \ \ \ 
{\Gamma}^{P,{\rm tree}}_{H^\pm ff^\prime}=\mp\frac{V_{f\fp}}{\sqrt{2}v}(m_f\zeta_f+m_{f^\prime}\zeta_{f^\prime}), \ \ \ \\
&{\Gamma}^{X,{\rm tree}}_{H^\pm ff^\prime}=0\ \ \ (X\neq S,P). 
\end{align}
The counterterms for $H^\pm f\fp$ vertices are presented by 
 \begin{align}
 \delta \Gamma^{S/P}_{\Hpm f\fp}=\frac{1}{2}(\delta\Gamma^R_{\Hpm f \fp}\pm\delta\Gamma^L_{\Hpm f \fp}),
 \end{align}
 with
 \begin{align}
 \label{eq:CTHPffpR}
 \delta \Gamma^{R}_{\Hpm f\fp}&=\mp\frac{\sqrt{2}m_{\fp}}{v}\zeta_{\fp}\left[\frac{\delta m_{\fp}}{m_{\fp}}-\frac{\delta v}{v}-\zeta_{\fp}\delta \beta+ \frac{\delta Z^{\fp}_R+\delta Z^{f}_L}{2}+\frac{\delta Z_{\Hpm}}{2} \right. \notag \\
 &+\left.\frac{1}{\zeta_{{\fp}}}\delta C_{\Hpm}-\left(\frac{1}{\zeta_{\fp}}+\zeta_{\fp}\right)\delta \beta^{\rm PT}\right], \\  
 \label{eq:CTHPffpL}
\delta \Gamma^{L}_{\Hpm f\fp}&=\pm\frac{\sqrt{2}m_{f}}{v}\zeta_{f}\left[\frac{\delta m_{f}}{m_{f}}-\frac{\delta v}{v}-\zeta_{f}\delta \beta+ \frac{\delta Z^{f}_R+\delta Z^{{\fp}}_L}{2}+\frac{\delta Z_{\Hpm}}{2} \right. \notag \\
&+\left. \frac{1}{\zeta_{f}}\delta C_{\Hpm}-\red{\left(\frac{1}{\zeta_{f}}+\zeta_{f}\right)}\delta \beta^{\rm PT}\right], 
 \end{align}
where concrete expressions for the counterterms for fermion masses $\delta m_{f^{(\prime)}}^{2}$, the wave function renormalization $\delta Z_{R/L}^{f^{(\prime)}}$, the electroweak VEV $\delta v$, and the mixing angle $\delta \beta$ are presented in Appendix C of Ref.~\cite{Kanemura:2017wtm}. 
 The wave function renormalization constants of the charged Higgs bosons are expressed by
 \begin{align}
 \delta Z_{\Hpm}&=-\left.\frac{d}{dq^{2}}\Pi_{H^{+}H^{-}}^{\rm 1PI}(q^{2})\right|_{q^{2}=\mHp^{2}},\\
 \delta C_{\Hpm}&=\delta \beta+\red{\frac{1}{\mHp^{2}}\left(\Pi_{H^{+}G^{-}}^{\rm 1PI}(0)
 +\sba\frac{T^{\rm 1PI}_{H}}{v}- \cba\frac{T^{\rm 1PI}_{h}}{v} \right)},
 \end{align}
 where 1PI diagram contributions to two point functions for the charged Higgs boson and $H^{+}$-$G^{-}$ mixing, $\Pi_{H^{+}H^{-}}^{\rm 1PI}(q^{2})$ and $\Pi_{H^{+}G^{-}}^{\rm 1PI}(q^{2})$ are given in Appendix \ref{sec: self}. 
 Those to the one-point functions $T^{\rm 1PI}_{H}$ and $T^{\rm 1PI}_{h}$ are presented in Appendix C of Ref.~\cite{Kanemura:2015mxa}. 
The pinch term for the mixing angle in the Feynman gauge $\delta \beta^{\rm PT}$ is given by
\begin{align}
\delta \beta^{\rm PT}&= -\frac{1}{2m_{A}^{2}}
\left(\Pi_{AG^{0}}^{\rm PT}(m_{A}^{2})+\Pi_{AG^{0}}^{\rm PT}(0)\right).
\end{align}
 The last term in Eqs.~\eqref{eq:CTHPffpR} and \eqref{eq:CTHPffpL} corresponds to a subtraction term due to the following reason. 
The pinch terms for the mixing angle are only required for the counterterms arising from the shift of the couplings, and, for those arising from the shift of scalar fields, we use the counterterm $\delta \beta$ without the pinch terms. 

\subsubsection{$H^{\pm} W^{\mp}\phi $ vertex}
\begin{figure}[t]
 \centering
 \includegraphics[scale=0.45]{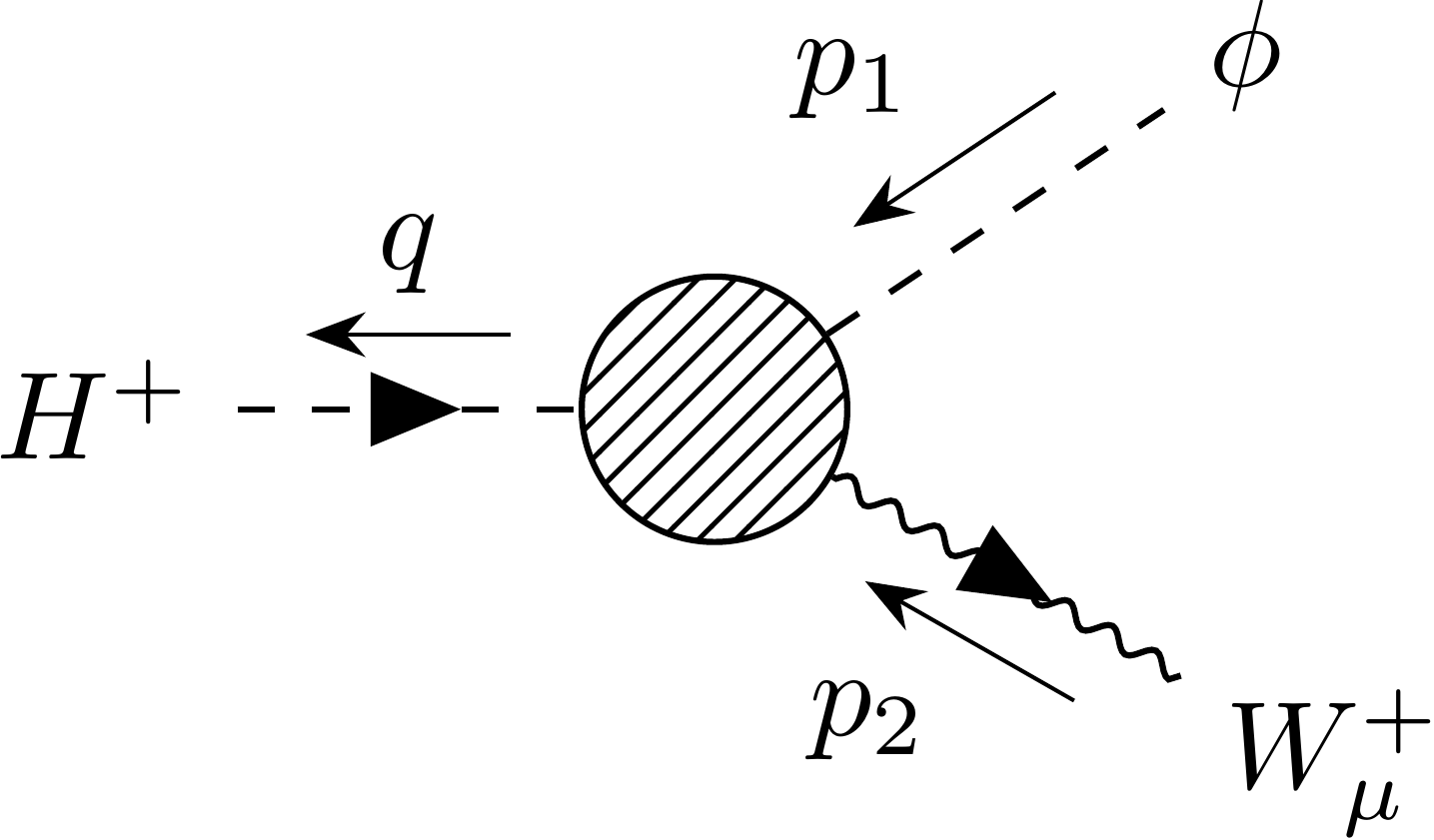}
 \caption{Momentum assignment for the renormalized $H^{+}W^{-\mu}\phi$ vertex.}
 \label{Mom_Hpm_to_PhiW}
\end{figure}

For calculations of charged Higgs boson decays into a vector boson and a scalar boson $H^\pm\to W^{\pm}\phi$ ($\phi=h,H,A$), renormalization of $H^{\pm} W^{\mp}\phi$ vertex are needed. 
the vertex functions can be commonly expressed as 
\begin{align}
\hat{\Gamma}^\mu_{H^{\pm} W^{\mp}\phi }(p_1^2,p^2_2,q^2)= (p_1+q)^\mu\hat{\Gamma}_{H^{\pm} W^{\mp}\phi }, \ \ \ \ 
\end{align}
where $p_1^\mu$ and $p_2^\mu$ denote the incoming momenta of a scalar boson $\phi$ and a $W^{\pm}$ bosons, respectively.
A momentum $q^\mu$ denotes the outgoing one of the charged Higgs boson (see Fig.~\ref{Mom_Hpm_to_PhiW}). 
Since we here assume that the external gauge bosons are on-shell, the term proportional to $p_2=q-p_1$ vanishes due to the orthogonality of the polarization vector.
Similar to the $\Hpm f\bar{f}^{\prime}$ vertex, the renormalized form factor can be decomposed as
\begin{align}
\hat{\Gamma}_{H^{\pm} W^{\mp}\phi}= \Gamma_{H^{\pm} W^{\mp}\phi}^{\rm tree}+\Gamma_{H^{\pm} W^{\mp}\phi}^{\rm loop}.
\end{align}
with $\Gamma_{H^{\pm} W^{\mp}\phi}^{\rm tree}=-i g_{\phi H^{\pm} W^{\mp} }$.  
The one-loop part $\Gamma_{H^{\pm} W^{\mp}\phi }^{\rm loop}$ is schematically expressed in terms of 1PI diagram $\Gamma_{H^{\pm} W^{\mp}\phi }^{\rm 1PI}$ and the counterterm $\delta \Gamma_{H^{\pm}\phi W^{\mp} }$ as
\begin{align}
\Gamma_{H^{\pm} W^{\mp}\phi}^{\rm loop}= \Gamma_{H^{\pm} W^{\mp}\phi }^{\rm 1PI} +\delta \Gamma_{H^{\pm} W^{\mp}\phi }. 
\end{align}
The 1PI diagram contributions to these vertex functions are given in Appendix \ref{sec:ApB}. 

The counterterms are expressed as 
\begin{align}
\notag
 \delta \Gamma_{ H^\pm W^{\mp} h}&= \mp\frac{m_W}{v}c_{\beta-\alpha}\Big[
\frac{\delta m_W^2}{2 m_W^2}-\frac{\delta v}{v}+\frac{1}{2}(\delta Z_W+\delta Z_{H^\pm}+\delta Z_h) \\ 
 &+\tan(\beta-\alpha)(\delta\beta^{\rm PT}-\delta\alpha^{\rm PT}+\delta C_{H^\pm}-\delta C_h)  
 \Big], \\ \notag
 \delta \Gamma_{ H^\pm W^{\mp} H}&= \pm\frac{m_W}{v}s_{\beta-\alpha}\Big[
\frac{\delta m_W^2}{2 m_W^2}-\frac{\delta v}{v}+\frac{1}{2}(\delta Z_W+\delta Z_{H^\pm}+\delta Z_H) \\ 
 &+\cot(\beta-\alpha)(\delta \beta^{\rm PT}+\delta \alpha^{\rm PT}-\delta C_{H^\pm}-\delta C_h)  
 \Big],\\
 \delta \Gamma_{ H^\pm W^{\mp} A}&= \red{+i}\frac{m_W}{v}\Big[
\frac{\delta m_W^2}{2 m_W^2}-\frac{\delta v}{v}+\frac{1}{2}(\delta Z_W+\delta Z_{H^\pm}+\delta Z_A)  
  \Big], 
\end{align}
where the $\delta \alpha^{\rm PT}$ denotes the pinch term contribution to $\delta \alpha$.  
In the Feynman gauge, it is expressed by
\begin{align}
\delta \alpha^{\rm PT}&= \frac{1}{2(m_{H}^{2}-m_{h}^{2})}
\left(\Pi_{Hh}^{\rm PT}(m_{h}^{2})+\Pi_{Hh}^{\rm PT}(m_{H}^{2})\right). 
\end{align}
Inclusion of the terms with $\delta \alpha^{\rm PT}$ and $\delta \beta^{\rm PT}$ has the same reason with the one presented in Sec.~\ref{sec:Hpffpvetex}. 
We again refer Appendix C of Ref.~\cite{Kanemura:2017wtm} for concrete expressions to 
the $W^{\pm}$ boson mass counterterm $\delta m_{W}$ and the
 wave function renormalization constants for the $W^{\pm}$ boson $\delta Z_{W}$ and the Higgs bosons
$\delta C_{h},~\delta Z_{\phi}$ $(\phi=h,H,A)$. 
\subsubsection{$H^{\pm}VW^{\mp}$ vertex} 
\begin{figure}[t]
 \centering
 \includegraphics[scale=0.45]{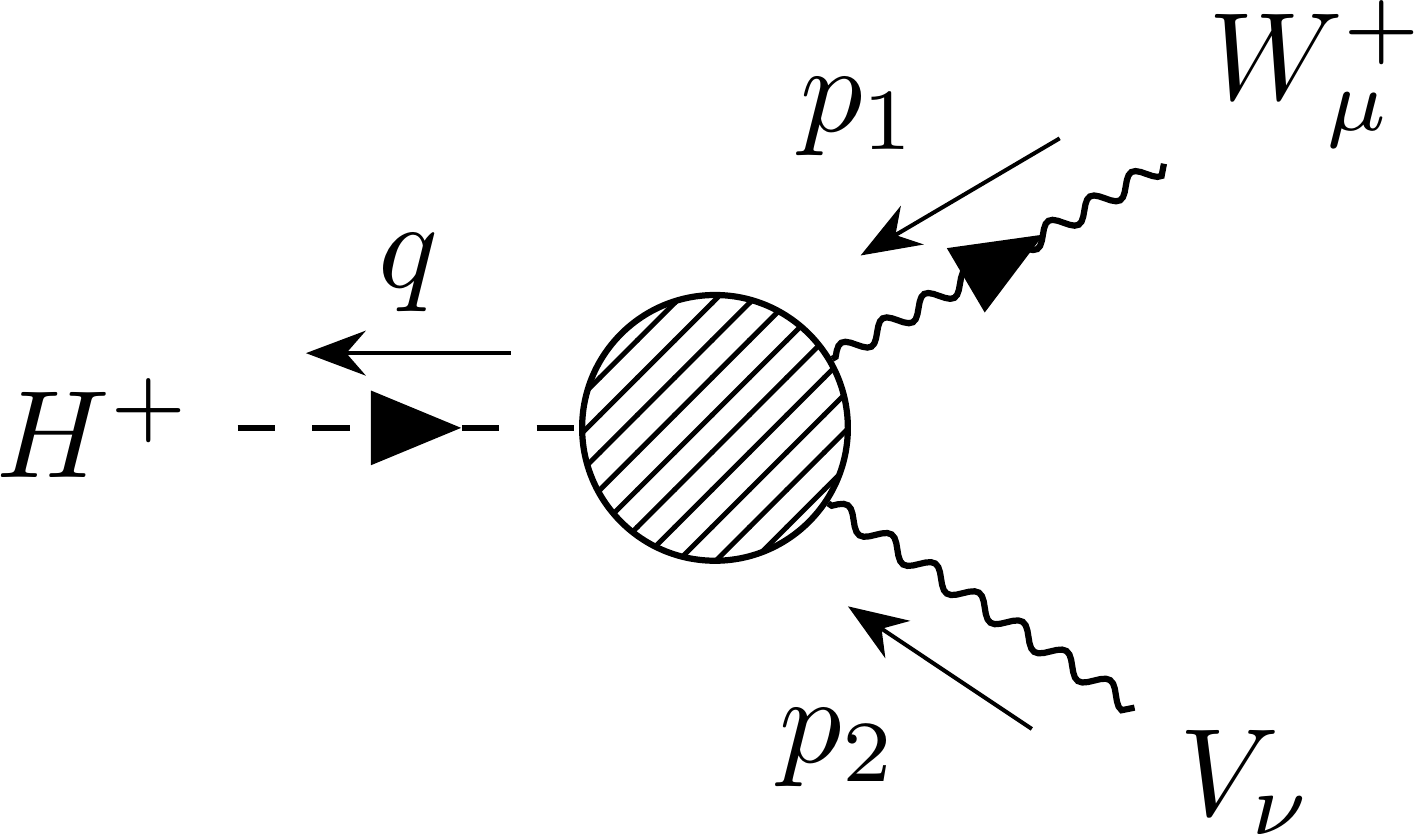}
 \caption{Momentum assignment for the renormalized $H^{+}V^{\mu}W^{-\nu}$ vertex.}
 \label{Mom_Hpm_to_WV}
\end{figure}

For computations of loop induced charged Higgs decays, i.e., $H^{\pm}\to ZW^{\pm}$ and $H^{\pm}\to \gamma W^{\pm}$, $H^{\pm}VW^{\mp}$ ($V=Z,\gamma$) vertex appears. 
The renormalized tensor vertex functions can be expressed by~\cite{CapdequiPeyranere:1990qk,Kanemura:1997ej}
\begin{align}
\label{eq:HVW}
\Gamma_{H^{\pm}VW^{\mp}}^{\mu\nu}=g^{\mu\nu}\Gamma_{H^{\pm}VW^{\mp}}^{1}+p_{1}^{\nu}p_{2}^{\mu}\Gamma_{H^{\pm}VW^{\mp}}^{2}
+i\epsilon^{\mu\nu\rho\sigma}p_{1\rho}^{}p_{2\sigma}\Gamma_{H^{\pm}VW^{\mp}}^{3},
\end{align}
where $p_{1}^{\mu}$ denotes an incoming momentum for a $W^{\pm}$ boson and $p_{2}^{\nu}$ is that of a Z boson or a photon, and $q$ is the outgoing momentum of the charged Higgs bosons (see Fig.~\ref{Mom_Hpm_to_WV}). 
In this expression, relations for on-shell vector bosons, $\epsilon_{\mu}(p_{1})p_{1}^{\mu}=\epsilon_{\nu}(p_{2})p_{2}^{\nu}=0$, have been applied. 
Since $H^{\pm}ZW^{^{\mp}}$ and $H^{\pm}\gamma W^{\mp}$ vertices do not exist at the tree level, form factors are written by the loop contributions,
\begin{align}
\Gamma_{H^{\pm}VW^{\mp}}^{i}=\Gamma_{H^{\pm}VW^{\mp}}^{i,{\rm loop}}=\Gamma_{H^{\pm}VW^{\mp}}^{i,{\rm 1PI}}+\delta \Gamma_{H^{\pm}VW^{\mp}}^{i}.
\end{align}
The 1PI diagram contributions are given in Appendix \ref{sec:ApB}. 
While there is no tree-level contributions, counterterms are introduced from $G^{\pm}\gamma W^{\mp}$ vertices though the mixing with charged Higgs bosons. It is expressed by
\begin{align}
\delta \Gamma_{H^{\pm}VW^{\mp}}^{1}=g_{G^{\pm}W^{\mp}V}(\delta C_{G^{+}H^{-}}+\delta \beta), \quad\quad
\delta \Gamma_{H^{\pm}VW^{\mp}}^{2,3}=0,
\end{align}
where the tree-level couplings for NG boson are $g_{G^{\pm}W^{\mp}\gamma}=eg_{}v/2$ and $g_{G^{\pm}W^{\mp}Z}=-gg_{Z}v s_{W}^{2}/2$.

\subsection{Decay rates of $H^{\pm}\to ff^{\prime}$}
The decay rates for charged Higgs boson decays into a pair of fermions with NLO EW corrections and QCD corrections can be written as 
\begin{align}\label{eq:hpffpNLO}
\Gamma(\Hpm\to f\bar{f}^{\prime})&= 
\frac{\red{N^{f}_{c}}\mHp |V_{f\fp }|^{2} }{8\pi v^{2}}\lambda^{\frac{1}{2}} (\mu_{f},\mu^{}_{\fp})
\Bigg[
(1-\mu_{f}-\mu_{\fp})\Big\{\red{m_{\fp}^{2}\zeta_{\fp}^{2}}\Big(1+\Delta_{RR}^{\rm QCD}
+\Delta_{RR}^{\rm EW}\Big) \notag \\
&+\red{m_{f}^{2}\zeta_{f}^{2}\Big(1+\Delta_{LL}^{\rm QCD}+\Delta_{LL}^{\rm EW} \Big)\Big\}}
+4\sqrt{\mu_{f}\mu_{\fp}}m_{f}m_{\fp}\zeta_{f}\zeta_{\fp}\Big(1+\Delta_{RL}^{\rm QCD}+\Delta_{RL}^{\rm EW}\Big) 
\Bigg] \notag \\
&+\Gamma(H^{\pm}\to f\bar{f}^{\prime} \gamma),
\end{align}
with $\mu_{f^{(\prime)}}=m_{f^{(\prime)}}^{2}/\mHp^{2}$ and the color factor $\red{N_{c}^{f}}= 3 ~(1)$ for quarks (leptons). 
The kinematical factor $\lambda(x,y)$ is defined by
\begin{align}
\lambda(x,y)=(1-x-y)^{2}-4xy.
\end{align}
 Factors $\Delta_{XX}^{\rm QCD}$ and $\Delta_{XX}^{\rm EW}$ $(XX=LL,\;RR,\;RL)$ denote QCD correction parts and EW correction parts, respectively. 
In this expression, the real photon emission contribution is included, by which the IR divergences are removed. 
The analytic formula is given in Appendix~C. 
The LO decay rate $\Gamma(H^{\pm}\to f\bar{f}^{\prime})_{\rm LO}$ is easily obtained by omitting all NLO contributions, $\Delta_{XX}^{\rm QCD/EW}\to 0$ and $\Gamma(H^{\pm}\to f\bar{f}^{\prime} \gamma)\to 0$.  

The EW corrections parts are written by
\begin{align}
\label{eq:DRR}
\Delta^{\rm EW}_{RR}&=-\frac{2v}{\sqrt{2} \red{m_{\fp}\zeta_{\fp}}V_{ff'}}{\rm Re}\big(G^{S, \mathrm{loop}}_{H^{\pm}ff'}+G^{P, \mathrm{loop}}_{H^{\pm}ff'}\big)
-\frac{2}{\red{\zeta_{\fp}}}\frac{{\rm Re}\hat{\Pi}_{H^{+}G^{-}}(\mHp^{2}) }{\mHp^{2}}- \Delta r, \\
\label{eq:DLL}
\Delta^{\rm EW}_{LL}&=\frac{2v}{\sqrt{2} \red{m_{f}\zeta_{f}}V_{ff'}}{\rm Re}\big(G^{S, \mathrm{loop}}_{H^{\pm}ff'}-G^{P, \mathrm{loop}}_{H^{\pm}ff'}\big)
 -\frac{2}{\red{\zeta_{f}}}\frac{{\rm Re}\hat{\Pi}_{H^{+}G^{-}}(\mHp^{2}) }{\mHp^{2}}- \Delta r, \\
 \label{eq:DRL}
\Delta^{\rm EW}_{RL}&=-\frac{v}{\sqrt{2} m_{f} m_{\fp}\zeta_{f} \zeta_{\fp} V_{ff'}}\Big[
\red{m_{f}\zeta_{f}} {\rm Re}\big(G^{S, \mathrm{loop}}_{H^{\pm}ff'}+G^{P, \mathrm{loop}}_{H^{\pm}ff'}\big)
-\red{m_{\fp}\zeta_{\fp}} {\rm Re}\big(G^{S, \mathrm{loop}}_{H^{\pm}ff'}-G^{P, \mathrm{loop}}_{H^{\pm}ff'}\big) 
\Big] \notag \\
& -\left(\frac{1}{\zeta_{f}}+\frac{1}{\zeta_{\fp}}\right)\frac{{\rm Re}\hat{\Pi}_{H^{+}G^{-}}(\mHp^{2}) }{\mHp^{2}}-\Delta r, 
\end{align}
where 
functions $G_{\Hpm f \fp}^{S,{\rm loop}}$ and $G_{\Hpm f \fp}^{P,{\rm loop}}$ are expressed in terms of form factors for the $\Hpm f \fp $ vertex as
\begin{align}
\label{eq:GSHpmf}
G_{\Hpm f \fp}^{S,{\rm loop}}&=\Gamma_{\Hpm f \fp}^{S,{\rm loop}}+\red{m_{\fp}}\Gamma_{\Hpm f \fp}^{V1,{\rm loop}}-\red{m_{f}}\Gamma_{\Hpm f \fp}^{V2,{\rm loop}}
+\mHp^{2}\red{\left(1-\frac{m_{-}^{2}}{\mHp^{2}}-\frac{m_{f}\mfp}{\mHp^{2}}\right)}\Gamma_{\Hpm f \fp}^{T,{\rm loop}}, \\ 
\label{eq:GPHpmf}
G_{\Hpm f \fp}^{P,{\rm loop}}&=\Gamma_{\Hpm f \fp}^{P,{\rm loop}}-\red{m_{\fp}}\Gamma_{\Hpm f \fp}^{A1,{\rm loop}}-\red{m_{f}}\Gamma_{\Hpm f \fp}^{A2,{\rm loop}}
+\mHp^{2}\red{\left(1-\frac{m_{+}^{2}}{\mHp^{2}}+\frac{m_{f}\mfp}{\mHp^{2}}\right)}\Gamma_{\Hpm f \fp}^{PT,{\rm loop}},
\end{align}
with $m_{+}=m_{f}+\mfp$ and $m_{-}=m_{f}-\mfp$. 
The last term in Eqs.~\eqref{eq:DRR}-\eqref{eq:DRL}, $\Delta r$, denotes the one-loop weak corrections to the muon decay, which is introduced due to resummation of universal leading higher-order corrections, such as large logarithms from light fermion masses and contributions with the squared mass of the top quark ~\cite{Sirlin:1980nh, Denner:1991kt}. 
 The contributions to the renormalized self-energy $\hat{\Pi}_{H^{+}G^{-}}$ come from $H^{+}W^{-}$ and $H^{+}G^{-}$ mixings, which are derived by using the Slavnov-Taylor identity~\cite{Williams:2011bu}.  

Expressions for the QCD correction parts are different depending on the final state fermions. 
For the decays into a pair of light quarks, we apply the QCD corrections at NNLO in the $\overline{\rm MS}$ scheme~\cite{Mihaila:2015lwa, Gorishnii:1990zu, Gorishnii:1991zr, Chetyrkin:1995pd, Larin:1995sq}.
The correction factors are expressed by a common factor, i.e., $\Delta^{\rm QCD}_{RR}=\Delta^{\rm QCD}_{LL}=\Delta^{\rm QCD}_{RL}=\Delta_{q}^{\Hpm}$;
\begin{align}
\Delta_{q}^{\Hpm}=\frac{\alpha_{s}(\mHp) }{\pi}C_{F}\frac{23}{4}
+\left(\frac{\alpha_{s}(\mHp) }{\pi}\right)^{2}(35.94-1.36N_{f}), 
\end{align}
where $\alpha_{s}(\mHp)$ denotes the strong coupling constant at the scale of $\mHp$, $C_{F}$ is the color factor $C_{F}=4/3$, and $N_{f}$ is the active flavor number. 
In the $\overline{\rm MS}$ scheme, the light quark masses in LO parts of Eq.~\eqref{eq:hpffpNLO}, which comes from the Yukawa couplings of $\Hpm$, are replaced by the corresponding running masses, $m_{q^{(\prime)}}$$\to \bar{m}_{q^{(\prime)}}(\mHp)$. 
Thereby, large logarithm contributions at the higher-order are absorbed by the quark masses~\cite{Braaten:1980yq, Sakai:1980fa, Inami:1980qp, Drees:1990dq}.
On the other hand, for the decay into quarks including the top quark, we apply both QCD corrections in the on-shell scheme and those of $\overline{\rm MS}$ scheme and interpolate them following Ref.~\cite{Djouadi:1997yw}. 
The reason is that the contributions of the top quark mass in the on-shell scheme are significant in the case of lighter charged Higgs bosons, $\mHp\sim m_{t}+m_{D}$ $(D=b, s ,d)$, whereas the logarithm contributions $\log(m_{t}^{2}/\mHp^{2})$ can dominate in case of $\mHp \gg m_{t}+m_{D}$. 
The QCD correction factors at NLO in the on-shell scheme are given by~\cite{Djouadi:1994gf,Djouadi:2005gj}
\begin{align}
\Delta^{\rm QCD}_{RR}=C_{F}\frac{\alpha_{S}(\mu) }{\pi}\Delta^{+}_{q\qp}, \quad\quad
\Delta^{\rm QCD}_{LL}=C_{F}\frac{\alpha_{S}(\mu) }{\pi}\Delta^{+}_{\qp q}, \quad\quad
\Delta^{\rm QCD}_{RL}=C_{F}\frac{\alpha_{S}(\mu) }{\pi}\Delta^{-}_{q\qp }, 
\end{align}
where
 \begin{align}
\Delta_{{f\fp}}^{+}&=\frac{9}{4}+\frac{3-2\mu_f+2\mu_{\fp}}{4}\ln\frac{\mu_f}{\mu_{\fp}}+\frac{(\frac{3}{2}-\mu_f-\mu_{\fp})\lambda_{f\fp}+5\mu_f\mu_{\fp}}{2\lambda_{f\fp}^{1/2}(1-\mu_f-\mu_{\fp})}\ln x_{f}x_{\fp}+B_{f\fp}, \\
\Delta_{{f\fp}}^{-}&=3+\frac{\mu_{\fp}-\mu_f}{2}\ln\frac{\mu_f}{\mu_{\fp}}+\frac{\lambda_{f\fp}+2(1-\mu_f-\mu_{\fp})}{2\lambda_{f\fp}^{1/2}}\ln x_{q}x_{\fp}+B_{f\fp},
 \end{align}
with $x_{f}=2\mu_{f}/(1-\mu_{f}-\mu_{\fp}+\lambda_{f\fp}^{1/2})$. 
The function $B_{f\fp}$ is given in Ref.~\cite{Djouadi:2005gj}. 
When we apply the OS QCD corrections, quark pole masses are used in the LO parts. 

\subsection{Decay rates of $H^{\pm}\to W^{\pm} \phi$}

We describe the one-loop corrected decay rates for the charged Higgs boson decay into the $W^{\pm}$ bosons and scalar bosons $\Hpm\to W^{\pm} \phi$ ($\phi=h, H, A$), with NLO EW corrections. 
They can be expressed by~\cite{Akeroyd:1998uw, Akeroyd:2000xa, Santos:1996hs, Krause:2016oke}
\begin{align}
\Gamma(\Hpm\to W^{\pm} \phi)=\Gamma(\Hpm\to W^{\pm} \phi)^{\rm LO}(1+\Delta^{\rm EW}_{\rm \phi})+\Gamma(\Hpm\to W^{\pm}\phi \gamma). 
\end{align}
The LO decay rate $\Gamma(\Hpm\to W^{\pm} \phi)^{\rm LO}$ is written by
\begin{align}
\red{\Gamma(\Hpm\to W^{\pm}\phi)^{\rm LO}}&=\frac{\mHp^{3}}{16\pi m^{2}_{W}}\lambda\left(\mu_{\phi},\mu_{W}\right)^{\frac{3}{2}}
|g_{\phi \Hpm W^{\mp}}^{}|^{2},
\end{align}
with $\mu_{\phi}=m_{\phi}^{2}/\mHp^{2} $ and $\mu_{W}=m_{W}^{2}/\mHp^{2}$. 
The NLO correction $\Delta_{\phi}^{\rm EW}$ is expressed by
\begin{align}
\label{eq:deltaWphi}
\Delta^{\rm EW}_{\rm \phi}=
\red{\frac{2{\rm Re}\left(\Gamma^{\rm tree}_{\Hpm W^{\mp} \phi}\Gamma^{\rm loop*}_{\Hpm W^{\mp} \phi}\right)}{\abs{\Gamma^{\rm tree}_{\Hpm W^{\mp} \phi}}^{2}}}
-2\frac{g_{\phi G^{\pm} W ^{\mp}} }{g_{\phi \Hpm  W^{\mp}}} \frac{{\rm Re}\hat{\Pi}_{H^{+}G^{-}} (\mHp^{2}) }{\mHp^{2}}-{\Delta r}-{\rm Re}\hat{\Pi}_{WW}^{\prime}(m_{W})^{},
\end{align}
where the tree-level couplings with the charged NG bosons are given by
\begin{align}
g_{h G^\pm W^{\mp} }^{}&= \mp i \frac{m_W}{v}s_{\beta-\alpha}, \ \ \ 
g_{H G^\pm W^{\mp} }^{}= \mp i \frac{m_W}{v}c_{\beta-\alpha},\ \ \ 
g_{A G^\pm W^{\mp} }^{} =0.
\end{align}
The term $\hat{\Pi}^{\prime}_{WW}(m_{W}^{2})$ arises because we do not impose that the residue of renormalized $W^{\pm}$ bosons propagator is unity. 

\subsection{Decay rates of $H^{\pm}\to W^{\pm} V$}
We present the loop induced decay rates for charged Higgs boson, $\Hpm\to W^{\pm} Z$ and $H^{\pm}\to W^{\pm} \gamma$. 
Using the form factors in Eq.~\eqref{eq:HVW}, the decay rate for $\Hpm\to W^{\pm} Z$ are expressed as~\cite{CapdequiPeyranere:1990qk,Kanemura:1997ej}
\begin{align}
\Gamma(\Hpm\to W^{\pm} Z)&=\frac{1 }{16\pi \mHp}\lambda^{\frac{1}{2} }\left(\mu_{W},\mu_{Z}\right)
\left(|\mathcal{M}_{TT}|^{2}+|\mathcal{M}_{LL}|^{2}\right), \\ 
|\mathcal{M}_{TT}|^{2}&=2|\Gamma^{1}_{H^{\pm}ZW^{\mp}}|^{2}+\frac{\mHp^{4}}{2}\lambda^{ }\left(\mu_{W},\mu_{Z}\right)|\Gamma^{3}_{H^{\pm}ZW^{\mp}}|^{2}, \\
|\mathcal{M}_{LL}|^{2}&=\frac{\mHp^{4}}{4m_{W}^{2}m_{Z}^{2}}\Bigg|
 \left(1-\mu_{W} -\mu_{Z}\right)\Gamma^{1}_{H^{\pm}ZW^{\mp}}+\frac{\mHp^{2}}{2}\lambda^{}\left(\mu_{W},\mu_{Z}\right)\Gamma^{2}_{H^{\pm}ZW^{\mp}}
\Bigg|^{2},
\end{align}
where $\mu_{W}= m_{W}^{2}/{\mHp^{2}}$ and $\mu_{Z}=m_{Z}^{2}/{\mHp^{2}}$. 
For $H^{\pm}\to W^{\pm} \gamma$, using the Ward identity, $\Gamma^{1}_{H^{\pm}\gamma W^{\mp}}=-p_{1}\cdot p_{2}\Gamma^{2}_{H^{\pm}\gamma W^{\mp}}$, 
one can obtain 
\begin{align}
\Gamma(\Hpm\to W^{\pm} \gamma)=
\frac{\mHp^{3}}{32\pi}\left(1-\frac{m_{W}^{2}}{\mHp^{2}}\right)^{3}
\left(|\Gamma^{2}_{H^{\pm}\gamma W^{\mp}}|^{2}+|\Gamma^{3}_{H^{\pm}\gamma W^{\mp}}|^{2}\right). 
\end{align}
Since there is no contribution from the longitudinal part in the process, this formula only involves the transverse part of gauge bosons in the final states.



%% file: 4_NLO_EW.tex

\section{Theoretical behaviors of charged Higgs boson decays with NLO corrections}\label{sec:IV}

\subsection{Impact of NLO EW corrections to the decay rates}
\label{sec:deltaEW}

\begin{figure}[t]
\includegraphics[width=1\textwidth,angle=0]{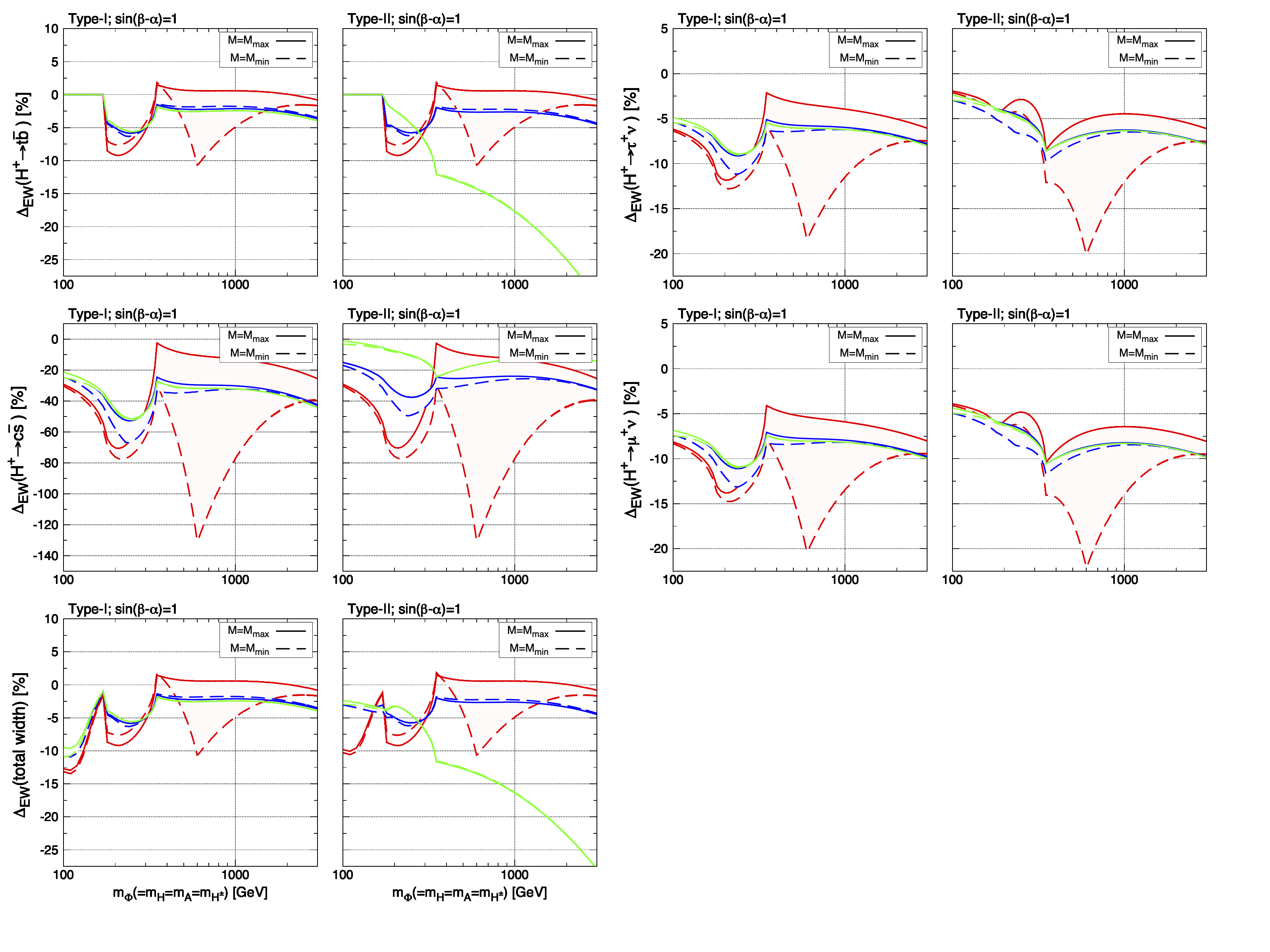}
\caption{
Magnitudes of NLO corrections to the decay widths for charged Higgs bosons in the alignment limit $s_{\beta-\alpha}=1$ with $\tan\beta=1\ {\rm (red)}, 3\ {\rm (blue)},\ 10\ {\rm (green)}$. 
We consider masses of additional Higgs bosons are degenerate, $m_{\Phi}\equiv m_{H^{\pm}}=m_{A}=m_{H}$. 
The dimensionful parameter $M_{\rm max}$ ($M_{\rm min}$) is the maximum (minimum) value of $M$ under the theoretical constraints.  
}
\label{FIG:BR1}
\end{figure}

\begin{figure}[t]
\includegraphics[width=1\textwidth,angle=0]{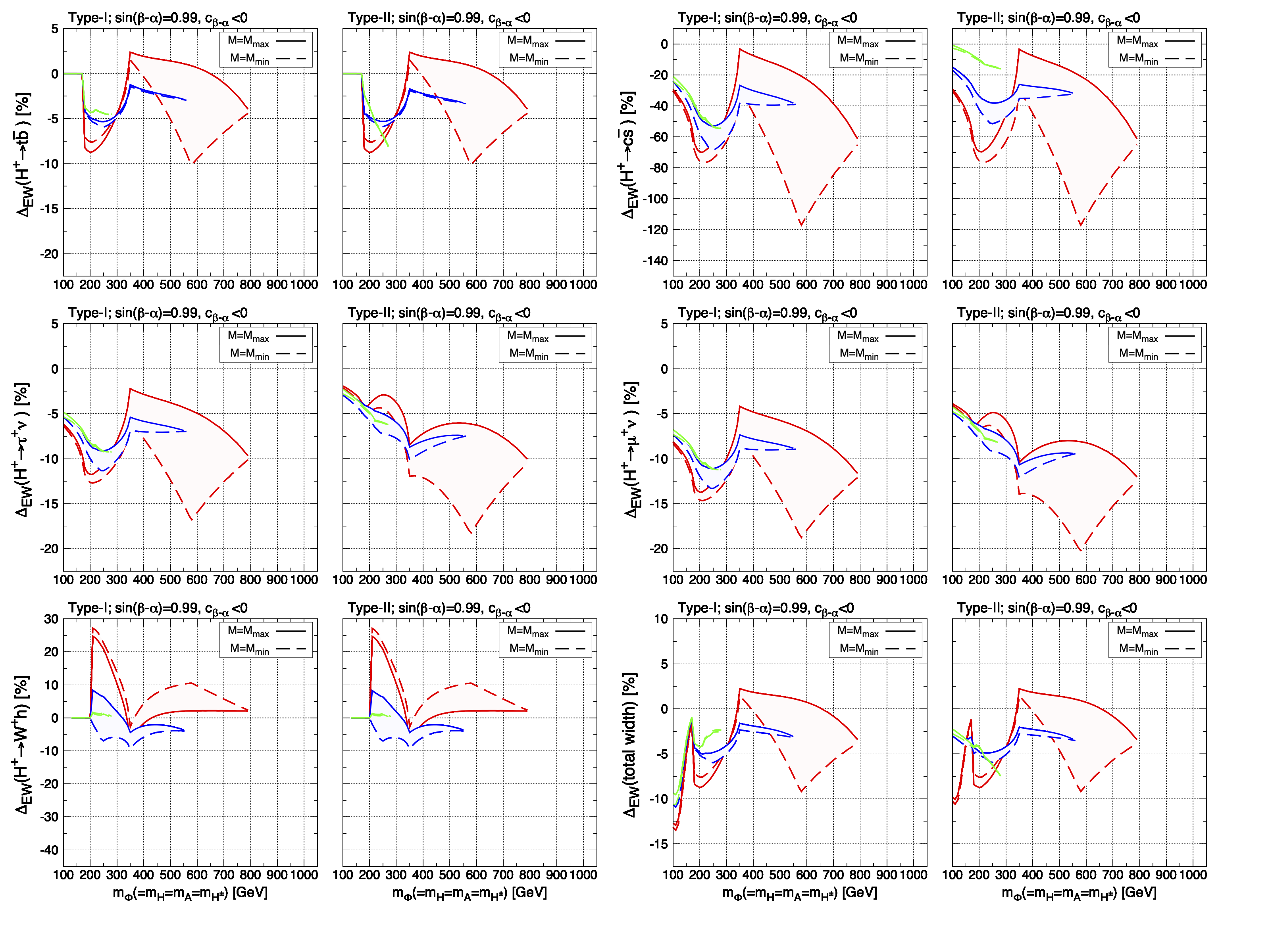}
\caption{
Magnitudes of NLO corrections to the decay widths for charged Higgs bosons in the case of $ s_{\beta-\alpha}=0.99$ with $c_{\beta-\alpha}<0$. 
The masses of the additional Higgs bosons are degenerate, $m_{\Phi}\equiv m_{H^{\pm}}=m_{A}=m_{H}$. 
The dimensionful parameter $M_{\rm max}$ ($M_{\rm min}$) is the maximum (minimum) value of $M$ under the theoretical constraints.  
}
\label{FIG:BR2}
\end{figure}

\begin{figure}[t]
\includegraphics[width=1\textwidth,angle=0]{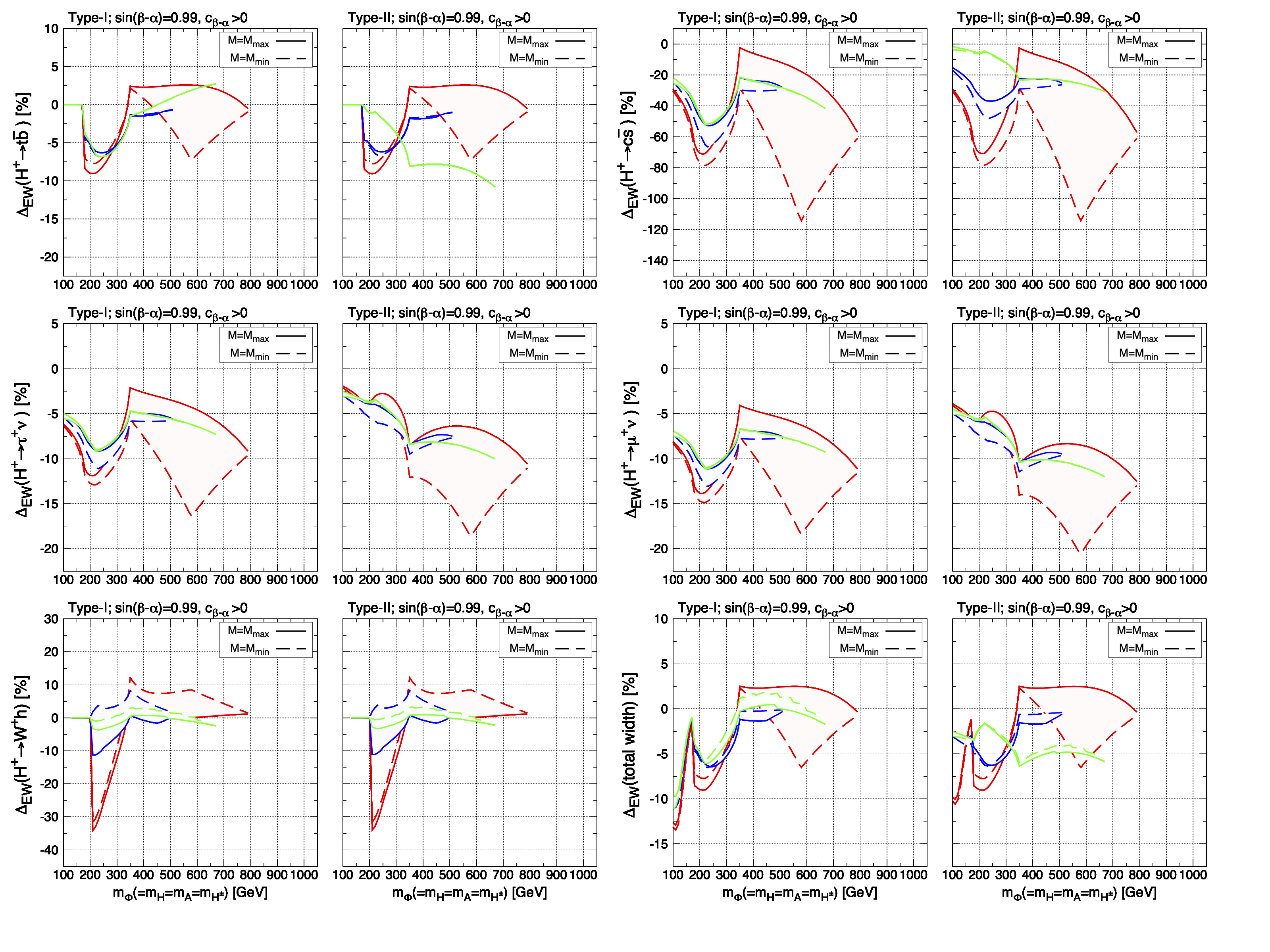}
\caption{
Magnitudes of NLO corrections to the decay widths for charged Higgs bosons in the case of $s_{\beta-\alpha}=0.99$ with $c_{\beta-\alpha}>0$. 
The masses of the additional Higgs bosons are degenerate, $m_{\Phi}\equiv m_{H^{\pm}}=m_{A}=m_{H}$. 
The dimensionful parameter $M_{\rm max}$ ($M_{\rm min}$) is the maximum (minimum) value of $M$ under the theoretical constraints.  
}
\label{FIG:BR3}
\end{figure}

In this section, we examine the impact of NLO EW corrections on the decay rates of the charged Higgs boson in Type-I and Type-II. 
We here omit to show the results for Type-X and Type-Y, because they are very similar to those of Type-I or Type-II. 
In the next subsection and next section, we compare the results for all types of the Yukawa interactions. 
We use the following quantities to describe the magnitudes of NLO EW corrections,
\begin{align}
\Delta_{\rm EW}(\Hp \to XY)=\frac{\Gamma^{\rm NLO\ EW}(\Hp \to XY)}{\Gamma^{\rm LO}(\Hp \to XY)}-1,
\end{align}
where $\Gamma^{\rm NLO\ EW}(\Hp \to XY)$ corresponds to the decay rates without QCD corrections. 
For the calculation of decay rates at LO $\Gamma^{\rm LO}$, we employ the quark running masses not the pole masses. 
We evaluate this quantity in both the alignment scenario, $\sba=1$, and a nearly alignment scenario, $\sba=0.99$. 
For each scenario, $\tan\beta$ is taken to be the following three values, $\tb=1, 5,$ and 10. 
The dimensionful parameter $M$ is scanned in the region of $0<M<1500~{\rm GeV}$. 
In this analysis, we impose the theoretical constraints discussed in Sec.~\ref{sec:constraint}, i.e., the perturbative unitarity and the vacuum stability.  
On the other hand, 
we here do not take into account the constraint from the flavor physics in order to compare the difference of $\Delta_{\rm EW}$ among all the types, while the mass of charged Higgs boson is strictly constrained by $B_{s}\to X_{s}\gamma$, especially for Type-II and Type-X. 
All results with the constraints including the flavor experiments are presented in the next section. 
 
In Fig.~\ref{FIG:BR1}, we show the EW corrections to various charged Higgs boson decays as a function of degenerated mass of the additional Higgs bosons, $m_{\Phi}\equiv\mHp=m_{H}=m_{A}$ in the alignment limit, $\sba=1$ with the different values of $\tb$, $\tb=1~{\rm (red)},~3~{\rm (blue)},$ and 10 (green). 
The solid (dashed) lines correspond to the results with a maximum (minimum) value of $M^{}$ satisfying the theoretical constraints, $M_{\rm max}$ ($M_{\rm min}$). 
In this case, charged Higgs decays into a pair of fermions are dominant. 
For the results of $H^{+}\to t\bar{b}$ with $\tb=1$, one can see that there are kinks at $m_{\Phi}\simeq m_{t}+m_{b},~2m_{t}$ and $600~{\rm GeV}$. 
The first one comes from the threshold of the $(t,b)$ loop diagram in the $H^{+}$-$H^{-}$ self-energy.
The second one comes from the threshold of the top loop diagram in the $A$-$G^{0}$ mixing self-energy, which appears in the counterterms for the $H^{+}f\fp$ vertex. 
The third kink corresponds to the points where the values of $M$ change from zero to non-zero due to the perturbative unitarity. 
At this point, the scalar couplings $\lambda_{H^{+}H^{-}\phi}$ ($\phi=h,H,A$), which are given in Appendix~\ref{sec:coup}, are maximized under the constraint from the perturbative unitarity. 
Non-decoupling effects of $h, H$ and $A$ loops in the $H^{+}$-$H^{-}$ self-energy are then  dominant, and $\Delta_{\rm EW}(\Hpm\to t\bar{b})$ can be almost 10\% for all the types of THDMs. 
On the other hand, even if $m_{\Phi}$ is sufficiently large, the EW corrections do not decouple. 
Namely, the decoupling theorem~\cite{Appelquist:1974tg} is not applicable in this case, and non-decoupling effects are significant. 

For the results with $\tb=3$, and 10, the possible values of $M$ are almost constants due to the strict theoretical constraints, i.e. $M\sim m_{\Phi}$.  
While for Type-I, one do not see large difference between $\tb=3$, and $\tb=10$,
for Type-II, the corrections can be sizable in the case of $\tb=10$, e.g., $\Delta_{\rm EW}(\Hp\to t\bar{b})\simeq-25~\%$ at $m_{\Phi}=2$ TeV.
We find that these behaviors can be explained by large negative contributions from the tensor form factors $\Gamma^{T}_{\Hp f\fp}$ and $\Gamma^{TP}_{\Hp f\fp}$, which give the contributions proportional to the square of the charged Higgs boson mass in the decay rate (see Eqs.~\eqref{eq:GSHpmf} and \eqref{eq:GPHpmf}). 

For other decay modes, one can see the similar behaviors described above. 
On the other hand, the remarkable thing is that the correction $\Delta_{\rm EW}(\Hp \to c\bar{s} )$ can be over $-100\%$ at $m_{\Phi}=600{\rm GeV}$.  
We note that $\Delta_{\rm EW}(\Hp \to c\bar{s} )$ tends to be larger than the other decay modes for the following reason. 
The decay rate with NLO EW corrections $\Gamma(\Hp \to c\bar{s} )$ is evaluated by using the pole masses for the charm quark and the strange quark while the LO decay rates are evaluated with the running masses at the scale of $ \mu =\mHp$. 
Consequently, the difference between the pole masses and the running masses enhances $\Delta_{\rm EW}(\Hp \to c\bar{s} )$~\red{\footnote{\red{For instance, the ratios of the running masses and pole masses are estimated as $m_{c}/\bar{m}_{c}(m_{H^{\pm}})=1.67\ {\rm GeV}/0.609\ {\rm GeV}=2.74$,  $m_{s}/\bar{m}_{s}(m_{H^{\pm}})=0.1\ {\rm GeV}/0.0491\ {\rm GeV}=2.03$. Magnitude of  $\Delta_{\rm EW}(\Hp \to c\bar{s} )$ is enlarged by these factors. The discussion does not depend on $s_{\beta-\alpha}$, so that the same holds in Fig.~\ref{FIG:BR2} and ~\ref{FIG:BR3}. Namely $\Delta_{\rm EW}(\Hp \to c\bar{s} )$ can also be over -100\% in case of $s_{\beta-\alpha}=0.99$.  } }}.

In Fig.~\ref{FIG:BR2}, the results in the nearly alignment scenario, $\sba=0.99$ with $\cba<0$, are shown as a function of the degenerate mass $m_{\Phi}$. 
In the non-alignment case, an upper bound of $m_{\Phi}$ is given for each value of $\tb$ because of the theoretical constraints. 
However, the maximum magnitudes of NLO EW corrections for the case of $\tb=1$ are almost unchanged from the scenario of the alignment limit. 
Apart from that, the charged Higgs bosons can decay into $W^{+}h$ in the nearly alignment scenario. 
For this decay mode, peaks appear at $\mHp\simeq m_{h}+ m_{W}$, which correspond to the thresholds of 1PI diagrams in the $\Hp W^{-} h$ vertex function such as $(W,\Hpm/G^{\pm}, h)$ and the $(h, h/H, W)$ loop diagrams. 
The maximum value of the corrections is $26\%$ in the case of $\tb=1$ for all the types of THDMs. 

In the Fig.~\ref{FIG:BR3}, we also show the results with $\sba=0.99$ and $\cba>0$. 
The remarkable difference from the results with $\cba<0$ is that the allowed regions for $\tb=10$ are broader than those for $\tb=3$. 
Hence, compared with $\cba<0$, the corrections $\Delta_{\rm EW}$ for $\tb=10$ can be larger. 
In addition, for $\Hp \to W^{+}h$, direction of the threshold peak at $m_{\phi}\simeq m_{h}+m_{W}$ is opposite from $\cba<0$ because the contributions from the $H^{+}W^{-}h$ vertex function depends on the tree-level coupling $g_{h H^{+}W^{-}}$ (see Eq.~\eqref{eq:deltaWphi}). 
On the other hand, one can see that the sign of $\Delta_{\rm EW}(\Hp \to W^{+}h)$ in the region $m_{\Phi}\gtrsim500$ with $\tan\beta=1$ are positive in the both cases of $\cba<0$ and $\cba>0$. 
The dominant contributions in this region mainly come from non-decoupling effects of additional Higgs bosons, i.e., pure scalar loop diagrams in $\delta C_{h}$ and $\delta C_{\Hp}$, which are proportional to the square of the scalar couplings $\lambda_{\phi_{i}\phi_{j} \phi_{k}}\lambda_{\phi_{i^{\prime}}\phi_{j^{\prime}} \phi_{k^{\prime}}}$ (They are defined in Appendix~\ref{sec:coup}). 
Among them, there are contributions that are not proportional to $\cba$, so that they do not depend on the sign of $\cba$. 


\subsection{Branching ratios}
\begin{figure}[t]
\includegraphics[width=1.\linewidth,angle=0]{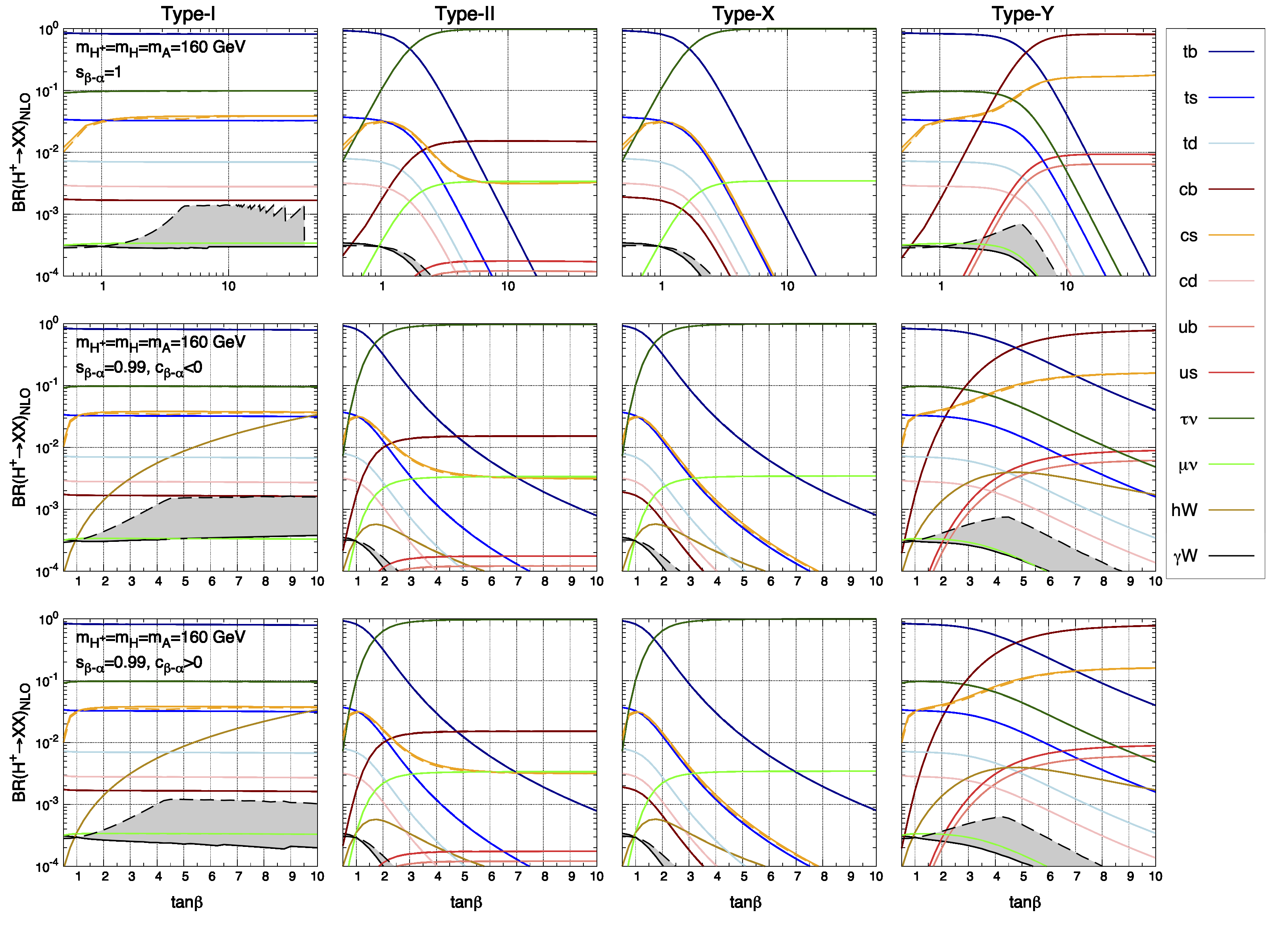}\hspace{0mm}
\caption{
Decay branching ratios for charged Higgs bosons as a function of $\tan\beta$ in the alignment limit $s_{\beta-\alpha}=1$ (top panels) and in the nearly alignment scenarios $s_{\beta-\alpha}=0.99$ with $\cba<0$ (middle panels) and $\cba>0$ (bottom panels), where the NLO EW and NNLO QCD corrections are included if they are applicable.  
Masses of the charged Higgs boson as well as the neutral Higgs bosons are taken to be degenerate, i.e., $\mHp=m_{H}=m_{A}=160{\rm GeV}$. 
Each decay mode is specified by color as given in the legend. 
Solid (dashed) lines correspond to the results with $M= M_{\rm max}$ $(M_{\rm min})$. 
The dimensionful parameter $M_{\rm max}$ ($M_{\rm min}$) is the maximum (minimum) value of $M$ under the theoretical constraints.  
The gray region shows predictions on $\BR(H^{+}\to W^{+}\gamma)$ in the region $M_{\rm min}<M<M_{\rm max}$. 
}
\label{FIG:BR160}
\end{figure}

\begin{figure}[t]
\includegraphics[width=1.\linewidth, angle=0]{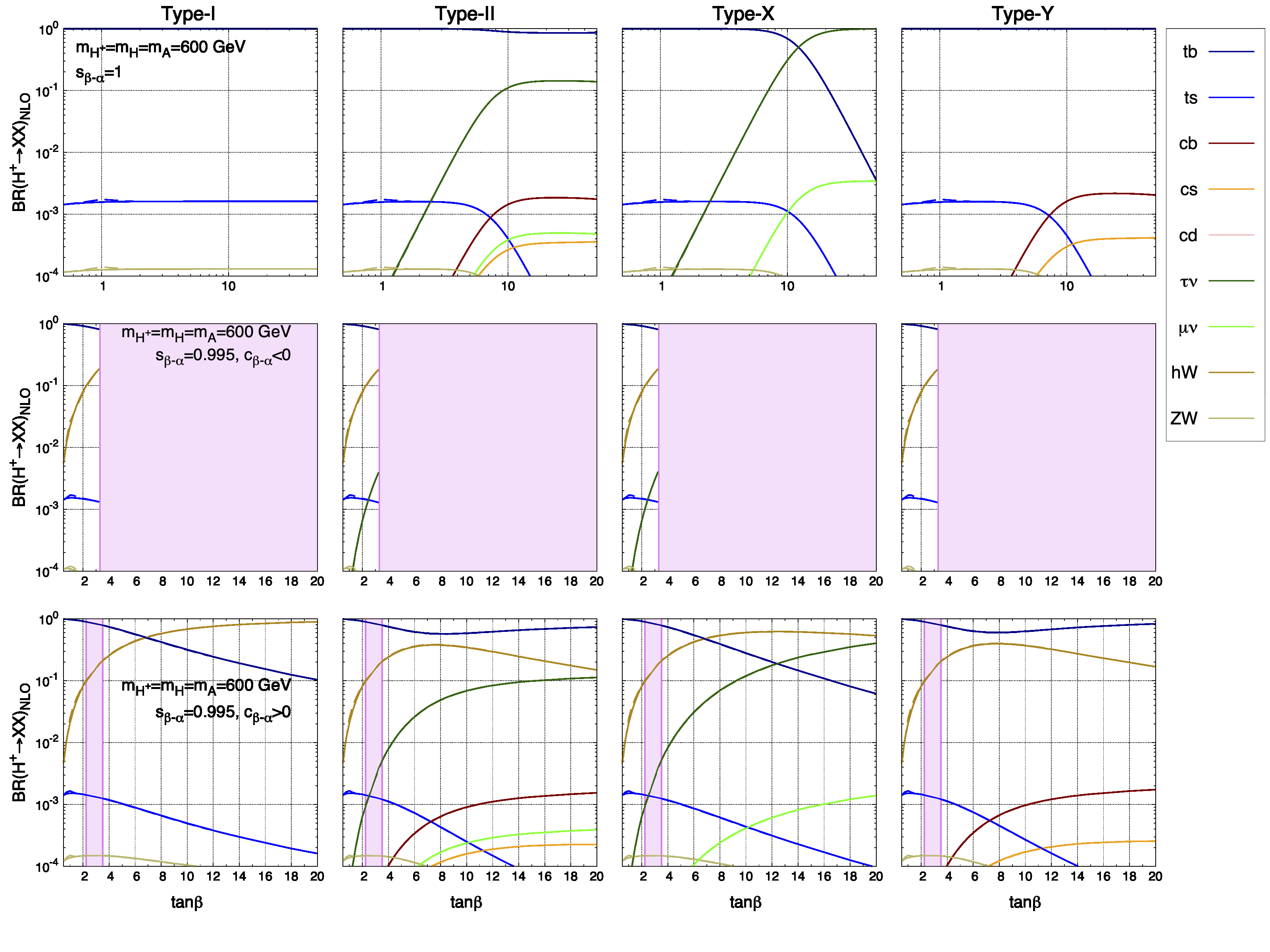}\hspace{0mm}
\caption{
Decay branching ratios for charged Higgs bosons as a function of $\tan\beta$ in the alignment limit $s_{\beta-\alpha}=1$ (top panels) and in the nearly alignment scenarios $s_{\beta-\alpha}=0.995$ with $\cba<0$ (middle panels) and $\cba>0$ (bottom panels), where the NLO EW and the NNLO QCD corrections are included if they are applicable.  
The masses of the charged Higgs boson as well as the neutral Higgs bosons are taken to be degenerate, i.e., $\mHp=m_{H}=m_{A}=600{\rm GeV}$. 
Each decay mode is specified by color as given in the legend. 
Solid (dashed) lines correspond to the results with $M= M_{\rm Max}$ $(M_{\rm min})$. 
The dimensionful parameter $M_{\rm max}$ ($M_{\rm min}$) is the maximum (minimum) value of $M$ under the theoretical constraints.  
The violet region corresponds to the one excluded by the theoretical constraints. }
\label{FIG:BR600}
\end{figure}

\begin{figure}[t]
\includegraphics[width=1.\linewidth, angle=0]{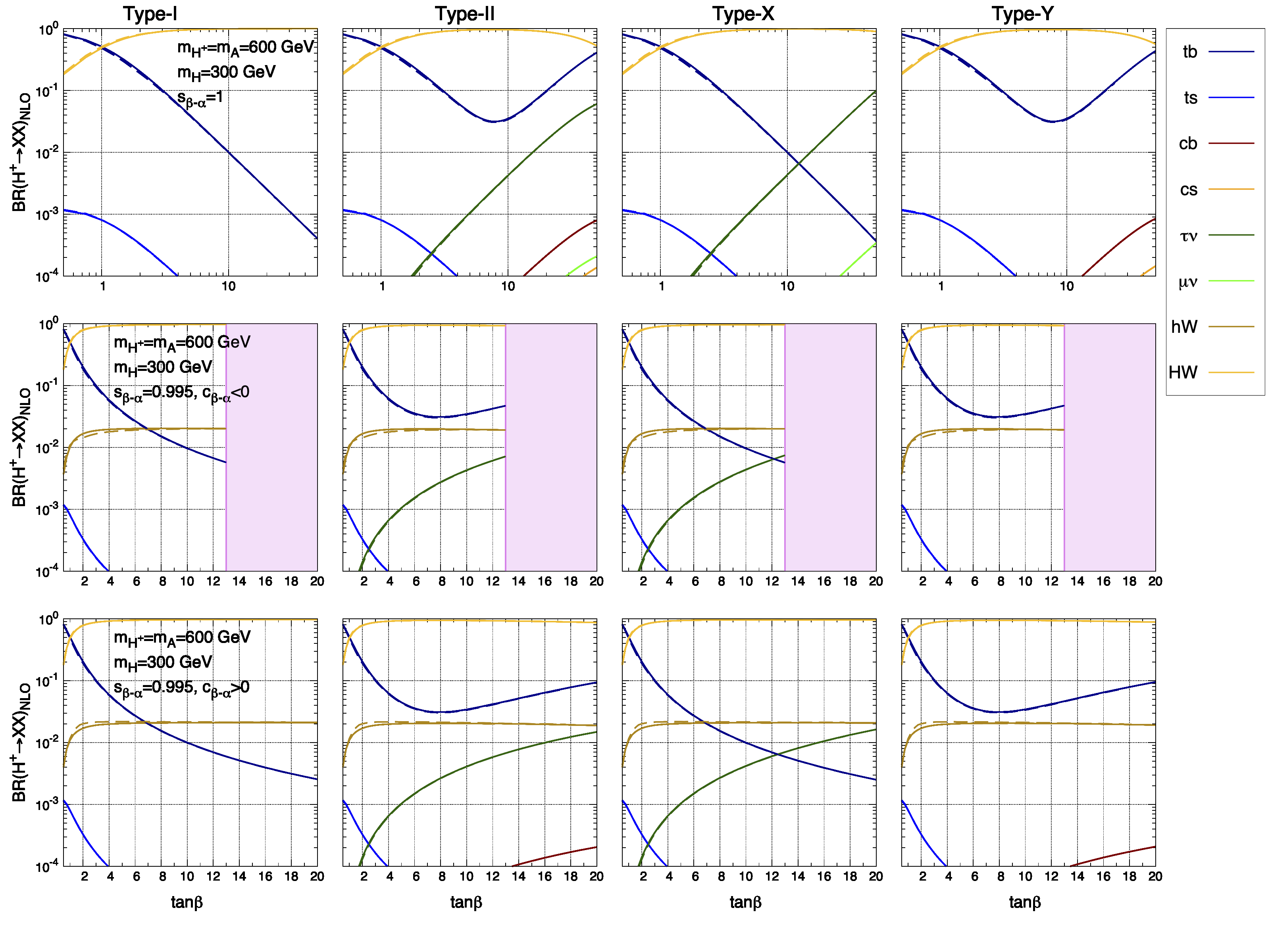}\hspace{0mm}
\caption{
Decay branching ratios for charged Higgs bosons as a function of $\tan\beta$ in the alignment limit $s_{\beta-\alpha}=1$ (top panels) and in the nearly alignment scenarios $s_{\beta-\alpha}=0.995$ with $\cba<0$ (middle panels) and $\cba>0$ (bottom panels), where the NLO EW and the NNLO QCD corrections are included if they are applicable.  
The masses of the charged Higgs boson and the neutral Higgs bosons are taken to be degenerate, i.e., $\mHp=m_{H}=m_{A}=600$ {\rm GeV} and $m_{H}=300~{\rm GeV}$. 
Each decay mode is specified by color as given in the legend. 
Solid (dashed) lines correspond to the results with $M= M_{\rm Max}$ $(M_{\rm min})$. 
The dimensionful parameter $M_{\rm max}$ ($M_{\rm min}$) is the maximum (minimum) value of $M$ under the theoretical constraints.  
The violet region corresponds to the one excluded by the theoretical constraints.
}
\label{FIG:BR600md300}
\end{figure}

In this subsection, we describe behaviors of the branching ratios with the NLO EW corrections as well as the QCD corrections.
Similar to the previous section, we evaluate them in both the alignment scenario and the nearly alignment scenario under the theoretical constraints and the S, T parameters. 
We here consider the following three cases with a different mass spectrum of the additional Higgs bosons:
\begin{itemize}
  \item[Case~1:] A relatively small mass of $H^{\pm}$, $\mHp=160\ {\rm GeV}$, 
  and degenerate masses of $H$ and $A$ with $H^{\pm}$, $m_{H}=m_{A}=\mHp$. In this case, the on-shell decay $\Hp\to t\bar{b}$ does not open. 
    \item[Case~2:] A relatively large mass of $H^{\pm}$, $\mHp=600\ {\rm GeV}$, 
  and degenerate masses of $H$ and $A$ with $H^{\pm}$, $m_{H}=m_{A}=\mHp$.
      \item[Case~3:] A relatively large mass of $H^{\pm}$, $\mHp=600\ {\rm GeV}$, degenerate masses of $A$ with $H^{\pm}$, $m_{A}=\mHp$, and lighter mass of $H$, $m_{H}=300\ {\rm GeV}$. In this case, the on-shell decay $H^{\pm}\to W^{+} H$ is kinematically allowed. 
\end{itemize} 
Whereas masses of the additional Higgs bosons are fixed for each case, $M$ and $\tb$ are commonly scanned in the following regions of $0<M<1500~{\rm GeV}$ and $0.5<\tb<50$. 
Taking into account the flavor constraints for Type-II and Type-Y, the mass of charged Higgs boson in each case would be too light. 
However, we dare to show the results for not only Type-I and Type-X but also Type-II and Type-Y for the comparison. 
For Case~2 and Case~3, we checked that behaviors of the charged Higgs bosons are similar if we change the mass of charged Higgs boson from $\mHp=600\ {\rm GeV}$ to $\mHp=800\ {\rm GeV}$. 

In Fig.~\ref{FIG:BR160}, we show the results in Case~1 for the alignment scenario, $\sba=1$, and the nearly alignment scenarios, $\sba=0.99$ with $\cba<0$ and $\cba>0$ from top panels to bottom panels. 
For the results of Type-I in $\sba=1$, all decay modes into quarks and leptons are proportional to $1/\tan^{2}\beta$. Hence, except for $\Hp\to \gamma W^{+}$, the branching ratios are almost constants without depending on $\tb$.
As the results, the decay $\Hp\to t^{\ast}\bar{b}$ dominates in the whole region of $\tb$. 
On the other hand, one can see that the branching ratio of $\Hp\to \gamma W^{+}$ can be much varied by the scale of $M$. 
This is due to contributions from the $(H,\Hpm,\Hpm)$ diagrams in the form factor $\Gamma^{2}_{\Hp \gamma W}$.  
We note that another pure scalar loop diagram disappears in the alignment limit. 
In the case of $M$=$M_{\rm min}$ (black dashed line) and $\tb\gg1$, the relevant scalar coupling for this diagram, $\lambda_{H^{+}H^{-}H}$ can be sizable. 
Hence, the decay $\Hp\to \gamma W^{+}$ is enhanced.  
For the results of Type-II and Type-X, the main decay mode becomes $\Hp\to \tau^{+}\nu$ in the large $\tb$ regions, since the tau Yukawa coupling is enhanced. 

For the results of the nearly alignment scenario $\sba=0.99$, behaviors of the charged Higgs boson decays are similar to those of the alignment scenario, while the value of the $\tb$ is bounded at $\tb\simeq 10$ because of the theoretical constraints. 
In these scenarios, the decay into $\Hp \to W^{+} h$ opens. 
The branching ratio can exceed 3\% when $\tb=10$ in Type-I. 
 
 In Fig.~\ref{FIG:BR600}, we show the results in Case~2 for the alignment scenario and the nearly alignment scenario from the top panels to the bottom panels. 
Here we take the value of $\sba$ in the nearly alignment scenario as $\sba=0.995$ in light of the severe theoretical constraints.  
In the bottom panels, the violet regions correspond to the ones excluded by the theoretical constraints. 
In Case~2, the on-shell decay into $t\bar{b}$ opens and it is the dominant decay mode in the alignment scenario expect for Type-X. 
However, the situation can be changed in the nearly alignment scenario with $\cba>0$. 
Namely, the additional decay mode $\Hp\to W^{+}h$ can overcome $\Hp \to t \bar{b}$ in Type- I and Type-X. 
Another remarkable behavior for Case~2 is that the EW corrections to the decay into $c\bar{s}$ can be considerably large at $\tb\simeq 1$ because of the non-decoupling effects of the additional Higgs bosons as already seen in Fig.~\ref{FIG:BR1}, while the magnitudes of the branching ratios are below $10^{-4}$. 

 In Fig~\ref{FIG:BR600md300}, the results of Case~3 are shown in the scenarios of $\sba=1$ and $\sba=0.995$.  
The feature of these scenarios is that the decay into $W^{+}H$ opens. 
The decay rate can be significant since it is proportional to the cube of $\mHp^{}$. 
In addition, the corresponding tree-level coupling is proportional to $\sba$, so that this decay mode appears even in the alignment scenario. 
Remarkably, if $\tb\gtrsim 1$, the decay into $\Hp \to W^{+} H$ dominates the branching ratios in both the alignment and nearly alignment scenarios for all the types of THDMs. 



%% file: 5_numerical_result.tex


\section{Phenomenological impact of the charged Higgs boson decays}\label{sec:V}

\subsection{Decay pattern of the charged Higgs bosons in the nearly alignment regions }\label{sec:Decaypattern}

\begin{figure}[t]
\includegraphics[width=1.\linewidth]{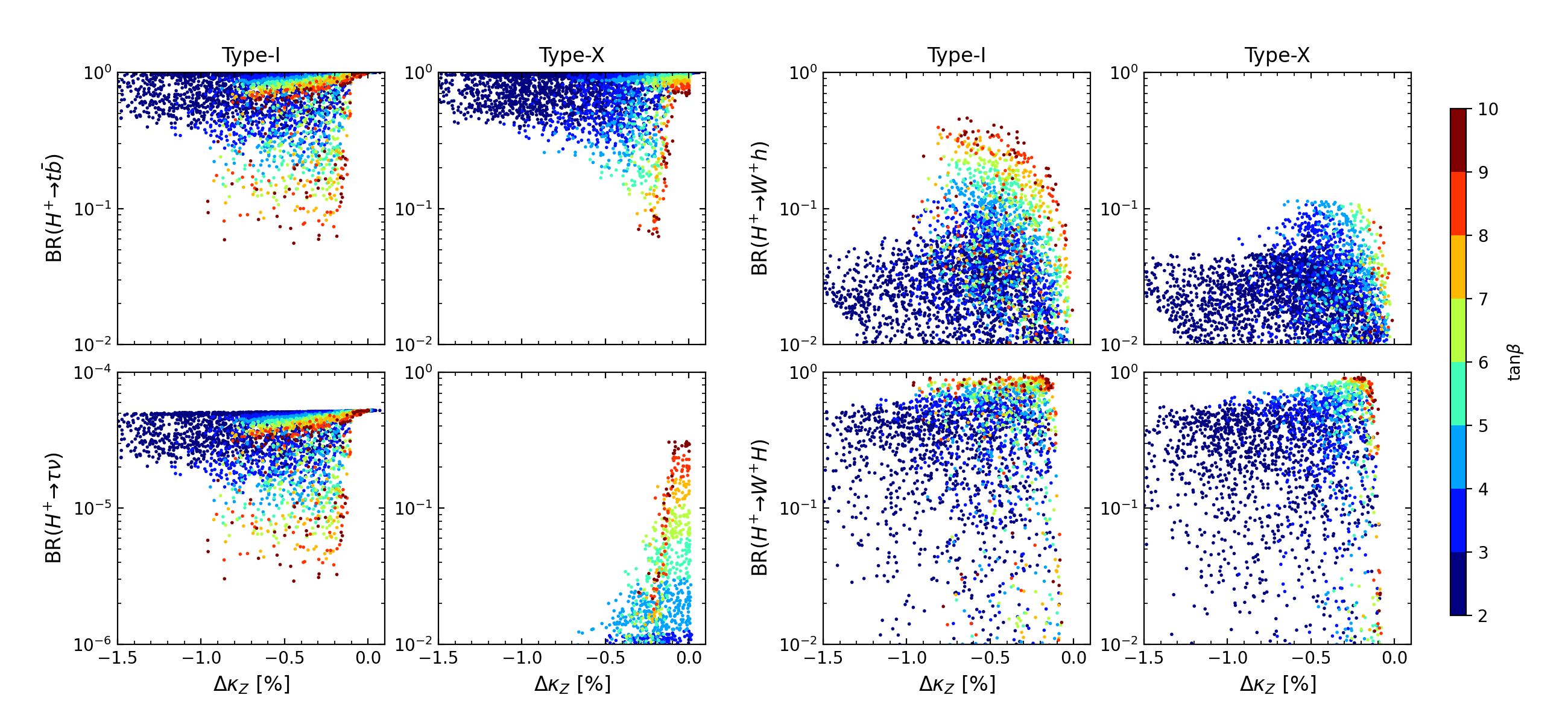}\hspace{0mm}
\caption{
Decay branching ratios for the charged Higgs bosons as a function of $\Delta \kappa_{Z}(\equiv \kappa_{Z}-1)$ in Scenario A, where colored points denote different values of $\tb$. 
Predictions on Type-I (Type-X) are shown in the first and third columns (the second and fourth columns). 
}
\label{FIG:SA}
\end{figure}

\begin{figure}[t]
\includegraphics[width=1.\linewidth]{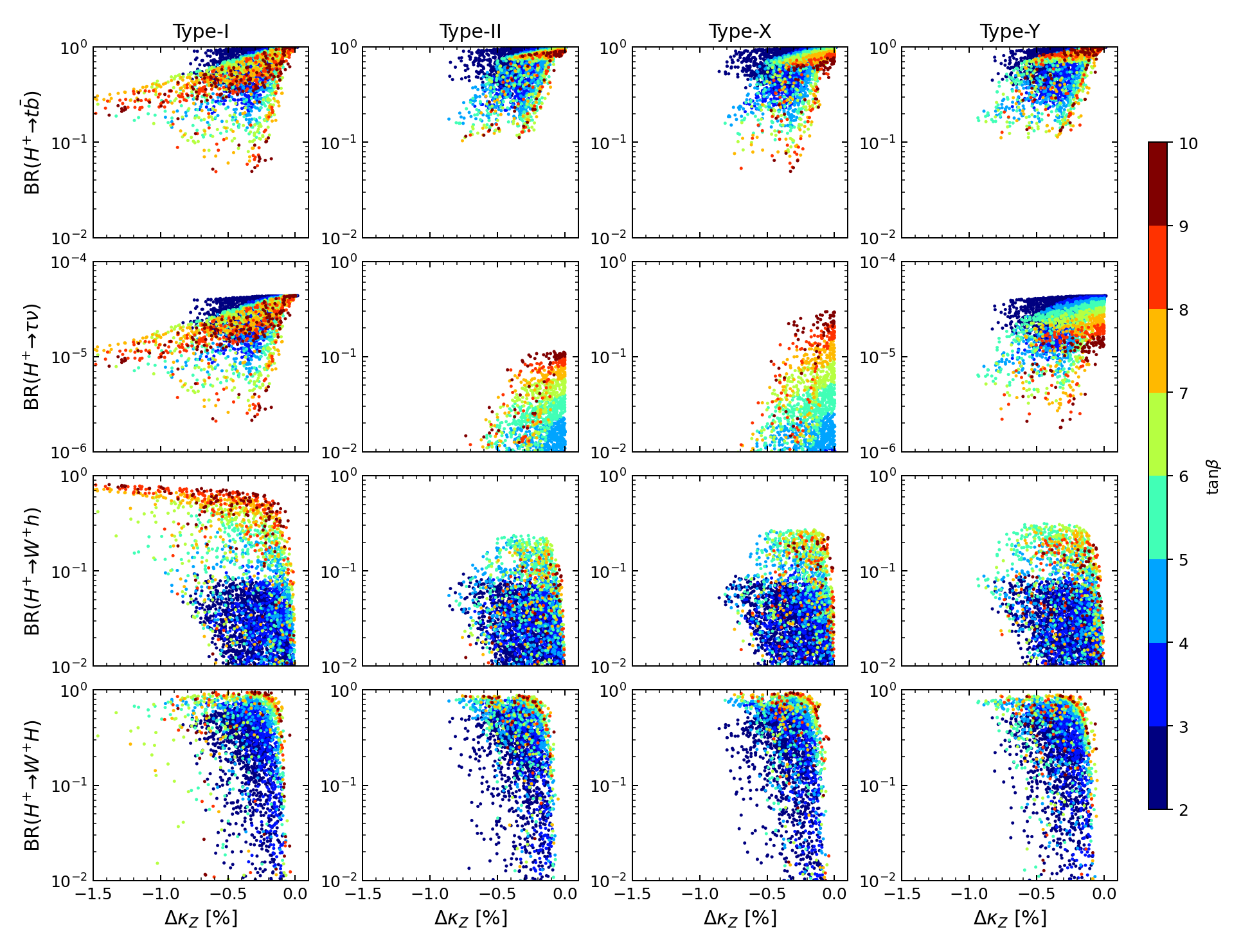}\hspace{0mm}
\caption{
Decay branching ratios for the charged Higgs bosons as a function of $\Delta \kappa_{Z}(\equiv \kappa_{Z}-1)$ in Scenario B, where colored points denote different values of $\tb$. 
Predictions on Type-I, II, X and Y are shown from the left panels to the right panels. 
}
\label{FIG:SB}
\end{figure}

\begin{figure}[t]
\includegraphics[trim=50 150 50 250, width=0.48\linewidth,clip]{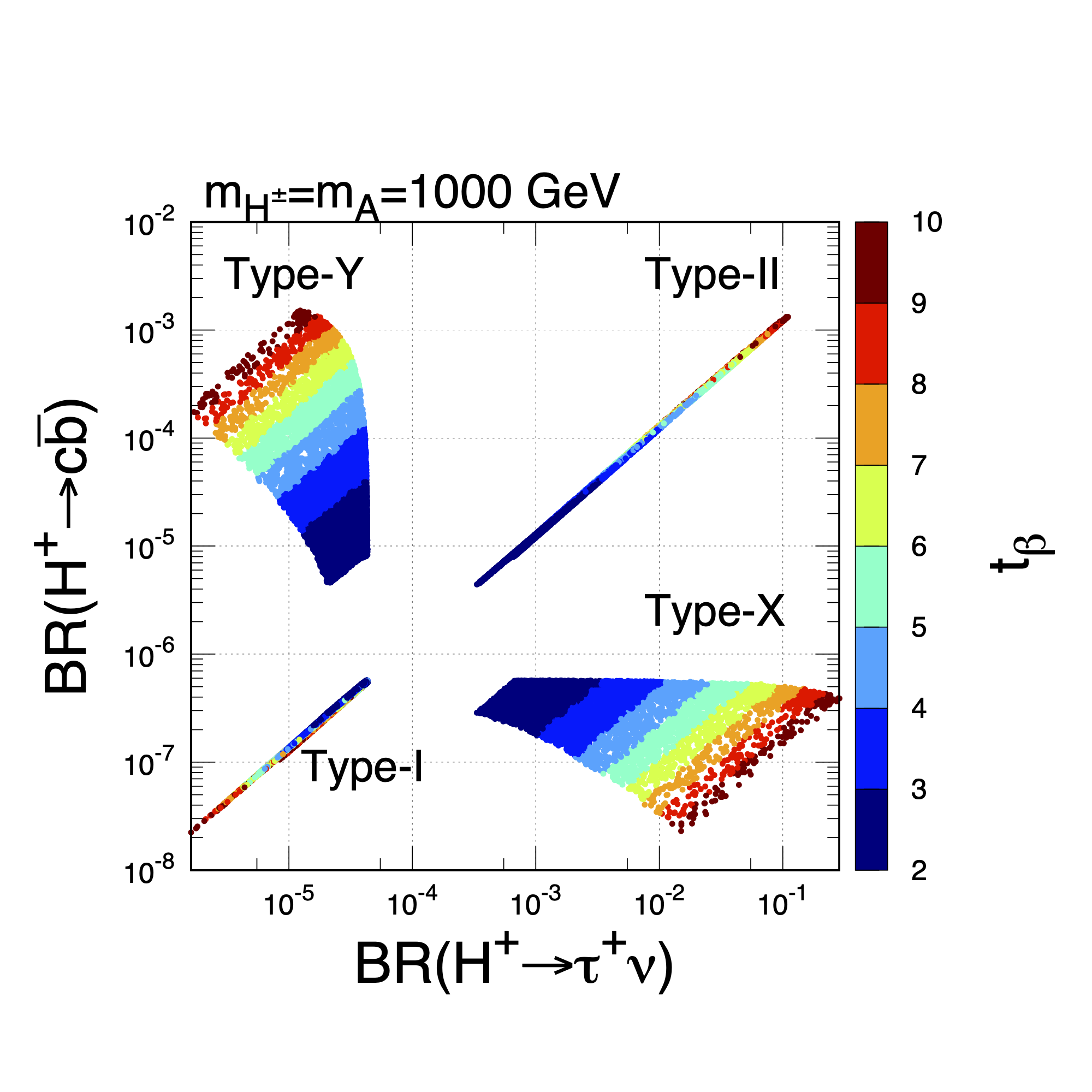}\hspace{0mm}
\includegraphics[trim=50 150 50 250, width=0.48\linewidth,clip]{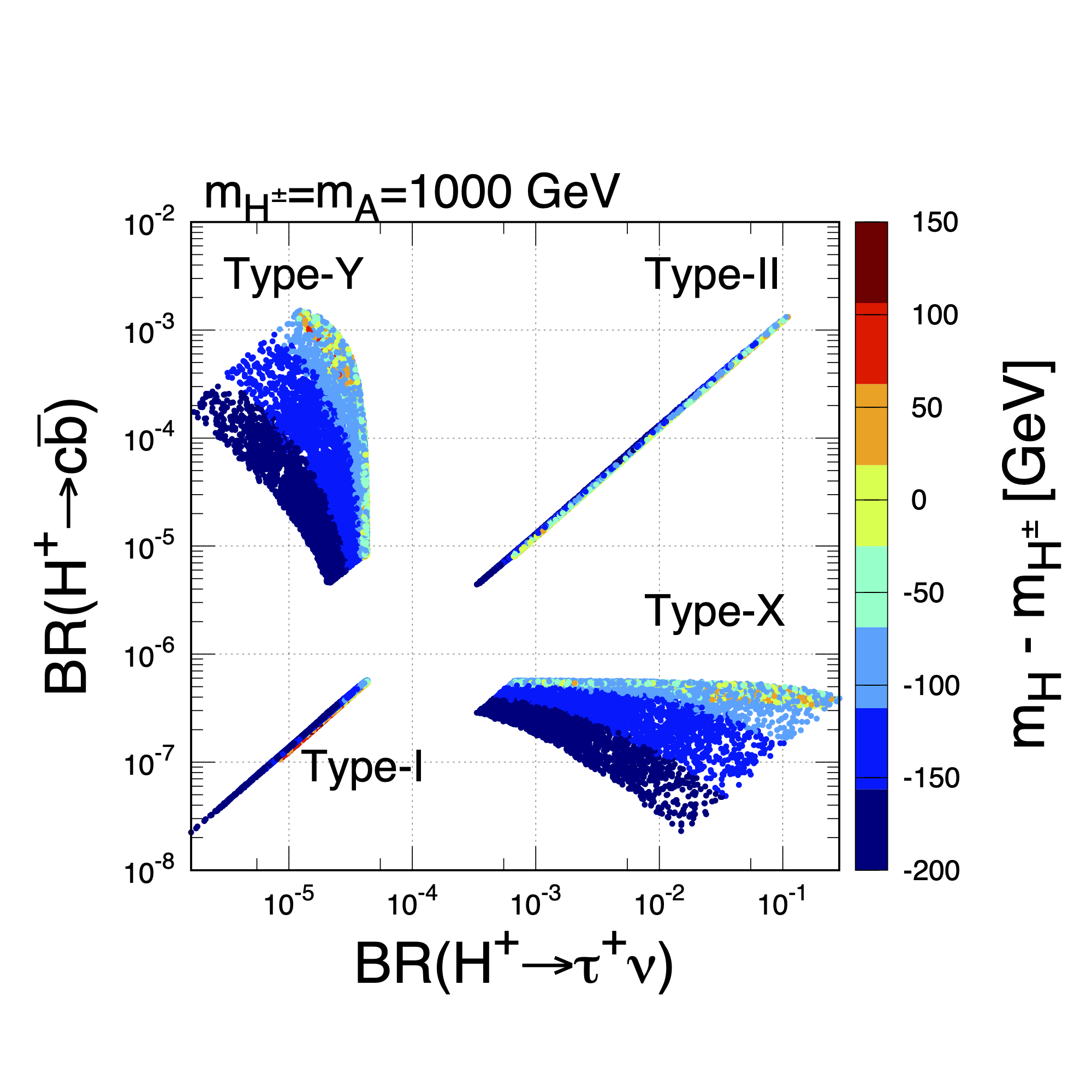}\hspace{0mm}
\includegraphics[trim=50 150 50 250, width=0.48\linewidth,clip]{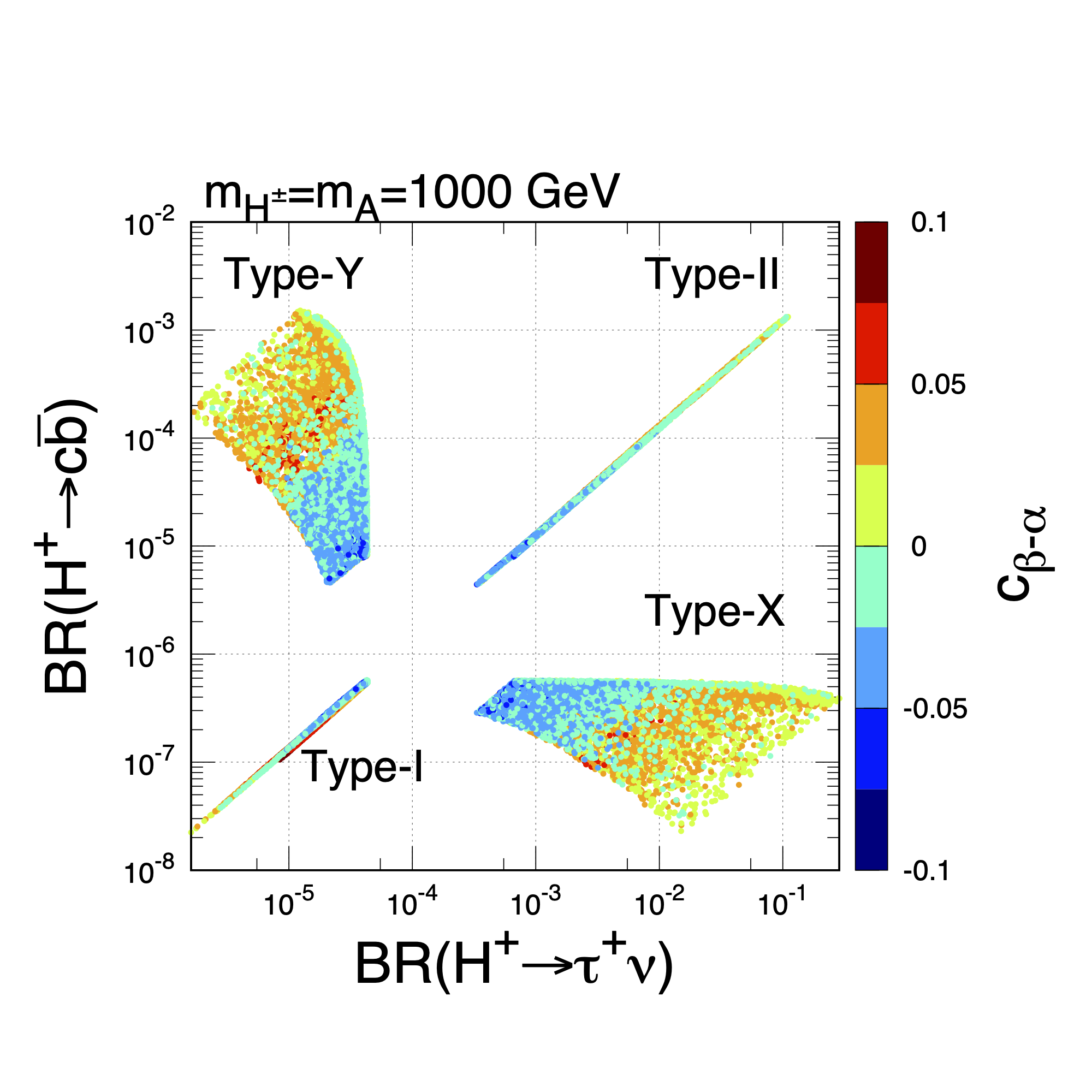}\hspace{0mm}
\caption{
Correlation between $\BR(\Hp\to\tau^{+}\nu)$ and $\BR(\Hp\to c\bar{b})$ for Scenario B in four types of THDMs. 
Colored dots correspond to different values of $\tb$, $m_{H}- \mHp$, and $c_{\beta-\alpha}$ in the left top panel, the right top panel and the bottom panel, respectively. 
}
\label{FIG:cbvstaunu}
\end{figure}

In this subsection, we discuss whether or not four types of THDMs can be discriminated by looking at decay patterns of the charged Higgs bosons, and also whether or not information of the inner parameters can be extracted. 
As already studied, the discrimination of THDMs can be accomplished by patterns of the deviations from the SM predictions for the couplings~\cite{Kanemura:2014bqa,Kanemura:2015mxa,Kanemura:2017wtm} and/or the branching ratios~\cite{Kanemura:2018yai,Kanemura:2019kjg} of the discovered Higgs boson if the deviations are actually found in the future collider experiments. 
In particular, four types of THDMs can be clearly separated by a correlation of the $hbb$ coupling and the $h\tau\tau$ coupling~\cite{Kanemura:2014bqa,Kanemura:2015mxa,Kanemura:2017wtm}. 
However, current experimental data from the LHC Run II favor the alignment regions, and such a desired situation would not be necessarily realized in the future. 
Hence, it would be worth investigating the impacts of discovery of the charged Higgs bosons for a test of THDMs
especially in the case that the significant deviations in the $h$ couplings are not detected in the future collider experiments. 

To this end, we consider two distinct scenarios for the mass of the charged Higgs boson,  
\begin{align}
\mbox{(Scenario A)}&:\quad \mHp=400~{\rm GeV} \\
\mbox{(Scenario B)}&:\quad \mHp=1000~{\rm GeV}, 
\end{align}
For Scenario A, Type II and Y are already excluded by the flavor constraint (see, e.g, Ref.~\cite{Haller:2018nnx}), so that we compare the difference of the branching ratios between Type-I and Type-X. 
For Scenario B, all the types of THDMs are not excluded by the flavor constraints. 
In order to avoid constraint from the T parameter, we set the mass of the CP-odd Higgs boson as $m_{A}=\mHp$. 
Whereas, the mass of the heavier CP-even Higgs boson is taken in the following range for each scenario, 
\begin{align}
\label{scanA:mH}
250~{\rm GeV}<m_{H}< 800~{\rm GeV} \quad\quad \mbox{for Scenario A}, \\
\label{scanB:mH}
800~{\rm GeV}<m_{H}< 1200~{\rm GeV} \quad\quad \mbox{for Scenario B}.
\end{align}
For the lower bound of Scenario A, we take into account the constraint from the direct search for $H\to ZZ^{\ast}$, by which $m_{H}\lesssim 250~{\rm GeV}$ and $\tb \lesssim 6~(5)$ are excluded in the case of $\sba=0.995$ with $\cba<0$ for Type-I (X)~\cite{Aiko:2020ksl}. 
The remaining parameters are scanned for both the scenarios as 
\begin{align}
\label{scan:other}
0.995<\sba<1,\quad 2<\tb<10, \quad 0<M< \mHp+500{\rm GeV}, 
\end{align}
considering both cases of $\cba<0$ and $\cba>0$. 
The lower bound of $\tb$ comes from the consideration of the constraint from $B_{d}\to\mu\mu$ for Scenario~A~\cite{Haller:2018nnx} and $ H \to hh$ in the case of $\sba=0.995$ with $\cba<0$ for Scenario~B~\cite{Aiko:2020ksl}. 
With these scan regions, we impose the theoretical constraints and the S,T parameters in the same way as Sec.~\ref{sec:IV}. 
Furthermore, we exclude parameter points that are not consistent with the current data of the Higgs signal strength at the LHC~in Ref.~\cite{ATLAS:2019nkf}. 
We calculate the decay rates for $h$ with NLO EW and NNLO QCD corrections by utilizing {\tt H-COUP v2}~\cite{Kanemura:2019slf} and evaluate the scaling factors $\kappa_{X}=\sqrt{\Gamma^{\rm THDM}_{h\to XX}/\Gamma^{\rm SM}_{h\to XX}}$ for each parameter point. 
We then remove parameter points if the calculated scaling factors deviate from the values presented in Table 11 (a) of Ref.~\cite{ATLAS:2019nkf} at 95~\% CL. 

While the alignment limit is defined by $\sba=1$ at tree level, this might not be valid beyond tree level. 
At loop levels, the quantum corrections by additional Higgs bosons can give non-zero contributions to $\Gamma_{hVV^{\ast}}$ even in $\sba=1$. 
Hence, we use the scaling factor $\kappa_{Z}$ and define the alignment limit as $\kappa_{Z}=1$ at loop levels. 
At the ILC 250, expected 1$\sigma$ (2 $\sigma$) accuracies of $\kappa_{Z}$ is 0.38\% (0.76\%)~\cite{Fujii:2017vwa}. 
Thus, we mainly discuss the behavior of the branching ratios of $\Hp$ for each type of THDMs within $\Delta \kappa_{Z} (\equiv \kappa_{Z}-1)\lesssim 0.76\%$, assuming situations that the deviations in the $h$ couplings are not found. 

In Fig.~\ref{FIG:SA}, we present the branching ratios at NLO in Type-I (top panels) and Type-X (bottom panels) for Scenario~A as a function of $\Delta \kappa_{Z}$, where the color points denote values of $\tb$. 
For $\BR(\Hp\to t\bar{b})$, one can see that the size of the branching ratio can reach almost 100$\%$ without depending on $\Delta \kappa_{Z}$ as well as types of THDMs. 
The reason is that such sizable $\BR(\Hp\to t\bar{b})$ is realized in the low $\tb$ region, where the top Yukawa coupling in the $H^{+}\bar{t}{b}$ vertex dominates for $\Hp\to t\bar{b}$. 
Thus, the difference between Type-I and Type-X does not appear. 
For $\BR(\Hp\to \tau^{+}{\nu})$, the prediction of Type-X is obviously larger than that of Type-I because of the $\tb$ enhancement for the $\tau$ Yukawa coupling in the $H^{+}\bar{\nu_{\tau}}{\tau}$ vertex for Type-X.  
Namely, in Scenario A, Type-X can be identified if $\BR(\Hp\to \tau^+\nu)$ is sizable for the discovered charged Higgs bosons. 
On the other hand, characteristic predictions of Type-I can be obtained in $\Hp\to W^+h$. 
The branching ratio $\BR(\Hp\to W^+h)$ in Type-I can exceed 20\%, while the prediction of Type-X is maximally around 11\%. 
Hence, a large $\BR(\Hp\to W^+h)$ is a clear signature in identifying Type-I. 
An intriguing point is that this signature can be mostly realized in the regions of $\Delta \kappa_{Z}\lesssim 0.76\%$. 
If the deviation in the $hZZ$ coupling is $\Delta \kappa_{Z}\lesssim -1\%$, $\BR(\Hp\to W^+h)$ is less than $5\%$ in both Type-I and Type-X. 
One can also see that these signatures of Type-I and Type-X contain information on the inner parameters of THDMs. 
Sizable values of $\BR(\Hp\to \tau\bar{\nu})$ in Type-X and $\BR(\Hp\to W^+h)$ in Type-I are caused in the large $\tb$ region. 
Therefore, information on $\tb$ can be extracted once these branching ratios are determined. 

In addition, we comment on behavior of the branching ratio for $\Hp\to W^{+(*)}H$. 
This decay mode kinematically opens when the heavier CP-even Higgs boson is lighter than the charged Higgs bosons. 
 The maximal size of $\BR(\Hp\to W^{+(*)}H)$ can reach almost 90\% in both Type-I and Type-X. 
As can be seen by comparing values of $\tan\beta$ and $\Delta \kappa_{Z}$, parameter points with huge values of $\BR(\Hp\to W^{+(*)}H)$ correspond to those with suppressed $\BR(\Hp\to t\bar{b})$. 

In Fig.~\ref{FIG:SB}, the branching ratios at NLO in Type-I, II, X and Y for Scenario~B are shown as a function of $\Delta \kappa_{Z}$ from left to right panels. 
Behavior of the $\BR(\Hp\to t\bar{b})$ are similar to Scenario~A, namely, the size of the branching ratio can be huge in the low $\tb$ region for all types of THDMs. 
Behavior of the $\BR(\Hp\to W^{+}H)$ also does not almost change from Scenario~A. 
In addition, for identification of Type-I and Type-X, one can rely on the processes $\Hp \to W^{+}h$ and $\Hp \to \tau^+\nu$ as same as Scenario~A. 
In Scenario~B, $\BR(\Hp \to W^{+}h)$ of Type-I can be considerably enhanced unlike other types of THDMs. 
The size of $\BR(\Hp \to\tau^{+}\nu)$ can reach 30\% only in Type-X. 
An interesting feature of Scenario B is that a sizable $\BR(\Hp \to W^{+}h)$ is only realized in the case of $\cba>0$ differently from Scenario A\footnote{The branching ratio $\BR(\Hp \to W^{+}h)$ can be enhanced in large $\tb$ regions, which does not occur in case of $\cba<0$ for Scenario B due to the theoretical constraints. The similar behavior can be seen in Figs.~\ref{FIG:BR600} and \ref{FIG:BR600md300}. }.
Hence, not only the size of $\tb$ but also the sign of $\cba$ can be extracted from the size of $\BR(\Hp \to W^{+}h)$. 
We note that all points with $\BR(\Hp \to W^{+}h)> 10\%$ correspond to $\cba>0$ for all types of THDMs. 

From the decay modes $\Hp\to t\bar{b}$, $\Hp\to \tau^{+}\nu$, $\Hp\to W^{+}h$ and $\Hp\to W^{+}H$, it would be difficult to separate Type-II and Type-Y. 
However, this can be performed by looking at the decay process $\Hp\to c\bar{b}$ as shown in Fig.~\ref{FIG:cbvstaunu}\footnote{For the evaluation of $\Gamma(\Hp\to c\bar{b})$, we only include QCD corrections. The EW corrections to this process are not implemented.}. 
For Type-II and Type-Y, $\BR(\Hp\to c\bar{b})$ can be larger than 0.1\% and one can distinguish these types from Type-I and Type-X by this decay mode. 
Furthermore, Type-II and Type-Y can be discriminated from the size of $\BR(\Hp \to\tau^{+}\nu)$. 
As seen from the left top panel of the Fig.~\ref{FIG:cbvstaunu}, enhancement of $\BR(\Hp\to c\bar{b})$ and/or $\BR(\Hp\to \tau^{+}\nu)$ is controlled by a value of $\tb$ inType-II, Type-X and Type-Y. 
From the right top panel of the figure, one can also see that there is a correlation between the branching ratios and the mass difference $m_{H}-\mHp$, in particular for Type-X and Type-Y. 
The reason for this can be understood as follows.  
When the mass difference is negatively large, ${\rm BR}(\Hp\to W^{+}H)$ becomes sizable without depending on the types of THDMs. This then reduces the size of $\BR(\Hp\to c\bar{b})$ and $\BR(\Hp\to \tau^{+}\nu)$. 
\red{We have studied on theoretical possibilities that Type II and Type X are separated from the other types of the THDMs by correlation  between $\BR(\Hp\to c\bar{b})$ and $\BR(\Hp \to\tau^{+}\nu)$. 
The predictions for $\BR(\Hp\to c\bar{b})$ in Type II and Type Y, which are  maximally $0.1\%$, are not large.
Phenomenological studies on expectation whether such a small branching ratio is measured at future colliders is beyond the scope of this paper. }

We give a comment on the results in another case of the degenerated mass of the additional Higgs bosons, i.e., $m_{H^{\pm}}=m_{H}$, where the T parameter constraint is satisfied when $\sba\simeq1$. 
We have performed the same analysis in this case, 
and obtained qualitatively similar results for magnitudes of the branching ratios while the allowed parameter regions after imposing the constraints from theoretical bounds and the electroweak oblique parameters are more strict than the case of $m_{H^{\pm}}=m_A$. 

Before we close this subsection, we mention the deviations in the $h$ couplings for Scenario A and for Scenario B. 
For both the scenarios all types of THDMs can be identified by looking at the branching ratios of the charged Higgs bosons even in the case of $\Delta \kappa_{Z}\lesssim 0.76\%$, where the deviation in the $hZZ$ coupling cannot be detected at the ILC~\cite{Fujii:2017vwa}. 
At the same time, even in this case, the deviations in other $h$ couplings like the Yukawa interactions can be sizable enough to be detected at the ILC 250 GeV~\cite{Fujii:2017vwa}. 
Namely, the deviations $\Delta \kappa_{b}$ and $\Delta \kappa_{\tau}$ for Type-II, $\Delta \kappa_{\tau}$ for Type-X, and $\Delta \kappa_{b}$ for Type-Y can still deviate significantly enough to be detected at the ILC 250 GeV. 
Therefore, a combination of the charged Higgs boson decays and the $h$ decays make it possible to identify details of THDMs. 

\subsection{ Impact of one-loop corrections to the branching ratios }
\begin{figure}[t]\centering
\includegraphics[width=0.5\linewidth]{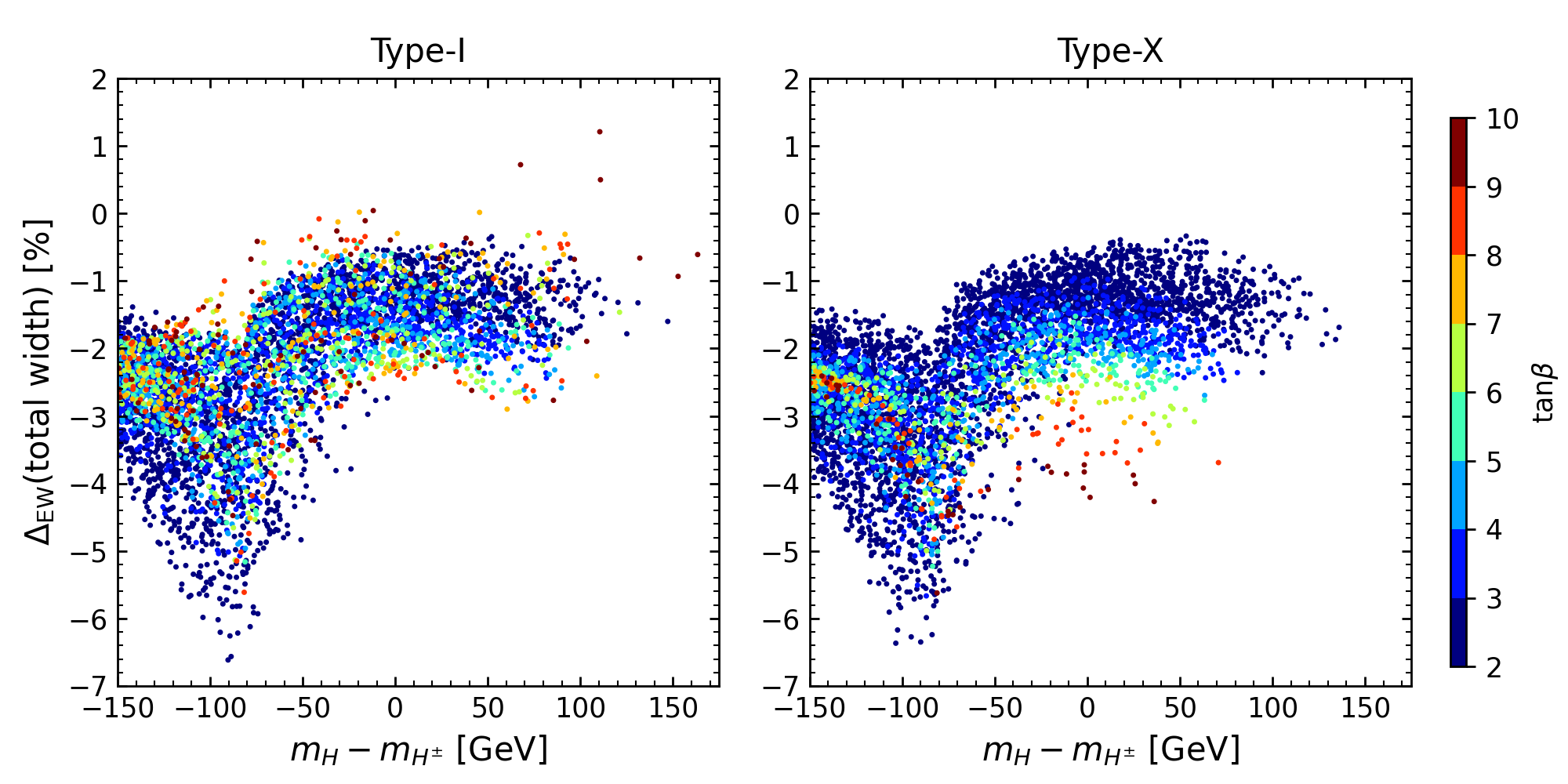}\hspace{0mm}
\caption{
The total decay width of the charged Higgs bosons as a function of the mass difference $m_{H}-\mHp$ in Scenario A for Type-I and Type-X. 
The colored dots correspond to different values of $\tb$. 
}
\label{FIG:SAtotal}
\end{figure}

\begin{figure}[t]\centering
\includegraphics[width=1.0\linewidth]{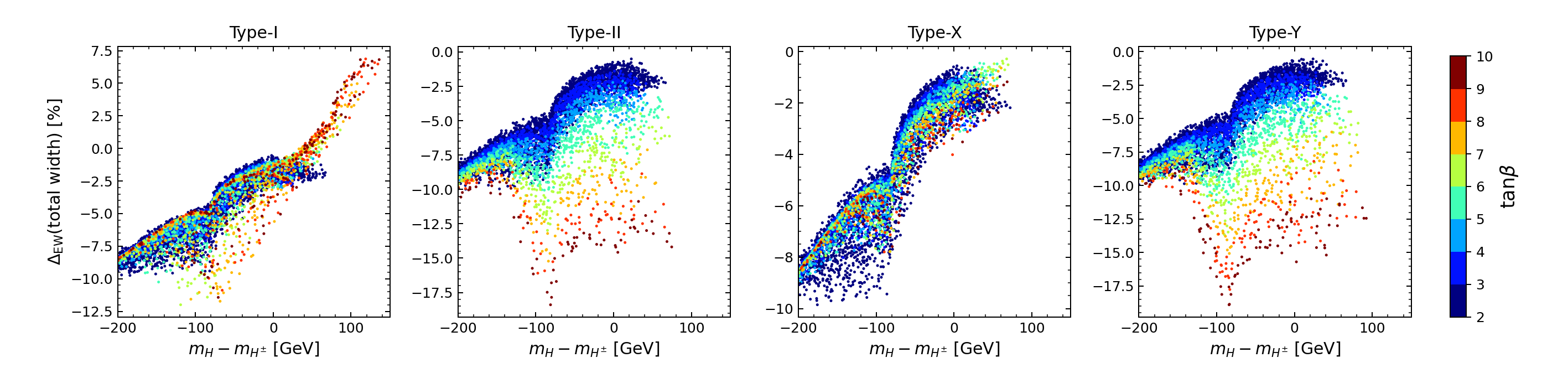}\hspace{0mm}
\caption{
The total decay width of the charged Higgs bosons as a function of the mass difference $m_{H}-\mHp$ in Scenario B for Type-I, II, X and Y. 
The colored dots correspond to different values of $\tb$. 
}
\label{FIG:SBtotal}
\end{figure}

\begin{figure}[t]\centering
\includegraphics[width=1\linewidth]{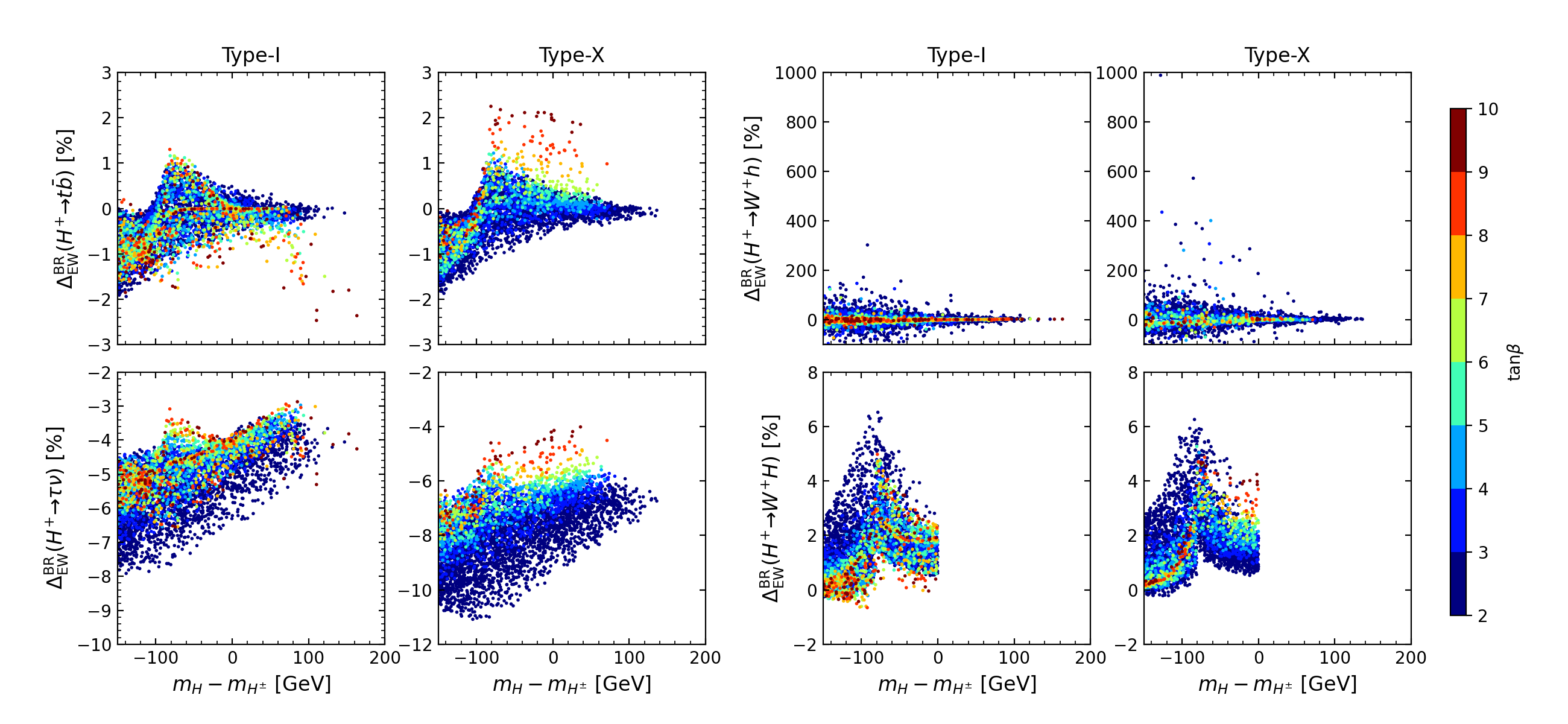}\hspace{0mm}
\caption{
Decay branching ratios of the charged Higgs bosons as a function of the mass difference $m_{H}-\mHp$ in Scenario A for Type-I and Type-X. 
The colored dots correspond to different values of $\tb$. 
}
\label{FIG:SABRs}
\end{figure}

\begin{figure}[t]\centering
\includegraphics[width=1\linewidth]{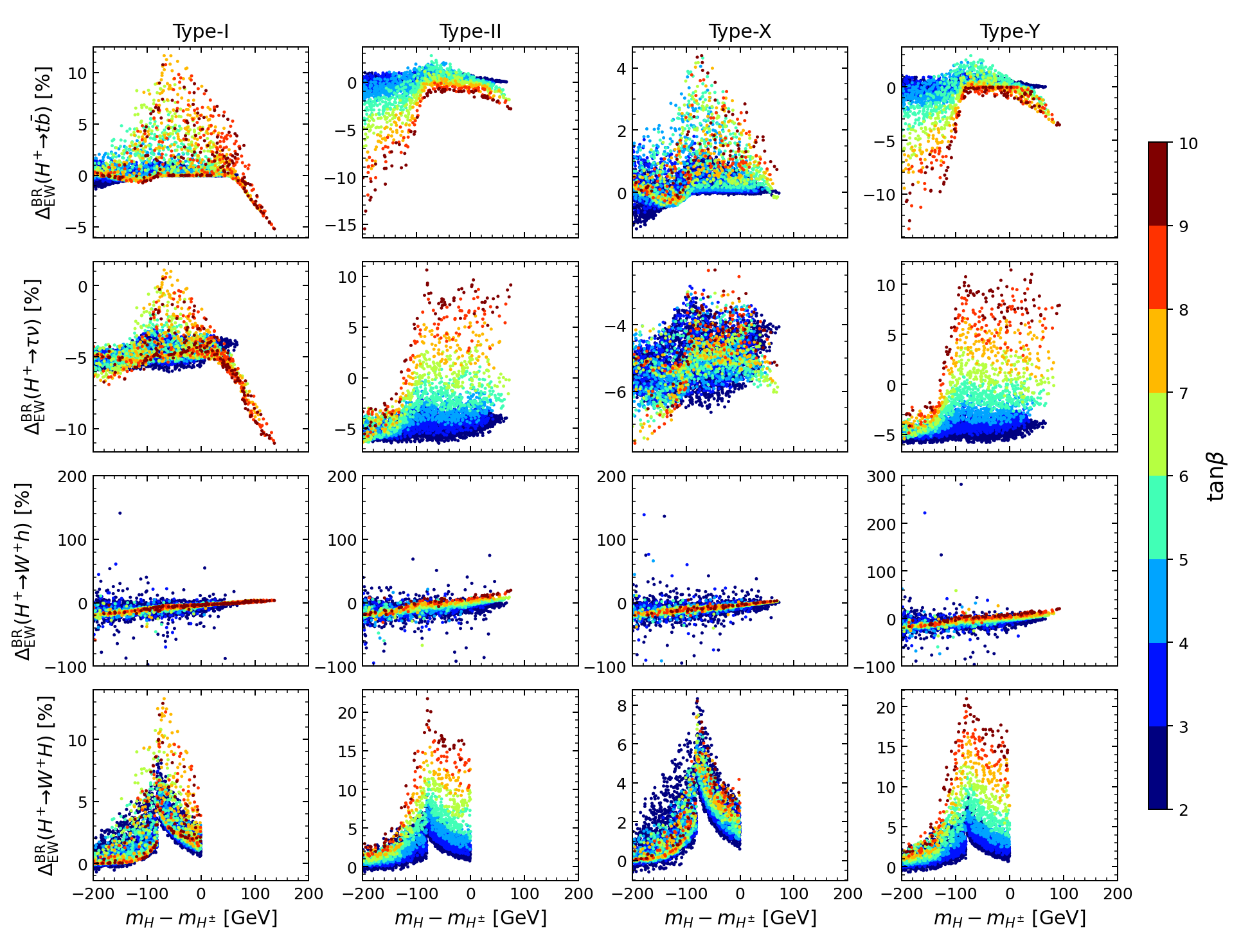}\hspace{0mm}
\caption{
Decay branching ratios of the charged Higgs bosons as a function of the mass difference $m_{H}-\mHp$ in Scenario B for Type-I, II, X and Y. 
The colored dots correspond to different values of $\tb$. 
}
\label{FIG:SBBRs}
\end{figure}

We next investigate the impact of NLO EW corrections on the branching ratios of $\Hp$. 
In particular, we discuss how the size of the corrections changes depending on the mass difference between the charged Higgs bosons and the additional neutral Higgs bosons. 
Focusing on Scenario~A and Scenario~B, we consider the case where the masses of CP-odd Higgs boson and that of the charged Higgs bosons are degenerate, $m_{A}=\mHp$. 
As pointed out in Ref.~\cite{Pomarol:1993mu}, in this case, the custodial symmetry is restored in the Higgs potential, so that the constraint from the T parameter is satisfied. 
We then scan $m_{H}$ in the regions given in Eqs.~\eqref{scanA:mH} and \eqref{scanB:mH}. 
 The other parameters are scanned as given in Eq.~\eqref{scan:other} for both scenarios. 
 In the following discussions, we introduce a quantity to describe magnitudes of the NLO EW corrections to the branching ratios, i.e., 
 \begin{align}
\Delta^{\rm BR}_{\rm EW}(\Hp \to XY)=\frac{{\rm BR}^{\rm NLO\ EW}(\Hp \to XY)}{{\rm BR}^{\rm LO}(\Hp \to XY)}-1, 
\end{align}
where ${\rm BR}^{\rm NLO\ EW}(\Hp \to XY)$ denotes the branching ratios with NLO EW corrections. 
In the evaluation of the branching ratio at LO ${\rm BR}^{\rm LO}(\Hp \to XY)$, the quark running masses are applied for the decays into quarks. 
We also describes the NLO EW corrections for the total decay width, which is defined by 
\begin{align}
\Delta_{\rm EW}({\rm total\ width})=\frac{\Gamma^{\rm NLO\ EW}_{\Hp}}{\Gamma^{\rm LO}_{\Hp}}-1,
\end{align}
with the total decay width for $\Hp$ at NLO EW (LO) being $\Gamma^{\rm NLO\ EW}_{\Hp}$ $(\Gamma^{\rm LO}_{\Hp})$. 
By definition, $\Delta^{\rm BR}_{\rm EW}$ can be reduced as $\Delta^{\rm BR}_{\rm EW}+1=(\Delta^{}_{\rm EW}-1)/(\Delta^{}_{\rm EW}({\rm total\ width})-1)$. 
Namely, it is controlled by the correction factor for the total decay width $\Delta_{\rm EW}({\rm total\ width})$ and the one for partial decay width $\Delta_{\rm EW}(\Hp \to XY)$.

In Fig.~\ref{FIG:SAtotal}, we present the correction factor for the total width in Scenario A as a function of the mass difference for the additional Higgs bosons. 
The color dots denote the values of $\tb$. 
Behavior of the corrections for Type-I (left panel) and the one for Type-X (right panel) is similar with each other in the regions $\tb\lesssim 4$, 
while these are different in $\tb\gtrsim 6$. 
When $m_{H}-m_{H^{\pm}}\simeq -80$ GeV, there appear thresholds, in which $H^{+}\to W^{+} H$ opens and the correction $\Delta^{}_{\rm EW}$(total width) reaches $-7\%$. 

In Fig.~\ref{FIG:SBtotal}, the results in Scenario B are shown for all types of THDMs. 
Similar to Scenario A, behavior for $\tb\lesssim 4$ does not change match for all the types. 
Clear difference among the types of THDMs arises for $\tb\gtrsim 4$. 
For Type-I, the allowed region of the mass difference $m_{H}-m_{\Hpm}$ is wider than the other types of THDMs. 
Consequently, the correction can be positive when $m_{H}-m_{\Hpm}\gtrsim 30\ {\rm GeV}$. 
On the other hand, for Type-II and Type-Y, the bulk of points with $\tb \gtrsim 5$ shows large negative corrections, compared with those for $\tb \lesssim 4$. 
This is because the bottom Yukawa coupling in $\Hp\bar{t}{b}$ vertex is enhanced by large $\tb$. 
The correction can reach $-20$ ($-25$) \% when $m_{H}-m_{\Hpm}\simeq -80 {\rm GeV}$ for Type-II (Type-Y) due to the effect of the threshold of the mode $\Hp\to W^+H$. 
For Type-X, the predictions in the high $\tb$ region almost do not deviate from those in the low $\tb$ region. 

We now move on discussions of the correction factor for the branching ratios. 
In Fig.~\ref{FIG:SABRs}, $\Delta_{\rm EW}^{\rm BR}$'s for the decays $\Hp \to t\bar{b},\ \Hp \to \tau^+\nu,\ \Hp \to W^+h,\ \Hp \to W^+H$ are shown. 
For $\Hp \to t\bar{b}$, The kink when $m_{H}-m_{H^{\pm}}\simeq 80$ GeV appears as with the correction for the total width (see Fig.~\ref{FIG:SAtotal}). 
The correction of the branching ratio distribute in narrow range, $-2.75\%\ (-1.9)\lesssim\Delta_{\rm EW}^{\rm BR} (\Hp \to t\bar{b})\lesssim +1.25\ (+2.3)\%$, for Type-I (Type-X). 
This can be understood as follows. 
A value of the correction factor for the partial width $\Delta_{\rm EW}(\Hp \to t\bar{b})$ is close to $\Delta_{\rm EW}({\rm total\ width})$ in the bulk of parameter points. 
Thereby, they are canceled with each other in the definition of $\Delta_{\rm EW}^{\rm BR}$. 
On the other hand, behavior of $\Delta_{\rm EW}^{\rm BR}(\Hp \to W^+H)$ is different from that of $\Delta_{\rm EW}^{\rm BR}(\Hp \to t\bar{b})$. 
We note that $\Delta_{\rm EW}^{}(\Hp \to W^+H)$ is relatively small, i.e., $-2.5\% \lesssim\Delta_{\rm EW}^{} \lesssim\ 0\%$, for both Type-I and Type-X \footnote{We have calculated the NLO EW corrections to the on-shell two-body decay of $\Hp \to W^+H$. In the range of $m_{W}<m_{H}-m_{\Hpm} <0\ {\rm GeV}$, where the off-shell decay $\Hp \to W^{\ast}H$ happens, the NLO EW corrections have not been implemented. }, so that $\Delta_{\rm EW}^{\rm BR}(\Hp \to W^+H)$ is dominated by the $\Delta_{\rm EW}^{}$ (total width). 
In fact, behavior of $\Delta_{\rm EW}^{\rm BR}$ is reversal of $\Delta_{\rm EW}$ (total width). 

One can also see that the size of the correction for $\Hp \to W^{+}h$ can be remarkably large. 
In the low $\tb$ region, $\Delta_{\rm EW}^{\rm BR}(\Hp\to W^+h)$ can exceed +100\%. 
We note that $c_{\beta-\alpha}$ is close to 0, $|c_{\beta-\alpha}|\lesssim 2.5\times 10^{-2}$, for all the parameter points with $\Delta_{\rm EW}^{\rm BR}(\Hp\to W^+h)\gtrsim +100\%$. 
In addition, the one-loop amplitude of $\Hp\to W^+h$ contains terms to be independent of $c_{\beta-\alpha}$. 
The counterterms $\delta C_{h}$ and $\delta C_{H^{\pm}}$ induce such terms, which can be enhanced by the non-decoupling effect of the additional Higgs bosons
 in case of $M\sim v$ .  
In this case, the one-loop amplitude can overcome the tree-level amplitude, and gives $\Delta_{\rm EW}^{\rm BR}(\Hp\to W^+h)\gtrsim 100\%$. 
Furthermore, we found that in some parameter points $\Delta_{\rm EW}^{\rm BR}(\Hp\to W^+h)$ can be smaller than $-100\%$.
The origin is considered due to the fact that terms of the squared one-loop amplitude are truncated in the calculation of the NLO corrections. 
The effect of the squared one-loop amplitude is discussed in the Sec.\ref{sec:sqWh}. 

In Fig.~\ref{FIG:SBBRs}, we show the results for the correction factors of the charged Higgs bosons in Scenario B for all types of THDMs. 
For $\Hp\to t\bar{b}$, the same picture described in the results for Scenario A holds for the low $\tb$ region. 
When $\tb\simeq 2$-$3$, the correction factor is close to zero, $-2.5\%\lesssim \Delta_{\rm EW}^{\rm BR}(\Hp\to t\bar{b})\lesssim +1.5\%$ for all types of THDMs. 
For high $\tb$ values the size of the correction can be much large. 
In Type-I, $\Delta_{\rm EW}^{\rm BR}(\Hp\to t\bar{b})$ can exceed 12\% near the threshold region $m_{H}-m_{H^{\pm}}\simeq -80$~GeV, while $\Delta_{\rm EW}^{\rm BR}(\Hp\to t\bar{b})$ can be negative in the case of $m_{H}-m_{H^{\pm}}\gtrsim 50$~GeV due the effect of $\Delta_{\rm EW}^{}({\rm total\ width})$. 
In Type-II and Y, $\Delta_{\rm EW}^{\rm BR}(\Hp\to t\bar{b})$ can be negatively large and can reach $-15.5\%$ due to the $\tb$ enhancement of the bottom Yukawa coupling in the $H^{+}\bar{t}b$ vertex for $\Delta_{\rm EW}(\Hp\to t\bar{b})$, which can be seen in Fig.~\ref{FIG:BR3}. 
For $\Hp \to \tau^{+}\nu$, one can see that behavior of $\Delta_{\rm EW}^{\rm BR}$ with the low $\tb$ value is similar without depending on the type of THDMs, but the difference can appear in the high $\tb$ region. 
For $\Hp \to W^{+}H$, we note that the correction factor for the partial width $\Delta_{\rm EW}(\Hp \to W^{+}H)$ monotonically decreases as the mass difference $m_{H}-m_{\Hp}$ becomes negatively large, e.g., $\Delta_{\rm EW}\sim -8\%$ $(-4\%)$ when $m_{H}-m_{\Hp}$ =200 GeV (100 GeV) for all types of THDMs.  
When the on-shell decay $\Hp \to W^{+}H$ is kinematically allowed, the correction $\Delta_{\rm EW}^{\rm BR}(\Hp \to W^{+}H)$ is determined by summation of $\Delta_{\rm EW}({\rm total\ width})$ and $\Delta_{\rm EW}(\Hp \to W^{+}H)$. 
The maximum value of the $\Delta_{\rm EW}^{\rm BR}(\Hp \to W^{+}H)$ is $+13\%$ and $+9\%$ for Type-I and Type-X, respectively, while that of Type-II and Type-Y is $+22 \%$. 

Finally, we comment on the results in the other case of the degenerate mass of the additional Higgs bosons, i.e., $ m_{H}=m_{H^{\pm}}$.  
We find that for Scenario A the correction factor of the total decay width $\Delta_{\rm EW}({\rm total\ width})$ shows a cusp structure at $m_{A}= 2m_{t}$, which is realized by the threshold of the top loop diagrams in $\delta \beta$, so that behavior is changed from the case the $ m_{A}=m_{H^{\pm}}$. 
However, the maximum and minimum values of $\Delta_{\rm EW}({\rm total\ width})$ are similar to the results of $ m_{A}=m_{H^{\pm}}$. 
Behavior of the $\Delta_{\rm EW}^{\rm BR}$ for the processes discussed above is somewhat different from the case of $ m_{A}=m_{H^{\pm}}$, while values of $\Delta_{\rm EW}^{\rm BR}$ distribute in the similar region to the case of $ m_{A}=m_{H^{\pm}}$. 
For Scenario B, we also note that the size of $\Delta_{\rm EW}^{\rm BR}$ for all the processes in the case of $ m_{H}=m_{H^{\pm}}$ tend to be smaller than that of the case $ m_{A}=m_{H^{\pm}}$ except for Type-I. 
For Type-I, the maximum value of $\Delta_{\rm EW}^{\rm BR}(\Hp\to t\bar{b})$ and $\Delta_{\rm EW}^{\rm BR}(\Hp\to \tau^{+}\nu)$ is $+12.7\%$ and $+2.6\%$, respectively, while the one for $\Delta_{\rm EW}^{\rm BR}(\Hp\to W^+h)$ is similar to the case of $ m_{A}=m_{H^{\pm}}$.

 \subsection{ Effect of the squared one-loop amplitude to $\Hp\to W^{+}h$ }\label{sec:sqWh}
\begin{figure}[t]\centering
\includegraphics[width=0.6\linewidth]{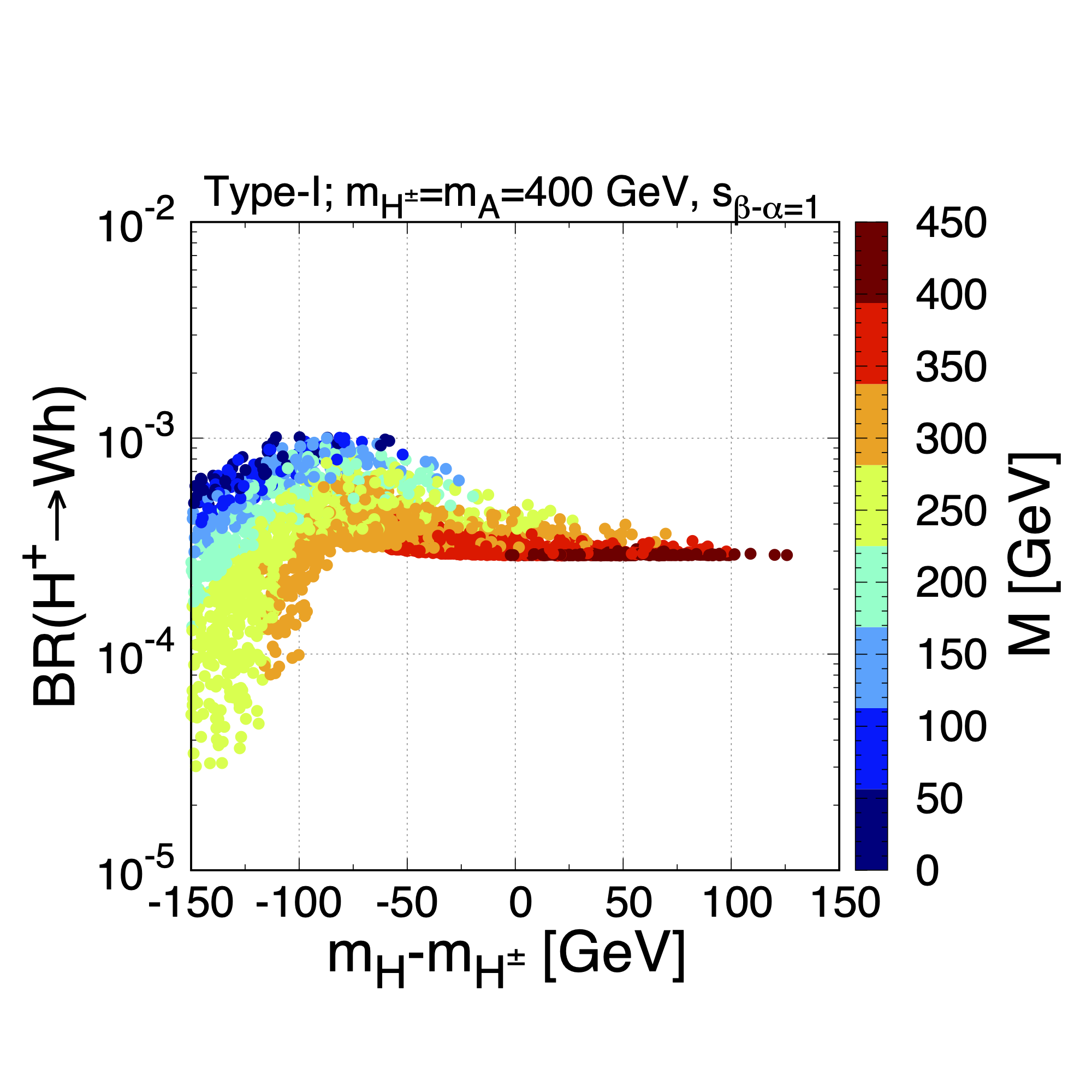}\hspace{0mm}
\caption{
The decay branching ratio for $\Hp\to W^{+}h$ for Scenario A of Type-I in the alignment limit, $\sba=1$, as a function of the mass difference $m_{H}-\mHp$, where the squared one-loop amplitude for $\Hp\to W^{+}h$ is included in the evaluation. 
The colored dots correspond to different values of $M$. 
}
\label{FIG:Wh1sq}
\end{figure}

Before we close this section, we discuss the effect of the squared one-loop amplitude for $\Hp\to W^{+}h$.
The squared amplitude for this process can be expressed by
\begin{align}\label{eq:ampsqWh}
|\mathcal{M}(\Hp\to W^+h)|^{2}&=C^{}_{\Hp W^- h}
\left(|\Gamma_{\Hp W^- h}^{\rm tree}|^{2}+2\Gamma^{\rm tree}_{\Hp W^- h}{\rm Re}\Gamma^{\rm loop}_{\Hp W^- h}+|\Gamma^{\rm loop}_{\Hp W^- h}|^{2}\right) \notag \\
&=C^{}_{\Hp W^- h}
\left(\frac{g^{2}}{4}\cba^{2}+g\cba{\rm Re}\Gamma^{\rm loop}_{\Hp W^- h}+|\Gamma^{\rm loop}_{\Hp W^- h}|^{2}\right),
\end{align}
where $C^{}_{\Hp W^- h}=\mHp^{4}/m_{W}^{2}\lambda(\mu_{h},\mu_{W})$.
The first (second) term corresponds to LO (NLO) contributions to $\Gamma(\Hp\to W^{+}h)$. 
We have involved up to the second term in above discussions. 
The third term $|\Gamma^{\rm loop}_{\Hp W^- h}|^{2}$ is the same order as contributions from the tree-level amplitude times two-loop amplitude, namely NNLO contributions. 
It is notable that this term contains contributions that are not proportional to $\cba$. 
Hence, the term $|\Gamma^{\rm loop}_{\Hp W^- h}|^{2}$ does not disappear even in the alignment limit. 
The term $|\Gamma^{\rm loop}_{\Hp W^- h}|^{2}$ can be identified as a leading contribution 
if one assumes $\cba$ to be tiny and expands the squared amplitude $|\mathcal{M}(\Hp\to W^+h)|^{2}$ into a power series of $\cba$. 
We notice from naive order estimation that the term $|\Gamma^{\rm loop}_{\Hp W^- h}|^{2}$ can be comparable with the first and second terms in Eq.~\eqref{eq:ampsqWh} 
under the situation where $|\cba|\lesssim 0.1$ and $M\lesssim v$. 
Therefore, the third term of Eq.~\eqref{eq:ampsqWh} could be significant in the alignment regions. 

To discuss  the effect of $|\Gamma^{\rm loop}_{\Hp W^- h}|^{2}$ in the region of 0.995$\lesssim\sba<1$ is beyond the scope of this paper. 
If $\sba\neq1$, the third term $|\Gamma^{\rm loop}_{\Hp W^- h}|^{2}$ contains the IR divergence and evaluation of real photon emissions at NNLO are required in order to obtain the IR finite result for $\Gamma(\Hp\to W^{+}h)$. 
As long as the case of the alignment limit $\sba=1$, the third term is IR finite. 
In addition, contributions to the squared amplitude from the tree-level amplitude times two-loop amplitude vanish in the alignment limit. 
Therefore, we only discuss two-loop corrections due to the third term in Eq.~\eqref{eq:ampsqWh} to the branching ratio for $\Hp\to W^{+}h$ only in the case of $\sba=1$, and how this is important. 

In Fig.~\ref{FIG:Wh1sq}, we show the branching ratio for $\Hp\to W^{+}h$ including the squared term $|\Gamma^{\rm loop}_{\Hp hW^{-}}|^{2}$ for Scenario A of Type-I in the alignment limit $\sba=1$ as a function of the mass difference $m_{H}-\mHp$. The other parameters $\tb$ and $M$ are scanned as in Eq.~\eqref{scan:other}. 
One can see that the branching ratio can be $0.1\%$ at most. 
The intriguing point is that the branching ratio is maximized when the soft-breaking parameter $M$ is small. 
We have also calculated the branching ratio in Scenario A for Type-X and Scenario B for all types of THDMs. 
For Scenario A of Type-X, we have got the almost same results with Type-I. 
On the other hand, for Scenario B, the soft-breaking parameter $M$ cannot be small under the theoretical consistencies such as the perturbative unitarity and the vacuum stability. 
Hence, the branching ratios are small, as compared to Scenario A. 
We find that the maximum value of the branching ratio is $0.004\%$ in Scenario B for all types of THDMs. 
\red{In short}, the non-decoupling effect of the additional Higgs bosons enhances the decay $\Hp\to W^{+}h$ through the contribution from the term of $|\Gamma^{\rm loop}_{\Hp hW^{-}}|^{2}$, and the branching ratio can be $0.1 \%$ when $\mHp=400$~GeV.

%% file: Apendix.tex

\section{Scalar couplings}\label{sec:coup}
From the Higgs potential, we obtain the scalar trilinear and the scalar quartic couplings.
We use the following notation for these couplings,
\begin{align}
\mathcal{L} = +\lambda_{\phi_{i}\phi_{j}\phi_{k}}\phi_{i}\phi_{j}\phi_{k}+\lambda_{\phi_{i}\phi_{j}\phi_{k}\phi_{l}}\phi_{i}\phi_{j}\phi_{k}\phi_{l}+\cdots.
\end{align}
In the following, we list the expression of the scalar trilinear and the scalar quartic couplings in terms of the coefficients of the Higgs potential in the Higgs basis.
In terms of the masses of Higgs bosons and mixing angles, the coefficients $Z_{1},..., Z_{7}$ are given as~\cite{Davidson:2005cw, Bernon:2015qea}
\begin{align}
%
Z_{1}v^{2} &= m_{H}^{2}\cos^{2}{(\beta-\alpha)} + m_{h}^{2}\sin^{2}{(\beta-\alpha)}, \\
Z_{2}v^{2} &= m_{H}^{2}\cos^{2}{(\beta-\alpha)} + m_{h}^{2}\sin^{2}{(\beta-\alpha)} + 4(m_{h}^{2}-m_{H}^{2})\cos{(\beta-\alpha)}\sin{(\beta-\alpha)}\cot{2\beta} \notag \\
&\quad
+ 4\left[(m_{H}^{2}-M^{2})\sin^{2}{(\beta-\alpha)} + (m_{h}^{2}-M^{2})\cos^{2}{(\beta-\alpha)}\right]\cot^{2}{2\beta}, \\
Z_{3}v^{2} &= m_{H}^{2}\cos^{2}{(\beta-\alpha)} + m_{h}^{2}\sin^{2}{(\beta-\alpha)}+ 2(m_{h}^{2}-m_{H}^{2})\cos{(\beta-\alpha)}\sin{(\beta-\alpha)}\cot{2\beta} \notag \\
&\quad
+ 2(m_{H^{\pm}}^{2} - M^{2}), \\
Z_{4}v^{2} &= m_{H}^{2}\sin^{2}{(\beta-\alpha)} + m_{h}^{2}\cos^{2}{(\beta-\alpha)} + m_{A}^{2} - 2m_{H^{\pm}}^{2}, \\
Z_{5}v^{2} &= m_{H}^{2}\sin^{2}{(\beta-\alpha)} + m_{h}^{2}\cos^{2}{(\beta-\alpha)} - m_{A}^{2}, \\
Z_{6}v^{2} &= (m_{h}^{2} - m_{H}^{2})\cos{(\beta-\alpha)}\sin{(\beta-\alpha)}, \\
Z_{7}v^{2} &= 2\left[(m_{H}^{2}-M^{2})\sin^{2}{(\beta-\alpha)} + (m_{h}^{2}-M^{2})\cos^{2}{(\beta-\alpha)}\right]\cot{2\beta} \notag \\
&\quad
+(m_{h}^{2}-m_{H}^{2})\cos{(\beta-\alpha)}\sin{(\beta-\alpha)}.
\end{align}

The Higgs trilinear couplings which are relevant for charged Higgs bosons decays are given by
\begin{align}
\lambda_{H^{+}H^{-}h} &= -(Z_{3}s_{\beta-\alpha}+Z_{7}c_{\beta-\alpha})v, \\
\lambda_{H^{\pm}G^{\mp}h} &= -\qty[Z_{6}s_{\beta-\alpha}+\frac{1}{2}(Z_{4}+Z_{5})c_{\beta-\alpha}]v, \\
\lambda_{G^{+}G^{-}h} &= -(Z_{1}s_{\beta-\alpha}+Z_{6}c_{\beta-\alpha})v, \\
\lambda_{H^{+}H^{-}H} &= -(Z_{3}c_{\beta-\alpha}-Z_{7}s_{\beta-\alpha})v, \\
\lambda_{H^{\pm}G^{\mp}H} &= -\qty[Z_{6}c_{\beta-\alpha}-\frac{1}{2}(Z_{4}+Z_{5})s_{\beta-\alpha}]v, \\
\lambda_{G^{+}G^{-}H} &= -(Z_{1}c_{\beta-\alpha}-Z_{6}s_{\beta-\alpha})v, \\
\lambda_{H^{\pm}G^{\mp}A} &= \pm\frac{i}{2}(Z_{4}-Z_{5})v.
\end{align}

The Higgs quartic couplings which are relevant for charged Higgs bosons decays are given by
\begin{align}
\lambda_{H^{+}H^{-}H^{+}H^{-}} &= \lambda_{H^{+}H^{-}AA} = -\frac{1}{2}Z_{2}, \\
\lambda_{H^{+}H^{-}G^{+}G^{-}} &= 2\lambda_{H^{+}H^{-}G^{0}G^{0}} = \red{-Z_{3}}, \\
\lambda_{G^{+}H^{-}G^{+}G^{-}} &= 2\lambda_{G^{+}H^{-}G^{0}G^{0}} = -Z_{6}, \\
\lambda_{G^{+}H^{-}H^{+}H^{-}} &= 2\lambda_{G^{+}H^{-}AA} = -Z_{7}, \\
\lambda_{H^{+}H^{-}hh} &= -\frac{1}{2}(Z_{2}c_{\beta-\alpha}^{2}+Z_{3}s_{\beta-\alpha}^{2}+2Z_{7}c_{\beta-\alpha}s_{\beta-\alpha}), \\
\lambda_{H^{+}H^{-}HH} &= -\frac{1}{2}(Z_{2}s_{\beta-\alpha}^{2}+Z_{3}c_{\beta-\alpha}^{2}-2Z_{7}c_{\beta-\alpha}s_{\beta-\alpha}),\\
\lambda_{G^{+}H^{-}hh} &= -\frac{1}{2}\qty[Z_{6}s_{\beta-\alpha}^{2}+Z_{7}c_{\beta-\alpha}^{2}+(Z_{4}+Z_{5})c_{\beta-\alpha}s_{\beta-\alpha}], \\
\lambda_{G^{+}H^{-}HH} &= -\frac{1}{2}\qty[Z_{6}c_{\beta-\alpha}^{2}+Z_{7}s_{\beta-\alpha}^{2}-(Z_{4}+Z_{5})c_{\beta-\alpha}s_{\beta-\alpha}].
\end{align}

\section{Analytic expressions for the 1PI diagrams}\label{sec:ApB}
We here give the analytic expressions for the 1PI diagram contributions in terms of the Passarino-Veltman functions defined in Ref.~\cite{Passarino:1978jh}.
We calculate 1PI diagrams in the 't Hooft-Feynman gauge, i.e., $\xi=1$.
In addition to the scalar couplings introduced in Appendix~\ref{sec:coup}, we use the following coupling constants,
\begin{align}
v_{f} &= \frac{1}{2}I_{f}-Q_{f}s_{W}^{2}, \quad
a_{f} = \frac{1}{2}I_{f}, \\
g^{h}_{f} &= -\frac{m_{f}\zeta_{h}^{f}}{v}, \quad
g^{H}_{f} = -\frac{m_{f}\zeta_{H}^{f}}{v}, \quad
g^{A}_{f} = 2iI_{f}\frac{m_{f}\zeta_{f}}{v},\quad
g^{G^{0}}_{f} = 2iI_{f}\frac{m_{f}}{v}, \\
g_{S}^{G^{\pm}} &= \red{\pm}\frac{m_{f}-m_{f'}}{\sqrt{2}v}, \quad
g_{P}^{G^{\pm}} = \mp\frac{m_{f}+m_{f'}}{\sqrt{2}v}, \\
g_{S}^{H^{\pm}} &= \red{\pm}\frac{m_{f}\zeta_{f}-m_{f'}\zeta_{f'}}{\sqrt{2}v}, \quad
g_{P}^{H^{\pm}} = \mp\frac{m_{f}\zeta_{f}+m_{f'}\zeta_{f'}}{\sqrt{2}v},
\end{align}
where $\zeta_{h}^{f}=s_{\beta-\alpha}+\zeta_{f}c_{\beta-\alpha}$ and  $\zeta_{H}^{f}=c_{\beta-\alpha}-\zeta_{f}c_{\beta-\alpha}$ with $\zeta_{f}$ defined in Table~\ref{tab:z2}.
We have used $I_{f}$ and $Q_{f}$ as the third component of the iso-spin and the electric charge of the fermion $f$, respectively.

\subsection{Self-energies for $H^{+}H^{-}$ and $H^{+}G^{-}$}\label{sec: self}
The 1PI diagram contributions to the scalar boson two-point functions are calculated as
\begin{align}
(16\pi^{2}){\Pi_{H^{+}H^{-}}^{{\rm 1PI}}(p^{2})}_{F}
&=
-2N_{c}^{f}
\bigg\{\frac{m_{u}^{2}\zeta_{u}^{2}+m_{d}^{2}\zeta_{d}^{2}}{v^{2}}\qty[A(u)+A(d)+(m_{u}^{2}+m_{d}^{2}-p^{2})B_{0}(p^{2}; u, d)] \notag \\
&\quad
-4m_{u}m_{d}\frac{m_{u}\zeta_{u}m_{d}\zeta_{d}}{v^{2}}B_{0}(p^{2}; u, d)\bigg\} \notag \\
&\quad
-2\frac{m_{\ell}^{2}\zeta_{\ell}^{2}}{v^{2}}[A(\ell)+(m_{\ell}^{2}-p^{2})B_{0}(p^{2}; 0, \ell)], \\
(16\pi^{2}){\Pi_{H^{+}G^{-}}^{{\rm 1PI}}(p^{2})}_{F}
&=
-2N_{c}^{f}
\bigg\{\frac{m_{u}^{2}\zeta_{u}+m_{d}^{2}\zeta_{d}}{v^{2}}[A(u)+A(d)+(m_{u}^{2}+m_{d}^{2}-p^{2})B_{0}(p^{2}; u, d)] \notag \\
&\quad
-2m_{u}m_{d}\frac{m_{u}m_{d}(\zeta_{u}+\zeta_{d})}{v^{2}}B_{0}(p^{2}; u, d)\bigg\} \notag \\
&\quad
-2\frac{m_{\ell}^{2}\zeta_{\ell}^{2}}{v^{2}}
[A(\ell)+(m_{\ell}^{2}-p^{2})B_{0}(p^{2}; 0, \ell)], \\
(16\pi^{2}){\Pi_{H^{+}H^{-}}^{{\rm 1PI}}(p^{2})}_{B}
&=
-\frac{g^{2}}{4}c_{\beta-\alpha}^{2}[2A(W)-A(h)+(2p^{2}+2m_{h}^{2}-m_{W}^{2})B_{0}(p^{2}; W, h)] \notag \\
&\quad
-\frac{g^{2}}{4}s_{\beta-\alpha}^{2}[2A(W)-A(H)+(2p^{2}+2m_{H}^{2}-m_{W}^{2})B_{0}(p^{2}; W, H)] \notag \\
&\quad
-\frac{g^{2}}{4}[2A(W)-A(A)+(2p^{2}+2m_{A}^{2}-m_{W}^{2})B_{0}(p^{2}; W, A)] \notag \\
&\quad
-\frac{g_{Z}^{2}}{4}c_{2W}^{2}[2A(Z)-A(H^{\pm})+(2p^{2}+2m_{H^{\pm}}^{2}-m_{Z}^{2})B_{0}(p^{2}; Z, H^{\pm})] \notag \\
&\quad
-e^{2}[2A(\gamma)-A(H^{\pm})+(2p^{2}+2m_{H^{\pm}}^{2}-m_{\gamma}^{2})B_{0}(p^{2}; \gamma, H^{\pm})] \notag \\
&\quad
+2g^{2}\left[A(W)-\frac{1}{2}m_{W}^{2}\right]+g_{Z}^{2}c_{2W}^{2}\left[A(Z)-\frac{1}{2}m_{Z}^{2}\right]
+4e^{2}\left[A(\gamma)-\frac{1}{2}m_{\gamma}^{2}\right] \notag \\
&\quad
+\lambda_{H^{+}H^{-}h}^{2}B_{0}(p^{2}; H^{\pm}, h)+\lambda_{H^{+}G^{-}h}^{2}B_{0}(p^{2}; G^{\pm}, h) \notag \\
&\quad
+\lambda_{H^{+}H^{-}H}^{2}B_{0}(p^{2}; H^{\pm}, H)+\lambda_{H^{+}G^{-}H}^{2}B_{0}(p^{2}; G^{\pm}, H) \notag \\
&\quad
+\lambda_{H^{+}G^{-}A}\lambda_{H^{-}G^{+}A}B_{0}(p^{2}; G^{\pm}, A) \notag \\
&\quad
-\lambda_{H^{+}H^{-}hh}A(h)-\lambda_{H^{+}H^{-}G^{0}G^{0}}A(G^{0})-\lambda_{H^{+}H^{-}G^{+}G^{-}}A(G^{\pm}) \notag \\
&\quad
-\lambda_{H^{+}H^{-}HH}A(H)-\lambda_{H^{+}H^{-}AA}A(A)-4\lambda_{H^{+}H^{-}H^{+}H^{-}}A(H^{\pm}), \\
(16\pi^{2}){\Pi_{H^{+}G^{-}}^{{\rm 1PI}}(p^{2})}_{B}
&=
-\frac{g^{2}}{4}s_{\beta-\alpha}c_{\beta-\alpha}[2A(W)-A(h)+(2p^{2}+2m_{h}^{2}-m_{W}^{2})B_{0}](p^{2}; W, h) \notag \\
&\quad
+\frac{g^{2}}{4}s_{\beta-\alpha}c_{\beta-\alpha}[2A(W)-A(H)+(2p^{2}+2m_{H}^{2}-m_{W}^{2})B_{0}](p^{2}; W, H) \notag \\
&\quad
+\lambda_{H^{-}G^{+}h}\lambda_{H^{+}H^{-}h}B_{0}(p^{2}; H^{\pm}, h)+\lambda_{G^{+}G^{-}h}\lambda_{H^{-}G^{+}h}B_{0}(p^{2}; G^{\pm}, h) \notag \\
&\quad
+\lambda_{G^{+}H^{-}H}\lambda_{H^{+}H^{-}H}B_{0}(p^{2}; H^{\pm}, H)+\lambda_{G^{+}G^{-}H}\lambda_{H^{-}G^{+}H}B_{0}(p^{2}; G^{\pm}, H) \notag \\
&\quad
-\lambda_{G^{+}H^{-}hh}A(h)-\lambda_{G^{+}H^{-}G^{0}G^{0}}A(G^{0})-2\lambda_{G^{+}H^{-}G^{+}G^{-}}A(G^{\pm}) \notag \\
&\quad
-\lambda_{G^{+}H^{-}HH}A(H)-\lambda_{G^{+}H^{-}AA}A(A)-2\lambda_{G^{+}H^{-}H^{+}H^{-}}A(H^{\pm}).
\end{align}

\subsection{Vertex functions for $H^{+}ff',\, H^{+}W^{-}\phi$ and $H^{+}VW^{-}$}\label{sec:vertex}
The 1PI diagrams for the $H^{+}ff'$ couplings are calculated as
\allowdisplaybreaks{
\begin{align}
16\pi^{2}\Gamma^{S}_{H^{+}ff'} &=
e^{2}Q_{f'}Q_{f}g^{H^{+}}_{S}C^{FVF}_{+}(f', \gamma, f) \notag \\
&\quad
+g_{Z}^{2}\Big[g^{H^{+}}_{S}(v_{f}v_{f'}-a_{f}a_{f'})C^{FVF}_{+}+g^{H^{+}}_{P}(a_{f}v_{f'}-v_{f}a_{f'})C^{FVF}_{-}\Big](f', Z, f) \notag \\
&\quad
+\frac{g^{2}}{4\sqrt{2}}\Big[
-g^{h}_{f}c_{\beta-\alpha}C^{VFS}_{S}(W, f, h)+g^{H}_{f}s_{\beta-\alpha}C^{VFS}_{S}(W, f, H)-ig^{A}_{f}C^{VFS}_{S}(W, f, A)\Big] \notag \\
&\quad
-\frac{g^{2}}{4\sqrt{2}}\Big[
-g^{h}_{f'}c_{\beta-\alpha}C^{SFV}_{S}(h, f', W)+g^{H}_{f'}s_{\beta-\alpha}C^{SFV}_{S}(H, f', W)+ig^{A}_{f'}C^{SFV}_{S}(A, f', W)\Big] \notag \\
&\quad
+e^{2}g_{S}^{H^{+}}\Big[Q_{f}C^{SFV}_{S}(H^{\pm}, f, \gamma)-Q_{f'}C^{VFS}_{S}(\gamma, f', H^{\pm})\Big] \notag \\
&\quad
+\frac{g_{Z}^{2}}{2}c_{2W}\Big[(v_{f}g_{S}^{H^{+}}+a_{f}g_{P}^{H^{+}})C^{SFV}_{S}(H^{\pm}, f, Z)-
(v_{f'}g_{S}^{H^{+}}-a_{f'}g_{P}^{H^{+}})C^{VFS}_{S}(Z, f', H^{\pm})\Big] \notag \\
&\quad
-g_{S}^{H^{+}}\Big[
g^{h}_{f}g^{h}_{f'}C^{FSF}_{S}(f', h, f)+g^{H}_{f}g^{H}_{f'}C^{FSF}_{S}(f', H, f) \notag \\
&\quad
+g^{A}_{f}g^{A}_{f'}C^{FSF}_{S}(f', A, f)+g^{G^{0}}_{f}g^{G^{0}}_{f'}C^{FSF}_{S}(f', G^{0}, f)\Big] \notag \\
&\quad
-m_{f}\Big\{
g_{S}^{G^{+}}[\lambda_{H^{+}G^{-}h}g^{h}_{f}C_{0}(G^{\pm}, f, h)+\lambda_{H^{+}G^{-}H}g^{H}_{f}C_{0}(G^{\pm}, f, H)] \notag \\
&\quad
+g_{S}^{H^{+}}[\lambda_{H^{+}H^{-}h}g^{h}_{f}C_{0}(H^{\pm}, f, h)+\lambda_{H^{+}H^{-}H}g^{H}_{f}C_{0}(H^{\pm}, f, H)] \notag \\
&\quad
+g_{P}^{G^{+}}\lambda_{H^{+}G^{-}A}g^{A}_{f}C_{0}(G^{\pm}, f, A)\Big\} \notag \\
&\quad
-m_{f'}\Big\{
g_{S}^{G^{+}}[\lambda_{H^{+}G^{-}h}g^{h}_{f'}C_{0}(h, f', G^{\pm})+\lambda_{H^{+}G^{-}H}g^{H}_{f'}C_{0}(H, f', G^{\pm})] \notag \\
&\quad
+g_{S}^{H^{+}}[\lambda_{H^{+}H^{-}h}g^{h}_{f'}C_{0}(h, f', H^{\pm})+\lambda_{H^{+}H^{-}H}g^{H}_{f'}C_{0}(H, f', H^{\pm})] \notag \\
&\quad
+g_{P}^{G^{+}}\lambda_{H^{+}G^{-}A}g^{A}_{f'}C_{0}(A, f', G^{\pm})\Big\}, \\
16\pi^{2}\Gamma^{P}_{H^{+}ff'} &=
e^{2}Q_{f'}Q_{f}g^{H^{+}}_{P}C^{FVF}_{-}(f', \gamma, f) \notag \\
&\quad
+g_{Z}^{2}\Big[g^{H^{+}}_{S}(a_{f}v_{f'}-v_{f}a_{f'})C^{FVF}_{+}+g^{H^{+}}_{P}(v_{f}v_{f'}-a_{f}a_{f'})C^{FVF}_{-}\Big](f', Z, f) \notag \\
&\quad
-\frac{g^{2}}{4\sqrt{2}}\Big[
-g^{h}_{f}c_{\beta-\alpha}C^{VFS}_{S}(W, f, h)+g^{H}_{f}s_{\beta-\alpha}C^{VFS}_{S}(W, f, H)-ig^{A}_{f}C^{VFS}_{S}(W, f, A)\Big] \notag \\
&\quad
-\frac{g^{2}}{4\sqrt{2}}\Big[
-g^{h}_{f'}c_{\beta-\alpha}C^{SFV}_{S}(h, f', W)+g^{H}_{f'}s_{\beta-\alpha}C^{SFV}_{S}(H, f', W)+ig^{A}_{f'}C^{SFV}_{S}(A, f', W)\Big] \notag \\
&\quad
+e^{2}g_{P}^{H^{+}}\Big[Q_{f}C^{SFV}_{S}(H^{\pm}, f, \gamma)-Q_{f'}C^{VFS}_{S}(\gamma, f', H^{\pm})\Big] \notag \\
&\quad
+\frac{g_{Z}^{2}}{2}c_{2W}\Big[(a_{f}g_{S}^{H^{+}}+v_{f}g_{P}^{H^{+}})C^{SFV}_{S}(H^{\pm}, f, Z)-(v_{f'}g_{P}^{H^{+}}-a_{f'}g_{S}^{H^{+}})C^{VFS}_{S}(Z, f', H^{\pm})\Big] \notag \\
&\quad
+g_{P}^{H^{+}}\Big[
g^{h}_{f}g^{h}_{f'}C^{FSF}_{P}(f', h, f)+g^{H}_{f}g^{H}_{f'}C^{FSF}_{P}(f', H, f) \notag \\
&\quad
+g^{A}_{f}g^{A}_{f'}C^{FSF}_{P}(f', A, f)+g^{G^{0}}_{f}g^{G^{0}}_{f'}C^{FSF}_{P}(f', G^{0}, f)\Big] \notag \\
&\quad
-m_{f}\Big\{
g_{P}^{G^{+}}[\lambda_{H^{+}G^{-}h}g^{h}_{f}C_{0}(G^{\pm}, f, h)+\lambda_{H^{+}G^{-}H}g^{H}_{f}C_{0}(G^{\pm}, f, H)] \notag \\
&\quad
+g_{P}^{H^{+}}[\lambda_{H^{+}H^{-}h}g^{h}_{f}C_{0}(H^{\pm}, f, h)+\lambda_{H^{+}H^{-}H}g^{H}_{f}C_{0}(H^{\pm}, f, H)] \notag \\
&\quad
+g_{S}^{G^{+}}\lambda_{H^{+}G^{-}A}g^{A}_{f}C_{0}(G^{\pm}, f, A)\Big\} \notag \\
&\quad
-m_{f'}\Big\{
g_{P}^{G^{+}}[\lambda_{H^{+}G^{-}h}g^{h}_{f'}C_{0}(h, f', G^{\pm})+\lambda_{H^{+}G^{-}H}g^{H}_{f'}C_{0}(H, f', G^{\pm})] \notag \\
&\quad
+g_{P}^{H^{+}}[\lambda_{H^{+}H^{-}h}g^{h}_{f'}C_{0}(h, f', H^{\pm})+\lambda_{H^{+}H^{-}H}g^{H}_{f'}C_{0}(H, f', H^{\pm})] \notag \\
&\quad
+g_{S}^{G^{+}}\lambda_{H^{+}G^{-}A}g^{A}_{f'}C_{0}(A, f', G^{\pm})\Big\}, \\
16\pi^{2}\Gamma^{V1}_{H^{+}ff'} &=
2e^{2}Q_{f'}Q_{f}g^{H^{+}}_{S}[m_{f}C_{11}+m_{f'}(C_{11}+C_{0})](f', \gamma, f) \notag \\
&\quad
+2g_{Z}^{2}\Big[g^{H^{+}}_{S}(v_{f}v_{f'}+a_{f}a_{f'})[m_{f}C_{11}+m_{f'}(C_{11}+C_{0})] \notag \\
&\quad
-g^{H^{+}}_{P}(v_{f}a_{f'}+a_{f}v_{f'})[m_{f}C_{11}-m_{f'}(C_{11}+C_{0})]\Big](f', Z, f) \notag \\
&\quad
+\frac{g^{2}}{4\sqrt{2}}m_{f}\Big[
-g^{h}_{f}c_{\beta-\alpha}(C_{11}+2C_{0})(W, f, h)+g^{H}_{f}s_{\beta-\alpha}(C_{11}+2C_{0})(W, f, H) \notag \\
&\quad
+ig^{A}_{f}(C_{11}+2C_{0})(W, f, A)\Big] \notag \\
&\quad
-\frac{g^{2}}{4\sqrt{2}}m_{f'}\Big[
-g^{h}_{f'}c_{\beta-\alpha}(C_{11}-C_{0})(h, f', W)+g^{H}_{f'}s_{\beta-\alpha}(C_{11}-C_{0})(H, f', W) \notag \\
&\quad
-ig^{A}_{f'}(C_{11}-C_{0})(A, f', W)\Big] \notag \\
&\quad
+e^{2}g_{S}^{H^{+}}\Big[Q_{f}m_{f}(C_{11}-C_{0})(H^{\pm}, f, \gamma)
-Q_{f'}m_{f'}(C_{11}+2C_{0})(\gamma, f', H^{\pm})\Big] \notag \\
&\quad
+\frac{g_{Z}^{2}}{2}c_{2W}\Big[m_{f}(v_{f}g_{S}^{H^{+}}-a_{f}g_{P}^{H^{+}})(C_{11}-C_{0})(H^{\pm}, f, Z) \notag \\
&\quad
-m_{f'}(v_{f'}g_{S}^{H^{+}}+a_{f'}g_{P}^{H^{+}})(C_{11}+2C_{0})(Z, f', H^{\pm})\Big]\notag \\
&\quad
+g_{S}^{H^{+}}\Big[
g^{h}_{f}g^{h}_{f'}[m_{f'}(C_{11}+C_{0})+m_{f}C_{11}](f', h, f)+g^{H}_{f}g^{H}_{f'}[m_{f'}(C_{11}+C_{0})+m_{f}C_{11}](f', H, f) \notag \\
&\quad
-g^{A}_{f}g^{A}_{f'}[m_{f'}(C_{11}+C_{0})+m_{f}C_{11}](f', A, f)-g^{G^{0}}_{f}g^{G^{0}}_{f'}[m_{f'}(C_{11}+C_{0})+m_{f}C_{11}](f', G^{0}, f)\Big] \notag \\
&\quad
-g_{S}^{G^{+}}[\lambda_{H^{+}G^{-}h}g^{h}_{f'}(C_{0}+C_{11})(h, f', G^{\pm})+\lambda_{H^{+}G^{-}H}g^{H}_{f'}(C_{0}+C_{11})(H, f', G^{\pm})] \notag \\
&\quad
-g_{S}^{H^{+}}[\lambda_{H^{+}H^{-}h}g^{h}_{f'}(C_{0}+C_{11})(h, f', H^{\pm})+\lambda_{H^{+}H^{-}H}g^{H}_{f'}(C_{0}+C_{11})(H, f', H^{\pm})] \notag \\
&\quad
+g_{P}^{G^{+}}\lambda_{H^{+}G^{-}A}g^{A}_{f'}(C_{0}+C_{11})(A, f', G^{\pm}) \notag \\
&\quad
-g_{S}^{G^{+}}[\lambda_{H^{+}G^{-}h}g^{h}_{f}(C_{0}+C_{11})(G^{\pm}, f, h)+\lambda_{H^{+}G^{-}H}g^{H}_{f}(C_{0}+C_{11})(G^{\pm}, f, H)] \notag \\
&\quad
-g_{S}^{H^{+}}[\lambda_{H^{+}H^{-}h}g^{h}_{f}(C_{0}+C_{11})(H^{\pm}, f, h)+\lambda_{H^{+}H^{-}H}g^{H}_{f}(C_{0}+C_{11})(H^{\pm}, f, H)] \notag \\
&\quad
+g_{P}^{G^{+}}\lambda_{H^{+}G^{-}A}g^{A}_{f}(C_{0}+C_{11})(G^{\pm}, f, A), \\
16\pi^{2}\Gamma^{V2}_{H^{+}ff'} &=
2e^{2}Q_{f'}Q_{f}g^{H^{+}}_{S}[m_{f}C_{12}+m_{f'}(C_{12}+C_{0})](f', \gamma, f) \notag \\
&\quad
+g_{Z}^{2}\Big[2g^{H^{+}}_{S}(v_{f}v_{f'}+a_{f}a_{f'})[m_{f}C_{12}+m_{f'}(C_{12}+C_{0})] \notag \\
&\quad
-2g^{H^{+}}_{P}(v_{f}a_{f'}+a_{f}v_{f'})[m_{f}C_{12}-m_{f'}(C_{12}+C_{0})]\Big](f', Z, f) \notag \\
&\quad
+\frac{g^{2}}{4\sqrt{2}}m_{f}\Big[
-g^{h}_{f}c_{\beta-\alpha}(C_{12}+2C_{0})(W, f, h)+g^{H}_{f}s_{\beta-\alpha}(C_{12}+2C_{0})(W, f, H) \notag \\
&\quad
+ig^{A}_{f}(C_{12}+2C_{0})(W, f, A)\Big] \notag \\
&\quad
-\frac{g^{2}}{4\sqrt{2}}m_{f'}\Big[
-g^{h}_{f'}c_{\beta-\alpha}(C_{12}-C_{0})(h, f', W)+g^{H}_{f'}s_{\beta-\alpha}(C_{12}-C_{0})(H, f', W) \notag \\
&\quad
-ig^{A}_{f'}(C_{12}-C_{0})(A, f', W)\Big] \notag \\
&\quad
+e^{2}g_{S}^{H^{+}}\qty[Q_{f}m_{f}(C_{12}-C_{0})(H^{\pm}, f, \gamma)-Q_{f'}m_{f'}(C_{12}+2C_{0})(\gamma, f', H^{\pm})] \notag \\
&\quad
+\frac{g_{Z}^{2}c_{2W}}{2}\Big[-m_{f'}(v_{f'}g_{S}^{H^{+}}+a_{f'}g_{P}^{H^{+}})(C_{12}+2C_{0})(Z, f', H^{\pm}) \notag \\
&\quad
+m_{f}(v_{f'}g_{S}^{H^{+}}-a_{f'}g_{P}^{H^{+}})(C_{12}-C_{0})(H^{\pm}, f, Z)\Big] \notag \\
&\quad
+g_{S}^{H^{+}}\Big[
g^{h}_{f}g^{h}_{f'}[m_{f'}(C_{12}+C_{0})+m_{f}C_{12}](f', h, f)+g^{H}_{f}g^{H}_{f'}[m_{f'}(C_{12}+C_{0})+m_{f}C_{12}](f', H, f) \notag \\
&\quad
-g^{A}_{f}g^{A}_{f'}[m_{f'}(C_{12}+C_{0})+m_{f}C_{12}](f', A, f)
-g^{G^{0}}_{f}g^{G^{0}}_{f'}[m_{f'}(C_{12}+C_{0})+m_{f}C_{12}](f', G^{0}, f) \Big] \notag \\
&\quad
-g_{S}^{G^{+}}[\lambda_{H^{+}G^{-}h}g^{h}_{f'}C_{12}(h, f', G^{\pm})+\lambda_{H^{+}G^{-}H}g^{H}_{f'}C_{12}(H, f', G^{\pm})] \notag \\
&\quad
-g_{S}^{H^{+}}[\lambda_{H^{+}H^{-}h}g^{h}_{f'}C_{12}(h, f', H^{\pm})+\lambda_{H^{+}H^{-}H}g^{H}_{f'}C_{12}(H, f', H^{\pm})] \notag \\
&\quad
+g_{P}^{G^{+}}\lambda_{H^{+}G^{-}A}g^{A}_{f'}C_{12}(A, f', G^{\pm}) \notag \\
&\quad
-g_{S}^{G^{+}}[\lambda_{H^{+}G^{-}h}g^{h}_{f}C_{12}(G^{\pm}, f, h)+\lambda_{H^{+}G^{-}H}g^{H}_{f}C_{12}(G^{\pm}, f, H)] \notag \\
&\quad
-g_{S}^{H^{+}}[\lambda_{H^{+}H^{-}h}g^{h}_{f}C_{12}(H^{\pm}, f, h)+\lambda_{H^{+}H^{-}H}g^{H}_{f}C_{12}(H^{\pm}, f, H)] \notag \\
&\quad
+g_{P}^{G^{+}}\lambda_{H^{+}G^{-}A}g^{A}_{f}C_{12}(G^{\pm}, f, A), \\
16\pi^{2}\Gamma^{A1}_{H^{+}ff'} &=
2e^{2}Q_{f}Q_{f'}g^{H^{+}}_{P}[m_{f}C_{11}-m_{f'}(C_{11}+C_{0})](f', \gamma, f) \notag \\
&\quad
-g_{Z}^{2}\Big[2g^{H^{+}}_{S}(a_{f}v_{f'}+v_{f}a_{f'})[m_{f}C_{11}+m_{f'}(C_{11}+C_{0})] \notag \\
&\quad
-2g^{H^{+}}_{P}(v_{f}v_{f'}+a_{f}a_{f'})[m_{f}C_{11}-m_{f'}(C_{11}+C_{0})]\Big](f', Z, f) \notag \\
&\quad
-\frac{g^{2}}{4\sqrt{2}}m_{f}\Big[
-g^{h}_{f}c_{\beta-\alpha}(C_{11}+2C_{0})(W, f, h)+g^{H}_{f}s_{\beta-\alpha}(C_{11}+2C_{0})(W, f, H) \notag \\
&\quad
+ig^{A}_{f}(C_{11}+2C_{0})(W, f, A)\Big] \notag \\
&\quad
+\frac{g^{2}}{4\sqrt{2}}m_{f'}\Big[
-g^{h}_{f'}c_{\beta-\alpha}(C_{11}-C_{0})(h, f', W)+g^{H}_{f'}s_{\beta-\alpha}(C_{11}-C_{0})(H, f', W) \notag \\
&\quad
-ig^{A}_{f'}(C_{11}-C_{0})(A, f', W)\Big] \notag \\
&\quad
+e^{2}g_{P}^{H^{+}}\qty[Q_{f'}m_{f'}(C_{11}+2C_{0})(\gamma, f', H^{\pm})
+Q_{f}m_{f}(C_{11}-C_{0})(H^{\pm}, f, \gamma)] \notag \\
&\quad
+\frac{g_{Z}^{2}c_{2W}}{2}\Big[m_{f'}(v_{f'}g_{P}^{H^{+}}+a_{f'}g_{S}^{H^{+}})(C_{11}+2C_{0})(Z, f', H^{\pm}) \notag \\
&\quad
+m_{f}(v_{f}g_{P}^{H^{+}}-a_{f}g_{S}^{H^{+}})(C_{11}-C_{0})(H^{\pm}, f, Z)\Big] \notag \\
&\quad
+g_{P}^{H^{+}}\Big[
g^{h}_{f}g^{h}_{f'}[m_{f'}(C_{11}+C_{0})-m_{f}C_{11}](f', h, f)+g^{H}_{f}g^{H}_{f'}[m_{f'}(C_{11}+C_{0})-m_{f}C_{11}](f', H, f) \notag \\
&\quad
-g^{A}_{f}g^{A}_{f'}[m_{f'}(C_{11}+C_{0})-m_{f}C_{11}](f', A, f)
-g^{G^{0}}_{f}g^{G^{0}}_{f'}[m_{f'}(C_{11}+C_{0})-m_{f}C_{11}](f', G^{0}, f)\Big] \notag \\
&\quad
+g_{P}^{G^{+}}[\lambda_{H^{+}G^{-}h}g^{h}_{f'}(C_{0}+C_{11})(h, f', G^{\pm})+\lambda_{H^{+}G^{-}H}g^{H}_{f'}(C_{0}+C_{11})(H, f', G^{\pm})] \notag \\
&\quad
+g_{P}^{H^{+}}[\lambda_{H^{+}H^{-}h}g^{h}_{f'}(C_{0}+C_{11})(h, f', H^{\pm})+\lambda_{H^{+}H^{-}H}g^{H}_{f'}(C_{0}+C_{11})(H, f', H^{\pm})] \notag \\
&\quad
-g_{S}^{G^{+}}\lambda_{H^{+}G^{-}A}g^{A}_{f'}(C_{0}+C_{11})(A, f', G^{\pm}) \notag \\
&\quad
-g_{P}^{G^{+}}[\lambda_{H^{+}G^{-}h}g^{h}_{f}(C_{0}+C_{11})(G^{\pm}, f, h)+\lambda_{H^{+}G^{-}H}g^{H}_{f}(C_{0}+C_{11})(G^{\pm}, f, H)] \notag \\
&\quad
-g_{P}^{H^{+}}[\lambda_{H^{+}H^{-}h}g^{h}_{f}(C_{0}+C_{11})(H^{\pm}, f, h)+\lambda_{H^{+}H^{-}H}g^{H}_{f}(C_{0}+C_{11})(H^{\pm}, f, H)] \notag \\
&\quad
+g_{S}^{G^{+}}\lambda_{H^{+}G^{-}A}g^{A}_{f}(C_{0}+C_{11})(G^{\pm}, f, A)\Big\}, \\
16\pi^{2}\Gamma^{A2}_{H^{+}ff'} &=
2e^{2}Q_{f}Q_{f'}g^{H^{+}}_{P}[m_{f}C_{12}-m_{f'}(C_{12}+C_{0})](f', \gamma, f) \notag \\
&\quad
-g_{Z}^{2}\Big[2g^{H^{+}}_{S}(a_{f}v_{f'}+v_{f}a_{f'})[m_{f}C_{12}+m_{f'}(C_{12}+C_{0})] \notag \\
&\quad
-2g^{H^{+}}_{P}(v_{f}v_{f'}+a_{f}a_{f'})[m_{f}C_{12}-m_{f'}(C_{12}+C_{0})]\Big](f', Z, f) \notag \\
&\quad
-\frac{g^{2}}{4\sqrt{2}}m_{f}\Big[
-g^{h}_{f}c_{\beta-\alpha}(C_{12}+2C_{0})(W, f, h)+g^{H}_{f}s_{\beta-\alpha}(C_{12}+2C_{0})(W, f, H) \notag \\
&\quad
+ig^{A}_{f}(C_{12}+2C_{0})(W, f, A)\Big] \notag \\
&\quad
+\frac{g^{2}}{4\sqrt{2}}m_{f'}\Big[
-g^{h}_{f'}c_{\beta-\alpha}(C_{12}-C_{0})(h, f', W)+g^{H}_{f'}s_{\beta-\alpha}(C_{12}-C_{0})(H, f', W) \notag \\
&\quad
-ig^{A}_{f'}(C_{12}-C_{0})(A, f', W)\Big] \notag \\
&\quad
+e^{2}g_{P}^{H^{+}}\Big[Q_{f'}m_{f'}(C_{12}+2C_{0})(\gamma, f', H^{\pm})
+Q_{f}m_{f}(C_{12}-C_{0})(H^{\pm}, f, \gamma)\Big], \\
&\quad
+\frac{g_{Z}^{2}c_{2W}}{2}\Big[m_{f}(v_{f}g_{P}^{H^{+}}-a_{f}g_{S}^{H^{+}})(C_{12}-C_{0})(H^{\pm}, f, Z) \notag \\
&\quad
+m_{f'}(v_{f'}g_{P}^{H^{+}}+a_{f'}g_{S}^{H^{+}})(C_{12}+2C_{0})(Z, f', H^{\pm})\Big] \notag \\
&\quad
+g_{P}^{H^{+}}\Big[
g^{h}_{f}g^{h}_{f'}[m_{f'}(C_{12}+C_{0})-m_{f}C_{12}](f', h, f)+g^{H}_{f}g^{H}_{f'}[m_{f'}(C_{12}+C_{0})-m_{f}C_{12}](f', H, f) \notag \\
&\quad
-g^{A}_{f}g^{A}_{f'}[m_{f'}(C_{12}+C_{0})-m_{f}C_{12}](f', A, f)
-g^{G^{0}}_{f}g^{G^{0}}_{f'}[m_{f'}(C_{12}+C_{0})-m_{f}C_{12}](f', G^{0}, f)\Big] \notag \\
&\quad
g_{P}^{G^{+}}[\lambda_{H^{+}G^{-}h}g^{h}_{f'}C_{12}(h, f', G^{\pm})+\lambda_{H^{+}G^{-}H}g^{H}_{f'}C_{12}(H, f', G^{\pm})] \notag \\
&\quad
+g_{P}^{H^{+}}[\lambda_{H^{+}H^{-}h}g^{h}_{f'}C_{12}(h, f', H^{\pm})+\lambda_{H^{+}H^{-}H}g^{H}_{f'}C_{12}(H, f', H^{\pm})] \notag \\
&\quad
-g_{S}^{G^{+}}\lambda_{H^{+}G^{-}A}g^{A}_{f'}C_{12}(A, f', G^{\pm}) \notag \\
&\quad
-g_{P}^{G^{+}}[\lambda_{H^{+}G^{-}h}g^{h}_{f}C_{12}(G^{\pm}, f, h)+\lambda_{H^{+}G^{-}H}g^{H}_{f}C_{12}(G^{\pm}, f, H)] \notag \\
&\quad
-g_{P}^{H^{+}}[\lambda_{H^{+}H^{-}h}g^{h}_{f}C_{12}(H^{\pm}, f, h)+\lambda_{H^{+}H^{-}H}g^{H}_{f}C_{12}(H^{\pm}, f, H)] \notag \\
&\quad
-g_{S}^{G^{+}}\lambda_{H^{+}G^{-}A}g^{A}_{f}C_{12}(G^{\pm}, f, A)\Big\}, \\
16\pi^{2}\Gamma^{T}_{H^{+}ff'} &=
\frac{g^{2}}{4\sqrt{2}}\Big[
-g^{h}_{f}c_{\beta-\alpha}(2C_{0}+2C_{11}-C_{12})(W, f, h)
+g^{H}_{f}s_{\beta-\alpha}(2C_{0}+2C_{11}-C_{12})(W, f, H) \notag \\
&\quad
-ig^{A}_{f}(2C_{0}+2C_{11}-C_{12})(W, f, A)\Big] \notag \\
&\quad
-\frac{g^{2}}{4\sqrt{2}}\Big[
-g^{h}_{f'}c_{\beta-\alpha}(C_{0}+C_{11}-2C_{12})(h, f', W)
+g^{H}_{f'}s_{\beta-\alpha}(C_{0}+C_{11}-2C_{12})(H, f', W) \notag \\
&\quad
+ig^{A}_{f'}(C_{0}+C_{11}-2C_{12})(A, f', W)\Big] \notag \\
&\quad
+e^{2}g_{S}^{H^{+}}\Big[
Q_{f}(C_{0}+C_{11}-2C_{12})(H^{\pm}, f, \gamma)
-Q_{f'}(2C_{0}+2C_{11}-C_{12})(\gamma, f', H^{\pm})\Big] \notag  \\
&\quad
+\frac{g_{Z}^{2}c_{2W}}{2}\Big[(v_{f}g_{S}^{H^{+}}+a_{f}g_{P}^{H^{+}})(C_{0}+C_{11}-2C_{12})(H^{\pm}, f, Z) \notag \\
&\quad
-(v_{f'}g_{S}^{H^{+}}-a_{f'}g_{P}^{H^{+}})(2C_{0}+2C_{11}-C_{12})(Z, f', H^{\pm})\Big] \notag \\
&\quad
+g_{S}^{H^{+}}\Big[
g^{h}_{f}g^{h}_{f'}(C_{11}-C_{12})(f', h, f)+g^{H}_{f}g^{H}_{f'}(C_{11}-C_{12})(f', H, f) \notag \\
&\quad
+g^{A}_{f}g^{A}_{f'}(C_{11}-C_{12})(f', A, f)+g^{G^{0}}_{f}g^{G^{0}}_{f'}(C_{11}-C_{12})(f', G^{0}, f)\Big], \\
16\pi^{2}\Gamma^{PT}_{H^{+}ff'} &=
-\frac{g^{2}}{4\sqrt{2}}\Big[
-g^{h}_{f}c_{\beta-\alpha}(2C_{0}+2C_{11}-C_{12})(W, f, h)
+g^{H}_{f}s_{\beta-\alpha}(2C_{0}+2C_{11}-C_{12})(W, f, H) \notag \\
&\quad
-ig^{A}_{f}(2C_{0}+2C_{11}-C_{12})(W, f, A)\Big] \notag \\
&\quad
-\frac{g^{2}}{4\sqrt{2}}\Big[
-g^{h}_{f'}c_{\beta-\alpha}(C_{0}+C_{11}-2C_{12})(h, f', W)
+g^{H}_{f'}s_{\beta-\alpha}(C_{0}+C_{11}-2C_{12})(H, f', W) \notag \\
&\quad
+ig^{A}_{f'}(C_{0}+C_{11}-2C_{12})(A, f', W)\Big] \notag \\
&\quad
+e^{2}g_{P}^{H^{+}}\Big[
Q_{f}(C_{0}+C_{11}-2C_{12})(H^{\pm}, f, \gamma)
-Q_{f'}(2C_{0}+2C_{11}-C_{12})(\gamma, f', H^{\pm})\Big] \notag  \\
&\quad
+\frac{g_{Z}^{2}c_{2W}}{2}\Big[(v_{f}g_{P}^{H^{+}}+a_{f}g_{S}^{H^{+}})(C_{0}+C_{11}-2C_{12})(H^{\pm}, f, Z) \notag \\
&\quad
-(v_{f'}g_{P}^{H^{+}}-a_{f'}g_{S}^{H^{+}})(2C_{0}+2C_{11}-C_{12})(Z, f', H^{\pm})\Big] \notag \\
&\quad
-g_{P}^{H^{+}}\Big[
g^{h}_{f}g^{h}_{f'}(C_{11}-C_{12})(f', h, f)+g^{H}_{f}g^{H}_{f'}(C_{11}-C_{12})(f', H, f) \notag \\
&\quad
+g^{A}_{f}g^{A}_{f'}(C_{11}-C_{12})(f', A, f)+g^{G^{0}}_{f}g^{G^{0}}_{f'}(C_{11}-C_{12})(f', G^{0}, f)\Big],
\end{align}}
where
\allowdisplaybreaks{\begin{align}
&C^{SFV}_{S}(X, Y, Z) \notag \\
&\quad=
B_{0}(p_{2}^{2}; Y, Z)+\Big[(m_{X}^{2}-q^{2}+p_{2}^{2})C_{0}-(q^{2}-p_{1}^{2}-p_{2}^{2})C_{11} \notag \\
&\qquad
+(q^{2}-p_{1}^{2}-2p_{2}^{2})C_{12}\Big](p_{1}^{2},p_{2}^{2},q^{2}; X, Y, Z), \\
&C^{VFS}_{S}(X, Y, Z) \notag \\
&\quad=
B_{0}(p_{2}^{2}; Y, Z)+\Big[(m_{X}^{2}+2p_{1}^{2})C_{0}+3p_{1}^{2}C_{11}+2(q^{2}-p_{1}^{2})C_{12}\Big](p_{1}^{2},p_{2}^{2},q^{2}; X, Y, Z), \\
&C^{FSF}_{S}(X, Y, Z) \notag \\
&\quad=
B_{0}(p_{2}^{2}; Y, Z)+\Big[(m_{X}^{2}+m_{f'}m_{f})C_{0}+(q^{2}-p_{2}^{2})C_{11}+p_{2}^{2}C_{12}\Big](p_{1}^{2},p_{2}^{2},q^{2}; X, Y, Z), \\
&C^{FSF}_{P}(X, Y, Z) \notag \\
&\quad=
B_{0}(p_{2}^{2}; Y, Z)+\Big[(m_{X}^{2}-m_{f'}m_{f})C_{0}+(q^{2}-p_{2}^{2})C_{11}+p_{2}^{2}C_{12}\Big](p_{1}^{2},p_{2}^{2},q^{2}; X, Y, Z), \\
&C^{FVF}_{+}(X, Y, Z) \notag \\
&\quad=
\Big[4p_{1}^{2}C_{21}+4p_{2}^{2}C_{22}+4(q^{2}-p_{1}^{2}-p_{2}^{2})C_{23}+16C_{24}-4 \notag \\
&\qquad
+2(q^{2}+p_{1}^{2}-p_{2}^{2})C_{11}+2(q^{2}-p_{1}^{2}+p_{2}^{2})C_{12}+4m_{f'}m_{f}C_{0}\Big](p_{1}^{2},p_{2}^{2},q^{2}; X, Y, Z), \\
&C^{FVF}_{-}(X, Y, Z) \notag \\
&\quad=
\Big[4p_{1}^{2}C_{21}+4p_{2}^{2}C_{22}+4(q^{2}-p_{1}^{2}-p_{2}^{2})C_{23}+16C_{24}-4 \notag \\
&\qquad
+2(q^{2}+p_{1}^{2}-p_{2}^{2})C_{11}+2(q^{2}-p_{1}^{2}+p_{2}^{2})C_{12}-4m_{f'}m_{f}C_{0}\Big](p_{1}^{2},p_{2}^{2},q^{2}; X, Y, Z).
\end{align}}

The 1PI diagrams for the $H^{+}W^{-}h$ couplings are calculated as
\allowdisplaybreaks{
\begin{align}
&(16\pi^{2}){\Gamma_{H^{+}W^{-}h}^{{\rm 1PI}}\red{(p_{1}^{2}, p_{2}^{2}, q^{2})}}_{F} \notag \\
&\quad=
-N_{c}^{f}g\bigg[
\frac{m_{u}\zeta_{h}^{u}}{v}C_{H^{+}W^{-}\phi}^{FFF}(u, u, d)
+\frac{m_{d}\zeta_{h}^{d}}{v}C_{H^{+}W^{-}\phi}^{FFF}(d, d, u)\bigg]
-g\frac{m_{\ell}\zeta_{h}^{\ell}}{v}C_{H^{+}W^{-}\phi}^{FFF}(\ell, \ell, \nu_{\ell}), \\
&(16\pi^{2}){\Gamma_{H^{+}W^{-}h}^{{\rm 1PI}}\red{(p_{1}^{2}, p_{2}^{2}, q^{2})}}_{B} \notag \\
&\quad=
-\frac{g^{3}}{8}c_{\beta-\alpha}\Big\{
C_{H^{+}W^{-}\phi}^{SVV}(A, Z, W)+c_{2W}C_{H^{+}W^{-}\phi}^{SVV}(H^{\pm}, W, Z)
+2s_{W}^{2}C_{H^{+}W^{-}\phi}^{SVV}(H^{\pm}, W, \gamma)\Big\} \notag \\
&\qquad
-\frac{g^{3}}{8}c_{\beta-\alpha}\Big\{
c_{\beta-\alpha}^{2}C_{H^{+}W^{-}\phi}^{VSS}(W, H^{\pm}, h)
+s_{\beta-\alpha}^{2}C_{H^{+}W^{-}\phi}^{VSS}(W, G^{\pm}, h) \notag \\
&\qquad
+s_{\beta-\alpha}^{2}C_{H^{+}W^{-}\phi}^{VSS}(W, H^{\pm}, H)
-s_{\beta-\alpha}^{2}C_{H^{+}W^{-}\phi}^{VSS}(W, G^{\pm}, H) \notag \\
&\qquad
-C_{H^{+}W^{-}\phi}^{VSS}(W, H^{\pm}, A)
-\frac{c_{2W}}{c_{W}^{2}}C_{H^{+}W^{-}\phi}^{VSS}(Z, A, H^{\pm})\Big\} \notag \\
&\qquad
+\frac{g^{3}}{8}c_{\beta-\alpha}\bigg\{
(B_{0}-B_{1})(q^{2}; h, W) -\frac{s_{W}^{2}}{c_{W}^{2}}c_{2W}(B_{0}-B_{1})(q^{2}; H^{\pm}, Z) \notag \\
&\qquad
+2s_{W}^{2}(B_{0}-B_{1})(q^{2}; H^{\pm}, \gamma)
+(B_{0}-B_{1})(q^{2}; H^{\pm}, W)
+\frac{s_{W}^{2}}{c_{W}^{2}}(B_{0}-B_{1})(p_{1}^{2}; A, Z) \bigg\} \notag \\   
&\qquad
+\frac{g^{3}}{4}m_{W}^{2}s_{\beta-\alpha}^{2}c_{\beta-\alpha}\Big\{
(C_{11}+2C_{0})(W, W, h)-(C_{11}+2C_{0})(W, W, H)\Big\} \notag \\
&\qquad
-\frac{g^{2}}{4}m_{W}\Big\{
i\frac{s_{W}^{2}}{c_{W}^{2}}\lambda_{H^{+}G^{-}A}c_{\beta-\alpha}C_{H^{+}W^{-}\phi}^{SVS}(A, Z, G^{\pm}) \notag \\
&\qquad
+\lambda_{H^{+}G^{-}h}s_{\beta-\alpha}^{2}C_{H^{+}W^{-}\phi}^{SVS}(G^{\pm}, W, h)
+\lambda_{H^{+}H^{-}h}s_{\beta-\alpha}c_{\beta-\alpha}C_{H^{+}W^{-}\phi}^{SVS}(H^{\pm}, W, h) \notag \\
&\qquad
+\lambda_{H^{+}G^{-}H}s_{\beta-\alpha}c_{\beta-\alpha}C_{H^{+}W^{-}\phi}^{SVS}(G^{\pm}, W, H)
+\lambda_{H^{+}H^{-}H}c_{\beta-\alpha}^{2}C_{H^{+}W^{-}\phi}^{SVS}(H^{\pm}, W, H)\Big\} \notag \\
&\qquad
-\frac{g^{2}}{4}m_{W}\Big\{
6\lambda_{hhh}s_{\beta-\alpha}c_{\beta-\alpha}C_{H^{+}W^{-}\phi}^{SSV}(h, h, W)
-2\lambda_{Hhh}s_{\beta-\alpha}^{2}C_{H^{+}W^{-}\phi}^{SSV}(H, h, W) \notag \\
&\qquad
-2\lambda_{HHh}s_{\beta-\alpha}c_{\beta-\alpha}C_{H^{+}W^{-}\phi}^{SSV}(H, H, W)
+2\lambda_{Hhh}c_{\beta-\alpha}^{2}C_{H^{+}W^{-}\phi}^{SSV}(h, H, W) \notag \\
&\qquad
-\frac{s_{W}^{2}}{c_{W}^{2}}c_{2W}\lambda_{H^{+}G^{-}h}C_{H^{+}W^{-}\phi}^{SSV}(H^{\pm}, G^{\mp}, Z)
+2s_{W}^{2}\lambda_{H^{+}G^{-}h}C_{H^{+}W^{-}\phi}^{SSV}(H^{\pm}, G^{\mp}, \gamma)\Big\} \notag \\
&\qquad
-\frac{g}{2}\Big\{
-6\lambda_{hhh}\lambda_{H^{+}H^{-}h}c_{\beta-\alpha}C_{H^{+}W^{-}\phi}^{SSS}(h, h, H^{\pm})
-2\lambda_{Hhh}\lambda_{H^{+}H^{-}H}c_{\beta-\alpha}C_{H^{+}W^{-}\phi}^{SSS}(H, h, H^{\pm}) \notag \\
&\qquad
+2\lambda_{Hhh}\lambda_{H^{+}H^{-}h}s_{\beta-\alpha}C_{H^{+}W^{-}\phi}^{SSS}(h, H, H^{\pm})
+2\lambda_{HHh}\lambda_{H^{+}H^{-}H}s_{\beta-\alpha}C_{H^{+}W^{-}\phi}^{SSS}(H, H, H^{\pm}) \notag \\
&\qquad 
-6\lambda_{hhh}\lambda_{H^{+}G^{-}h}s_{\beta-\alpha}C_{H^{+}W^{-}\phi}^{SSS}(h, h, G^{\pm})
-2\lambda_{Hhh}\lambda_{H^{+}G^{-}H}s_{\beta-\alpha}C_{H^{+}W^{-}\phi}^{SSS}(H, h, G^{\pm}) \notag \\
&\qquad
-2\lambda_{Hhh}\lambda_{H^{+}G^{-}h}c_{\beta-\alpha}C_{H^{+}W^{-}\phi}^{SSS}(h, H, G^{\pm})
-2\lambda_{HHh}\lambda_{H^{+}G^{-}H}c_{\beta-\alpha}C_{H^{+}W^{-}\phi}^{SSS}(H, H, G^{\pm}) \notag \\
&\qquad
+\lambda_{H^{+}H^{-}h}^{2}c_{\beta-\alpha}C_{H^{+}W^{-}\phi}^{SSS}(H^{\pm}, H^{\pm}, h)
-\lambda_{H^{+}H^{-}H}\lambda_{H^{+}H^{-}h}s_{\beta-\alpha}C_{H^{+}W^{-}\phi}^{SSS}(H^{\pm}, H^{\pm}, H) \notag \\
&\qquad
+\lambda_{H^{+}G^{-}h}^{2}c_{\beta-\alpha}C_{H^{+}W^{-}\phi}^{SSS}(G^{\pm}, H^{\pm}, h)
-\lambda_{H^{+}G^{-}H}\lambda_{H^{+}G^{-}h}s_{\beta-\alpha}C_{H^{+}W^{-}\phi}^{SSS}(G^{\pm},H^{\pm}, H) \notag \\
&\qquad
-i\lambda_{H^{+}G^{-}A}\lambda_{H^{+}G^{-}h}C_{H^{+}W^{-}\phi}^{SSS}(G^{\pm},H^{\pm},A) \notag \\
&\qquad 
+\lambda_{H^{+}G^{-}h}\lambda_{H^{+}H^{-}h}s_{\beta-\alpha}C_{H^{+}W^{-}\phi}^{SSS}(H^{\pm}, G^{\pm}, h) \notag \\
&\qquad
+\lambda_{H^{+}G^{-}h}\lambda_{H^{+}H^{-}H}c_{\beta-\alpha}C_{H^{+}W^{-}\phi}^{SSS}(H^{\pm}, G^{\pm}, H) \notag \\
&\qquad
+\lambda_{G^{+}G^{-}h}\lambda_{H^{+}G^{-}h}s_{\beta-\alpha}C_{H^{+}W^{-}\phi}^{SSS}(G^{\pm}, G^{\pm}, h) \notag \\
&\qquad
+\lambda_{G^{+}G^{-}h}\lambda_{H^{+}G^{-}H}c_{\beta-\alpha}C_{H^{+}W^{-}\phi}^{SSS}(G^{\pm}, G^{\pm}, H)\Big\},
\end{align}}
where we defined
\begin{align}
C_{H^{+}W^{-}\phi}^{FFF}(f, f, f')
&=
\frac{m_{f}\zeta_{f}}{v}B_{0}(p_{2}^{2}; f f')
+\Big\{\frac{m_{f}\zeta_{f}}{v}[m_{f}^{2}(2C_{11}+3C_{0})+p_{1}^{2}(C_{11}-C_{12})+q^{2}C_{12}] \notag \\
&\quad
-m_{f}m_{f'}\frac{m_{f'}\zeta_{f'}}{v}(2C_{11}+C_{0})\Big\} (p_{1}^{2}, p_{2}^{2}, q^{2}; f, f, f'), \\
C_{H^{+}W^{-}\phi}^{SVV}(X, Y, Z)
&=
4B_{0}(p_{2}^{2}; Y, Z) -\Big[4C_{24}+(3p_{1}^{2}-p_{2}^{2}+q^{2})C_{21}+2(q^{2}-p_{1}^{2})C_{23} \notag \\
&\quad
+(p_{1}^{2}+2p_{2}^{2}-q^{2})C_{11}+2(q^{2}-p_{1}^{2})C_{12} \notag \\
&\quad
-(p_{2}^{2}+4m_{X}^{2})C_{0}\Big](p_{1}^{2}, p_{2}^{2}, q^{2}; X, Y, Z), \\
C_{H^{+}W^{-}\phi}^{VSS}(X, Y, Z)
&=
\Big[4C_{24}+(3p_{1}^{2}-p_{2}^{2}+q^{2})C_{21}+2(q^{2}-p_{1}^{2})C_{23} +(5p_{1}^{2}-3p_{2}^{2}+3q^{2}+m_{X}^{2})C_{11} \notag \\
&\quad
+2(q^{2}-p_{1}^{2})C_{12} + (2p_{1}^{2}-2p_{2}^{2}+2q^{2}+m_{X}^{2})C_{0}\Big](p_{1}^{2}, p_{2}^{2}, q^{2}; X, Y, Z), \\
C_{H^{+}W^{-}\phi}^{SVS}(X, Y, Z)&=C_{H^{+}W^{-}\phi}^{SSV}(X, Y, Z)
=
[C_{0}-C_{11}](p_{1}^{2}, p_{2}^{2}, q^{2}; X, Y, Z), \\
C_{H^{+}W^{-}\phi}^{SSS}(X, Y, Z)
&=
[C_{0}+C_{11}](p_{1}^{2}, p_{2}^{2}, q^{2}; X, Y, Z).
\end{align}

The 1PI diagrams for the $H^{+}W^{-}H$ couplings are calculated as
\allowdisplaybreaks{
\begin{align}
&(16\pi^{2}){\Gamma_{H^{+}W^{-}H}^{{\rm 1PI}}\red{(p_{1}^{2}, p_{2}^{2}, q^{2})}}_{F} \notag \\
&\quad=
-N_{c}^{f}g\bigg[
\frac{m_{u}\zeta_{H}^{u}}{v}C_{H^{+}W^{-}\phi}^{FFF}(u, u, d)
+\frac{m_{d}\zeta_{H}^{d}}{v}C_{H^{+}W^{-}\phi}^{FFF}(d, d, u)\bigg]
-g\frac{m_{\ell}\zeta_{H}^{\ell}}{v}C_{H^{+}W^{-}\phi}^{FFF}(\ell, \ell, \nu_{\ell}), \\
&(16\pi^{2}){\Gamma_{H^{+}W^{-}H}^{{\rm 1PI}}\red{(p_{1}^{2}, p_{2}^{2}, q^{2})}}_{B} \notag \\
&\quad=
\frac{g^{3}}{8}s_{\beta-\alpha}\Big\{
C_{H^{+}W^{-}\phi}^{SVV}(A, Z, W)
+c_{2W}C_{H^{+}W^{-}\phi}^{SVV}(H^{\pm}, W, Z)
+2s_{W}^{2}C_{H^{+}W^{-}\phi}^{SVV}(H^{\pm}, W, \gamma)\Big\} \notag \\
&\qquad
+\frac{g^{3}}{8}s_{\beta-\alpha}\Big\{
c_{\beta-\alpha}^{2}C_{H^{+}W^{-}\phi}^{VSS}(W, H^{\pm}, h)
-c_{\beta-\alpha}^{2}C_{H^{+}W^{-}\phi}^{VSS}(W, G^{\pm}, h) \notag \\
&\qquad
+s_{\beta-\alpha}^{2}C_{H^{+}W^{-}\phi}^{VSS}(W, H^{\pm}, H)
+c_{\beta-\alpha}^{2}C_{H^{+}W^{-}\phi}^{VSS}(W, G^{\pm}, H) \notag \\
&\qquad
-C_{H^{+}W^{-}\phi}^{VSS}(W, H^{\pm}, A)
-\frac{c_{2W}}{c_{W}^{2}}C_{H^{+}W^{-}\phi}^{VSS}(Z, A, H^{\pm})\Big\} \notag \\
&\qquad
-\frac{g^{3}}{8}s_{\beta-\alpha}\bigg\{
(B_{0}-B_{1})(q^{2}; H, W) -\frac{s_{W}^{2}}{c_{W}^{2}}c_{2W}(B_{0}-B_{1})(q^{2}; H^{\pm}, Z) \notag \\
&\qquad
+2s_{W}^{2}(B_{0}-B_{1})(q^{2}; H^{\pm}, \gamma)
+(B_{0}-B_{1})(q^{2}; H^{\pm}, W)
+\frac{s_{W}^{2}}{c_{W}^{2}}(B_{0}-B_{1})(p_{1}^{2}; A, Z) \bigg\} \notag \\   
&\qquad
+\frac{g^{3}}{4}m_{W}^{2}c_{\beta-\alpha}^{2}s_{\beta-\alpha}\Big\{
(C_{11}+2C_{0})(W, W, h)-(C_{11}+2C_{0})(W, W, H)\Big\} \notag \\
&\qquad
-\frac{g^{2}}{4}m_{W}\Big\{
-i\frac{s_{W}^{2}}{c_{W}^{2}}\lambda_{H^{+}G^{-}A}s_{\beta-\alpha}C_{H^{+}W^{-}\phi}^{SVS}(A, Z, G^{\pm}) \notag \\
&\qquad
+\lambda_{H^{+}G^{-}h}s_{\beta-\alpha}c_{\beta-\alpha}C_{H^{+}W^{-}\phi}^{SVS}(G^{\pm}, W, h)
-\lambda_{H^{+}H^{-}h}s_{\beta-\alpha}^{2}C_{H^{+}W^{-}\phi}^{SVS}(H^{\pm}, W, h) \notag \\
&\qquad
+\lambda_{H^{+}G^{-}H}c_{\beta-\alpha}^{2}C_{H^{+}W^{-}\phi}^{SVS}(G^{\pm}, W, H)
-\lambda_{H^{+}H^{-}H}s_{\beta-\alpha}c_{\beta-\alpha}C_{H^{+}W^{-}\phi}^{SVS}(H^{\pm}, W, H)\Big\} \notag \\
&\qquad
-\frac{g^{2}}{4}m_{W}\Big\{
2\lambda_{Hhh}s_{\beta-\alpha}c_{\beta-\alpha}C_{H^{+}W^{-}\phi}^{SSV}(h, h, W)
-2\lambda_{HHh}s_{\beta-\alpha}^{2}C_{H^{+}W^{-}\phi}^{SSV}(H, h, W) \notag \\
&\qquad
-6\lambda_{HHH}s_{\beta-\alpha}c_{\beta-\alpha}C_{H^{+}W^{-}\phi}^{SSV}(H, H, W)
+2\lambda_{HHh}c_{\beta-\alpha}^{2}C_{H^{+}W^{-}\phi}^{SSV}(h, H, W) \notag \\
&\qquad
-\frac{s_{W}^{2}}{c_{W}^{2}}c_{2W}\lambda_{H^{+}G^{-}H}C_{H^{+}W^{-}\phi}^{SSV}(H^{\pm}, G^{\mp}, Z)
+2s_{W}^{2}\lambda_{H^{+}G^{-}H}C_{H^{+}W^{-}\phi}^{SSV}(H^{\pm}, G^{\mp}, \gamma)\Big\} \notag \\
&\qquad
-\frac{g}{2}\Big\{
-2\lambda_{Hhh}\lambda_{H^{+}H^{-}h}c_{\beta-\alpha}C_{H^{+}W^{-}\phi}^{SSS}(h, h, H^{\pm})
-2\lambda_{HHh}\lambda_{H^{+}H^{-}H}c_{\beta-\alpha}C_{H^{+}W^{-}\phi}^{SSS}(H, h, H^{\pm}) \notag \\
&\qquad
+2\lambda_{HHh}\lambda_{H^{+}H^{-}h}s_{\beta-\alpha}C_{H^{+}W^{-}\phi}^{SSS}(h, H, H^{\pm})
+6\lambda_{HHH}\lambda_{H^{+}H^{-}H}s_{\beta-\alpha}C_{H^{+}W^{-}\phi}^{SSS}(H, H, H^{\pm}) \notag \\
&\qquad 
-2\lambda_{Hhh}\lambda_{H^{+}G^{-}h}s_{\beta-\alpha}C_{H^{+}W^{-}\phi}^{SSS}(h, h, G^{\pm})
-2\lambda_{HHh}\lambda_{H^{+}G^{-}H}s_{\beta-\alpha}C_{H^{+}W^{-}\phi}^{SSS}(H, h, G^{\pm}) \notag \\
&\qquad
-2\lambda_{HHh}\lambda_{H^{+}G^{-}h}c_{\beta-\alpha}C_{H^{+}W^{-}\phi}^{SSS}(h, H, G^{\pm})
-6\lambda_{HHH}\lambda_{H^{+}G^{-}H}c_{\beta-\alpha}C_{H^{+}W^{-}\phi}^{SSS}(H, H, G^{\pm}) \notag \\
&\qquad
+\lambda_{H^{+}H^{-}h}\lambda_{H^{+}H^{-}H}c_{\beta-\alpha}C_{H^{+}W^{-}\phi}^{SSS}(H^{\pm}, H^{\pm}, h)
-\lambda_{H^{+}H^{-}H}^{2}s_{\beta-\alpha}C_{H^{+}W^{-}\phi}^{SSS}(H^{\pm}, H^{\pm}, H) \notag \\
&\qquad
+\lambda_{H^{+}G^{-}h}\lambda_{H^{+}G^{-}H}c_{\beta-\alpha}C_{H^{+}W^{-}\phi}^{SSS}(G^{\pm}, H^{\pm}, h)
-\lambda_{H^{+}G^{-}H}^{2}s_{\beta-\alpha}C_{H^{+}W^{-}\phi}^{SSS}(G^{\pm},H^{\pm}, H) \notag \\
&\qquad
-i\lambda_{H^{+}G^{-}A}\lambda_{H^{+}G^{-}H}C_{H^{+}W^{-}\phi}^{SSS}(G^{\pm},H^{\pm},A) \notag \\
&\qquad 
+\lambda_{H^{+}G^{-}H}\lambda_{H^{+}H^{-}h}s_{\beta-\alpha}C_{H^{+}W^{-}\phi}^{SSS}(H^{\pm}, G^{\pm}, h)  \notag \\
&\qquad
+\lambda_{H^{+}G^{-}H}\lambda_{H^{+}H^{-}H}c_{\beta-\alpha}C_{H^{+}W^{-}\phi}^{SSS}(H^{\pm}, G^{\pm}, H) \notag \\
&\qquad
+\lambda_{G^{+}G^{-}H}\lambda_{H^{+}G^{-}h}s_{\beta-\alpha}C_{H^{+}W^{-}\phi}^{SSS}(G^{\pm}, G^{\pm}, h)  \notag \\
&\qquad
+\lambda_{G^{+}G^{-}H}\lambda_{H^{+}G^{-}H}c_{\beta-\alpha}C_{H^{+}W^{-}\phi}^{SSS}(G^{\pm}, G^{\pm}, H)\Big\}.
\end{align}}

The 1PI diagrams for the $H^{+}W^{-}A$ couplings are calculated as
\allowdisplaybreaks{
\begin{align}
&(16\pi^{2}){\Gamma_{H^{+}W^{-}A}^{{\rm 1PI}}\red{(p_{1}^{2}, p_{2}^{2}, q^{2})}}_{F} \notag \\
&\quad=
iN_{c}^{f}g\bigg[
\frac{m_{u}\zeta_{u}}{v}C_{H^{+}W^{-}\phi}^{FFF}(u, u, d)
+\frac{m_{d}\zeta_{d}}{v}C_{H^{+}W^{-}\phi}^{FFF}(d, d, u)\bigg]
+ig\frac{m_{\ell}\zeta_{\ell}}{v}C_{H^{+}W^{-}\phi}^{FFF}(\ell, \ell, \nu_{\ell}), \\
&(16\pi^{2}){\Gamma_{H^{+}W^{-}A}^{{\rm 1PI}}\red{(p_{1}^{2}, p_{2}^{2}, q^{2})}}_{B} \notag \\
&\quad=
i\frac{g^{3}}{8}\Big\{
c_{\beta-\alpha}^{2}C_{H^{+}W^{-}\phi}^{SVV}(h, Z, W)
+s_{\beta-\alpha}^{2}C_{H^{+}W^{-}\phi}^{SVV}(H, Z, W) \notag \\
&\qquad
+c_{2W}C_{H^{+}W^{-}\phi}^{SVV}(H^{\pm}, W, Z)
+2s_{W}^{2}C_{H^{+}W^{-}\phi}^{SVV}(H^{\pm}, W, \gamma)\Big\} \notag \\
&\qquad
-i\frac{g^{3}}{8}\Big\{
c_{\beta-\alpha}^{2}C_{H^{+}W^{-}\phi}^{VSS}(W, H^{\pm}, h)
+s_{\beta-\alpha}^{2}C_{H^{+}W^{-}\phi}^{VSS}(W, H^{\pm}, H)
-C_{H^{+}W^{-}\phi}^{VSS}(W, H^{\pm}, A) \notag \\
&\qquad
+\frac{c_{2W}}{c_{W}^{2}}c_{\beta-\alpha}^{2}C_{H^{+}W^{-}\phi}^{VSS}(Z, h, H^{\pm})
+\frac{c_{2W}}{c_{W}^{2}}s_{\beta-\alpha}^{2}C_{H^{+}W^{-}\phi}^{VSS}(Z, H, H^{\pm})
\Big\} \notag \\
&\qquad
-i\frac{g^{3}}{8}\Big\{
(B_{0}-B_{1})(q^{2}; A, W)
-\frac{s_{W}^{2}}{c_{W}^{2}}c_{2W}(B_{0}-B_{1})(q^{2}; H^{\pm}, Z) 
+2s_{W}^{2}(B_{0}-B_{1})(q^{2}; H^{\pm}, \gamma) \notag \\
&\qquad
+(B_{0}-B_{1})(p_{1}^{2}; H^{\pm}, W)
+\frac{s_{W}^{2}}{c_{W}^{2}}c_{\beta-\alpha}^{2}(B_{0}-B_{1})(p_{1}^{2}; h, Z)
+\frac{s_{W}^{2}}{c_{W}^{2}}s_{\beta-\alpha}^{2}(B_{0}-B_{1})(p_{1}^{2}; H, Z)
\Big\} \notag \\   
&\qquad
+i\frac{g^{2}}{4}m_{W}\Big\{
\frac{s_{W}^{2}}{c_{W}^{2}}\lambda_{H^{+}G^{-}h}c_{\beta-\alpha}C_{H^{+}W^{-}\phi}^{SVS}(h, Z, G^{\pm})
-\frac{s_{W}^{2}}{c_{W}^{2}}\lambda_{H^{+}G^{-}H}s_{\beta-\alpha}C_{H^{+}W^{-}\phi}^{SVS}(H, Z, G^{\pm}) \notag \\
&\qquad
+\lambda_{H^{+}H^{-}h}s_{\beta-\alpha}C_{H^{+}W^{-}\phi}^{SVS}(H^{\pm}, W, h)
+\lambda_{H^{+}H^{-}H}c_{\beta-\alpha}C_{H^{+}W^{-}\phi}^{SVS}(H^{\pm}, W, H)\Big\} \notag \\
&\qquad
+i\frac{g^{2}}{4}m_{W}\Big\{
2\lambda_{AAh}s_{\beta-\alpha}C_{H^{+}W^{-}\phi}^{SSV}(A, h, W)
+2\lambda_{AAH}c_{\beta-\alpha}C_{H^{+}W^{-}\phi}^{SSV}(A, H, W) \notag \\
&\qquad
-i\frac{s_{W}^{2}}{c_{W}^{2}}c_{2W}\lambda_{H^{+}G^{-}A}C_{H^{+}W^{-}\phi}^{SSV}(H^{\pm}, G^{\mp}, Z)
+2is_{W}^{2}\lambda_{H^{+}G^{-}A}C_{H^{+}W^{-}\phi}^{SSV}(H^{\pm}, G^{\mp}, \gamma)\Big\} \notag \\
&\qquad
-i\frac{g}{2}\Big\{
2\lambda_{hAA}\lambda_{H^{+}H^{-}h}C_{H^{+}W^{-}\phi}^{SSS}(h, A, H^{\pm})
+2\lambda_{HAA}\lambda_{H^{+}H^{-}H}C_{H^{+}W^{-}\phi}^{SSS}(H, A, H^{\pm}) \notag \\
&\qquad 
+2i\lambda_{AAh}\lambda_{H^{+}G^{-}A}s_{\beta-\alpha}C_{H^{+}W^{-}\phi}^{SSS}(A, h, G^{\pm})
+2i\lambda_{AAH}\lambda_{H^{+}G^{-}A}c_{\beta-\alpha}C_{H^{+}W^{-}\phi}^{SSS}(A, H, G^{\pm})
\notag \\
&\qquad
-i\lambda_{H^{-}G^{+}A}\lambda_{H^{+}G^{-}h}c_{\beta-\alpha}C_{H^{+}W^{-}\phi}^{SSS}(G^{\pm}, H^{\pm}, h)
+i\lambda_{H^{-}G^{+}A}\lambda_{H^{+}G^{-}H}s_{\beta-\alpha}C_{H^{+}W^{-}\phi}^{SSS}(G^{\pm}, H^{\pm}, H) \notag \\
&\qquad
-\lambda_{H^{-}G^{+}A}\lambda_{H^{+}G^{-}A}C_{H^{+}W^{-}\phi}^{SSS}(G^{\pm},H^{\pm},A) \notag \\
&\qquad
+\lambda_{AGh}\lambda_{H^{+}G^{-}h}C_{H^{+}W^{-}\phi}^{SSS}(h, G^{0}, G^{\pm})
+\lambda_{AGH}\lambda_{H^{+}G^{-}H}C_{H^{+}W^{-}\phi}^{SSS}(H, G^{0}, G^{\pm}) \notag \\
&\qquad
-i\lambda_{H^{+}G^{-}A}\lambda_{H^{+}H^{-}h}s_{\beta-\alpha}C_{H^{+}W^{-}\phi}^{SSS}(H^{\pm}, G^{\pm}, h)  \notag \\
&\qquad
-i\lambda_{H^{+}G^{-}A}\lambda_{H^{+}H^{-}H}c_{\beta-\alpha}C_{H^{+}W^{-}\phi}^{SSS}(H^{\pm}, G^{\pm}, H)\Big\}.
\end{align}}

The 1PI diagrams for the $H^{+}ZW^{-}$ couplings are calculated as
\allowdisplaybreaks{
\begin{align}
&(16\pi^{2}){\Gamma^{1, {\rm 1PI}}_{H^{+}ZW^{-}}\red{(p_{1}^{2}, p_{2}^{2}, q^{2})}}_{F} \notag \\
&\quad=
\sum_{f}4I_{f}N_{c}^{f}\frac{gg_{Z}}{v}\bigg\{
-m_{f}^{2}\zeta_{f}(v_{f'}+a_{f'})\Big[B_{0}(p_{2}^{2}; f', f')-2C_{24}
+\frac{1}{2}(q^{2}+p_{1}^{2}-p_{2}^{2}+2m_{f}^{2})C_{0} \notag \\
&\qquad
+\frac{1}{2}(q^{2}+3p_{1}^{2}-p_{2}^{2})C_{11}
+(q^{2}-p_{1}^{2})C_{12}\Big]
+m_{f}^{2}m_{f'}^{2}\zeta_{f}(v_{f'}-a_{f'})C_{0} \notag \\
&\qquad
+m_{f'}^{2}\zeta_{f'}(v_{f'}+a_{f'})\Big[B_{0}(p_{2}^{2}; f', f')-2C_{24}+m_{f}^{2}C_{0}
+p_{1}^{2}C_{11}+\frac{1}{2}(q^{2}-p_{1}^{2}-p_{2}^{2})C_{12}\Big] \notag \\
&\qquad
-m_{f'}^{2}\zeta_{f'}(v_{f'}-a_{f'})\Big[B_{0}(p_{2}^{2}; f', f')+m_{f}^{2}C_{0}
+\frac{1}{2}(q^{2}+p_{1}^{2}-p_{2}^{2})C_{11}  \notag \\
&\qquad
+\frac{1}{2}(q^{2}-p_{1}^{2}+p_{2}^{2})C_{12}\Big]\bigg\} (f, f', f')
+(f\leftrightarrow f'), \\
&(16\pi^{2}){\Gamma^{2, {\rm 1PI}}_{H^{+}ZW^{-}}\red{(p_{1}^{2}, p_{2}^{2}, q^{2})}}_{F} \notag \\
&\quad=
\sum_{f} 4I_{f}N_{c}^{f}\frac{gg_{Z}}{v}\bigg\{
m_{f}^{2}\zeta_{f}(v_{f'}+a_{f'})\Big(C_{11}+2C_{12}+C_{0}+2C_{23}\Big) \notag \\
&\qquad
-m_{f'}^{2}\zeta_{f'}(v_{f'}+a_{f'})\Big(C_{12}+2C_{23}\Big)
+m_{f'}^{2}\zeta_{f'}(v_{f'}-a_{f'})\Big(C_{11}-C_{12}\Big)\bigg\} (f, f', f') \notag \\
&\qquad
+(f\leftrightarrow f'), \\
&(16\pi^{2}){\Gamma^{3, {\rm 1PI}}_{H^{+}ZW^{-}}\red{(p_{1}^{2}, p_{2}^{2}, q^{2})}}_{F} \notag \\
&\quad=
\sum_{f} 2N_{c}^{f}\frac{gg_{Z}}{v}\bigg\{
-m_{f}^{2}\zeta_{f}(v_{f'}+a_{f'})\Big(C_{0}+C_{11}\Big) 
+m_{f'}^{2}\zeta_{f'}(v_{f'}+a_{f'})C_{12} \notag \\
&\qquad
+m_{f'}^{2}\zeta_{f'}(v_{f'}-a_{f'})\Big(C_{11}-C_{12}\Big)\bigg\} (f, f', f')
+(f\leftrightarrow f'), \\
&(16\pi^{2}){\Gamma^{1, {\rm 1PI}}_{H^{+}ZW^{-}}\red{(p_{1}^{2}, p_{2}^{2}, q^{2})}}_{B} \notag \\
&\quad=
\frac{gg_{Z}}{v}\Big[
\lambda_{H^{+}H^{-}h}vc_{\beta-\alpha}C_{24}(H^{\pm}, A, h)
-\lambda_{H^{+}H^{-}H}vs_{\beta-\alpha}C_{24}(H^{\pm}, A, H) \notag \\
&\qquad
-\lambda_{H^{+}H^{-}h}vc_{\beta-\alpha}c_{2W}C_{24}(h, H^{\pm}, H^{\pm})
+\lambda_{H^{+}H^{-}H}vs_{\beta-\alpha}c_{2W}C_{24}(H, H^{\pm}, H^{\pm}) \notag \\
&\qquad
-s_{\beta-\alpha}c_{\beta-\alpha}(m_{H^{\pm}}^{2}-m_{A}^{2})C_{24}(G^{\pm}, h, A)
+s_{\beta-\alpha}c_{\beta-\alpha}(m_{H^{\pm}}^{2}-m_{A}^{2})C_{24}(G^{\pm}, H, A)  \notag \\
&\qquad
+s_{\beta-\alpha}c_{\beta-\alpha}(m_{H^{\pm}}^{2}-m_{h}^{2})C_{24}(G^{\pm}, G^{0}, h)
-s_{\beta-\alpha}c_{\beta-\alpha}(m_{H^{\pm}}^{2}-m_{H}^{2})C_{24}(G^{\pm}, G^{0}, H) \notag \\
&\qquad
-s_{\beta-\alpha}c_{\beta-\alpha}(m_{H^{\pm}}^{2}-m_{h}^{2})c_{2W}C_{24}(h, G^{\pm}, G^{\pm})
+s_{\beta-\alpha}c_{\beta-\alpha}(m_{H^{\pm}}^{2}-m_{H}^{2})c_{2W}C_{24}(H, G^{\pm}, G^{\pm}) \notag \\
&\qquad
-m_{W}^{2}s_{\beta-\alpha}c_{\beta-\alpha}C_{24}(W^{\pm}, h, A)
+m_{W}^{2}s_{\beta-\alpha}c_{\beta-\alpha}C_{24}(W^{\pm}, H, A) \notag \\
&\qquad
+m_{Z}^{2}c_{2W}s_{\beta-\alpha}c_{\beta-\alpha}C_{24}(W^{\pm}, h, Z)
-m_{Z}^{2}c_{2W}s_{\beta-\alpha}c_{\beta-\alpha}C_{24}(W^{\pm}, H, Z) \notag \\
&\qquad
+m_{W}^{2}s_{W}^{2}s_{\beta-\alpha}c_{\beta-\alpha}C_{24}(h, G^{\pm}, W^{\pm})
-m_{W}^{2}s_{W}^{2}s_{\beta-\alpha}c_{\beta-\alpha}C_{24}(H, G^{\pm}, W^{\pm}) \notag \\
&\qquad
-m_{W}^{2}s_{W}^{2}(m_{H^{\pm}}^{2}-m_{h}^{2})s_{\beta-\alpha}c_{\beta-\alpha}C_{0}(h, W^{\pm}, G^{\pm}) \notag \\
&\qquad
+m_{W}^{2}s_{W}^{2}(m_{H^{\pm}}^{2}-m_{H}^{2})s_{\beta-\alpha}c_{\beta-\alpha}C_{0}(H, W^{\pm}, G^{\pm}) \notag \\
&\qquad
-m_{Z}^{2}s_{W}^{2}(m_{H^{\pm}}^{2}-m_{h}^{2})s_{\beta-\alpha}c_{\beta-\alpha}C_{0}(G^{\pm}, Z, h) \notag \\
&\qquad
+m_{Z}^{2}s_{W}^{2}(m_{H^{\pm}}^{2}-m_{H}^{2})s_{\beta-\alpha}c_{\beta-\alpha}C_{0}(G^{\pm}, Z, H) \notag \\
&\qquad
+m_{W}^{2}s_{\beta-\alpha}c_{\beta-\alpha}F_{VVS}(W^{\pm}, Z, h)
-m_{W}^{2}s_{\beta-\alpha}c_{\beta-\alpha}F_{VVS}(W^{\pm}, Z, H) \notag \\
&\qquad
-m_{W}^{2}c_{W}^{2}s_{\beta-\alpha}c_{\beta-\alpha}F_{SVV}(h, W^{\pm}, W^{\pm})
+m_{W}^{2}c_{W}^{2}s_{\beta-\alpha}c_{\beta-\alpha}F_{SVV}(H, W^{\pm}, W^{\pm}) \notag \\
&\qquad
-\frac{1}{2}s_{W}^{2}\lambda_{H^{+}H^{-}h}v c_{\beta-\alpha}B_{0}(q^{2}; h, H^{\pm})
+\frac{1}{2}s_{W}^{2}\lambda_{H^{+}H^{-}H}v s_{\beta-\alpha}B_{0}(q^{2}; H, H^{\pm}) \notag \\
&\qquad
-\frac{1}{2}s_{W}^{2}s_{\beta-\alpha}c_{\beta-\alpha}(m_{H^{\pm}}^{2}-m_{h}^{2})B_{0}(q^{2}; h, G^{\pm})
+\frac{1}{2}s_{W}^{2}s_{\beta-\alpha}c_{\beta-\alpha}(m_{H^{\pm}}^{2}-m_{H}^{2})B_{0}(q^{2}; H, G^{\pm}) \notag \\
&\qquad
+m_{W}^{2}s_{W}^{2}s_{\beta-\alpha}c_{\beta-\alpha}B_{0}(p_{1}^{2}; W^{\pm}, h)
-m_{W}^{2}s_{W}^{2}s_{\beta-\alpha}c_{\beta-\alpha}B_{0}(p_{1}^{2}; W^{\pm}, H) \notag \\
&\qquad
+m_{Z}^{2}s_{W}^{2}s_{\beta-\alpha}c_{\beta-\alpha}B_{0}(p_{2}^{2}; Z, h)
-m_{Z}^{2}s_{W}^{2}s_{\beta-\alpha}c_{\beta-\alpha}B_{0}(p_{2}^{2}; Z, H)
\Big], \\
&(16\pi^{2}){\Gamma^{2, {\rm 1PI}}_{H^{+}ZW^{-}}\red{(p_{1}^{2}, p_{2}^{2}, q^{2})}}_{B} \notag \\
&\quad=
\frac{gg_{Z}}{v}\Big[
\lambda_{H^{+}H^{-}h}vc_{\beta-\alpha}(C_{12}+C_{23})(H^{\pm}, A, h)
-\lambda_{H^{+}H^{-}H}vs_{\beta-\alpha}(C_{12}+C_{23})(H^{\pm}, A, H) \notag \\
&\qquad
-\lambda_{H^{+}H^{-}h}vc_{\beta-\alpha}c_{2W}(C_{12}+C_{23})(h, H^{\pm}, H^{\pm})
+\lambda_{H^{+}H^{-}H}vs_{\beta-\alpha}c_{2W}(C_{12}+C_{23})(H, H^{\pm}, H^{\pm}) \notag \\
&\qquad
-s_{\beta-\alpha}c_{\beta-\alpha}(m_{H^{\pm}}^{2}-m_{A}^{2})(C_{12}+C_{23})(G^{\pm}, h, A) \notag \\
&\qquad
+s_{\beta-\alpha}c_{\beta-\alpha}(m_{H^{\pm}}^{2}-m_{A}^{2})(C_{12}+C_{23})(G^{\pm}, H, A) \notag \\
&\qquad
+s_{\beta-\alpha}c_{\beta-\alpha}(m_{H^{\pm}}^{2}-m_{h}^{2})(C_{12}+C_{23})(G^{\pm}, G^{0}, h) \notag \\
&\qquad
-s_{\beta-\alpha}c_{\beta-\alpha}(m_{H^{\pm}}^{2}-m_{H}^{2})(C_{12}+C_{23})(G^{\pm}, G^{0}, H) \notag \\
&\qquad
-s_{\beta-\alpha}c_{\beta-\alpha}(m_{H^{\pm}}^{2}-m_{h}^{2})c_{2W}(C_{12}+C_{23})(h, G^{\pm}, G^{\pm}) \notag \\
&\qquad
+s_{\beta-\alpha}c_{\beta-\alpha}(m_{H^{\pm}}^{2}-m_{H}^{2})c_{2W}(C_{12}+C_{23})(H, G^{\pm}, G^{\pm}) \notag \\
&\qquad
-m_{W}^{2}s_{\beta-\alpha}c_{\beta-\alpha}(2C_{0}+2C_{11}+C_{12}+C_{23})(W^{\pm}, h, A) \notag \\
&\qquad
+m_{W}^{2}s_{\beta-\alpha}c_{\beta-\alpha}(2C_{0}+2C_{11}+C_{12}+C_{23})(W^{\pm}, H, A) \notag \\
&\qquad
+m_{Z}^{2}c_{2W}s_{\beta-\alpha}c_{\beta-\alpha}(C_{23}-C_{12})(H^{\pm}, h, Z)
-m_{Z}^{2}c_{2W}s_{\beta-\alpha}c_{\beta-\alpha}(C_{23}-C_{12})(H^{\pm}, H, Z) \notag \\
&\qquad
+m_{W}^{2}s_{W}^{2}s_{\beta-\alpha}c_{\beta-\alpha}(C_{23}-C_{12})(h, G^{\pm}, W^{\pm})
-m_{W}^{2}s_{W}^{2}s_{\beta-\alpha}c_{\beta-\alpha}(C_{23}-C_{12})(H, G^{\pm}, W^{\pm}) \notag \\
&\qquad
+m_{W}^{2}s_{\beta-\alpha}c_{\beta-\alpha}(2C_{0}-2C_{11}+5C_{12}+C_{23})(W^{\pm}, Z, h) \notag \\
&\qquad
-m_{W}^{2}s_{\beta-\alpha}c_{\beta-\alpha}(2C_{0}-2C_{11}+5C_{12}+C_{23})(W^{\pm}, Z, H) \notag \\
&\qquad
-m_{W}^{2}c_{W}^{2}s_{\beta-\alpha}c_{\beta-\alpha}(C_{23}+3C_{12}-4C_{11})(h, W^{\pm}, W^{\pm}) \notag \\
&\qquad
+m_{W}^{2}c_{W}^{2}s_{\beta-\alpha}c_{\beta-\alpha}(C_{23}+3C_{12}-4C_{11})(H, W^{\pm}, W^{\pm}) 
\Big], \\
&(16\pi^{2}){\Gamma^{3, {\rm 1PI}}_{H^{+}ZW^{-}}\red{(p_{1}^{2}, p_{2}^{2}, q^{2})}}_{B}=0,
\end{align}}
where
\begin{align}
F_{VVS}(V_{1},V_{2},S) &= -B_{0}(p_{2}^{2}; V_{2}, S)+\big[C_{24}
-(q^{2}+3p_{1}^{2}-p_{2}^{2})C_{11}-2(q^{2}-p_{1}^{2})C_{12} \notag \\
&\quad
-(2q^{2}+2p_{1}^{2}-2p_{2}^{2}+m_{V_{1}}^{2})C_{0}\big](p_{1}^{2}, p_{2}^{2}, q^{2}; V_{1},V_{2},S), \\
F_{SVV}(S,V_{1},V_{2}) &= -B_{0}(p_{2}^{2}; V_{1}, V_{2})+\big[C_{24}
+(q^{2}-p_{1}^{2}-p_{2}^{2})C_{11}+2p_{2}^{2}C_{12} \notag \\
&\quad
+(p_{1}^{2}-p_{2}^{2}-m_{S}^{2})C_{0}\big](p_{1}^{2}, p_{2}^{2}, q^{2}; S, V_{1},V_{2}).
\end{align}

The 1PI diagrams for the $H^{+}\gamma W^{-}$ couplings are calculated as
\allowdisplaybreaks{
\begin{align}
&(16\pi^{2}){\Gamma^{1, {\rm 1PI}}_{H^{+}\gamma W^{-}}\red{(p_{1}^{2}, p_{2}^{2}, q^{2})}}_{F} \notag \\
&\quad=
\sum_{f}4I_{f}N_{c}^{f}Q_{f'}\frac{eg}{v}\bigg\{
-m_{f}^{2}\zeta_{f}\Big[B_{0}(p_{2}^{2}; f', f')-2C_{24}
+\frac{1}{2}(q^{2}+p_{1}^{2}-p_{2}^{2}+2m_{f}^{2}-2m_{f'}^{2})C_{0} \notag \\
&\qquad
+\frac{1}{2}(q^{2}+3p_{1}^{2}-p_{2}^{2})C_{11}
+\frac{1}{2}(q^{2}-p_{1}^{2})C_{12}\Big] \notag \\
&\qquad
+m_{f'}^{2}\zeta_{f'}\Big[-2C_{24}-\frac{1}{2}(q^{2}-p_{1}^{2}-p_{2}^{2})C_{11}-p_{2}^{2}C_{12}\Big]\bigg\}(f, f', f')
+(f\leftrightarrow f'), \\
&(16\pi^{2}){\Gamma^{2, {\rm 1PI}}_{H^{+}\gamma W^{-}}\red{(p_{1}^{2}, p_{2}^{2}, q^{2})}}_{F} \notag \\
&\quad=
\sum_{f} 4I_{f}N_{c}^{f}Q_{f'}\frac{eg}{v}\bigg\{
m_{f}^{2}\zeta_{f}\Big(C_{11}+2C_{12}+C_{0}+2C_{23}\Big) \notag \\
&\qquad
+m_{f'}^{2}\zeta_{f'}\Big(C_{11}-2C_{12}-2C_{23}\Big)\bigg\} (f, f', f')
+(f\leftrightarrow f'), \\
&(16\pi^{2}){\Gamma^{3, {\rm 1PI}}_{H^{+}\gamma W^{-}}\red{(p_{1}^{2}, p_{2}^{2}, q^{2})}}_{F} \notag \\
&\qquad=
\sum_{f}2N_{c}^{f}Q_{f'}\frac{eg}{v}\bigg\{
-m_{f}^{2}\zeta_{f}\Big(C_{0}+C_{11}\Big)
+m_{f'}^{2}\zeta_{f'}C_{11}\bigg\} (f, f', f')
+(f\leftrightarrow f'), \\
&(16\pi^{2}){\Gamma^{1, {\rm 1PI}}_{H^{+}\gamma W^{-}}\red{(p_{1}^{2}, p_{2}^{2}, q^{2})}}_{B} \notag \\
&\quad=
\frac{2eg}{v}\Big[
-\lambda_{H^{+}H^{-}h}vc_{\beta-\alpha}C_{24}(h, H^{\pm}, H^{\pm})
+\lambda_{H^{+}H^{-}H}vs_{\beta-\alpha}C_{24}(H, H^{\pm}, H^{\pm}) \notag \\
&\qquad
-s_{\beta-\alpha}c_{\beta-\alpha}(m_{H^{\pm}}^{2}-m_{h}^{2})C_{24}(h, G^{\pm}, G^{\pm})
+s_{\beta-\alpha}c_{\beta-\alpha}(m_{H^{\pm}}^{2}-m_{H}^{2})C_{24}(H, G^{\pm}, G^{\pm}) \notag \\
&\qquad
-\frac{m_{W}^{2}}{2}s_{\beta-\alpha}c_{\beta-\alpha}C_{24}(h, G^{\pm}, W^{\pm})
+\frac{m_{W}^{2}}{2}s_{\beta-\alpha}c_{\beta-\alpha}C_{24}(H, G^{\pm}, W^{\pm}) \notag \\
&\qquad
+\frac{m_{W}^{2}}{2}(m_{H^{\pm}}^{2}-m_{h}^{2})s_{\beta-\alpha}c_{\beta-\alpha}C_{0}(h, W^{\pm}, G^{\pm})
-\frac{m_{W}^{2}}{2}(m_{H^{\pm}}^{2}-m_{H}^{2})s_{\beta-\alpha}c_{\beta-\alpha}C_{0}(H, W^{\pm}, G^{\pm}) \notag \\
&\qquad
-\frac{m_{W}^{2}}{2}s_{\beta-\alpha}c_{\beta-\alpha}F_{SVV}(h, W^{\pm}, W^{\pm})
+\frac{m_{W}^{2}}{2}s_{\beta-\alpha}c_{\beta-\alpha}F_{SVV}(H, W^{\pm}, W^{\pm}) \notag \\
&\qquad
+\frac{1}{4}\lambda_{H^{+}H^{-}h}v c_{\beta-\alpha}B_{0}(q^{2}; h, H^{\pm})
-\frac{1}{4}\lambda_{H^{+}H^{-}H}v s_{\beta-\alpha}B_{0}(q^{2}; H, H^{\pm}) \notag \\
&\qquad
+\frac{1}{4}s_{\beta-\alpha}c_{\beta-\alpha}(m_{H^{\pm}}^{2}-m_{h}^{2})B_{0}(q^{2}; h, G^{\pm}) 
-\frac{1}{4}s_{\beta-\alpha}c_{\beta-\alpha}(m_{H^{\pm}}^{2}-m_{H}^{2})B_{0}(q^{2}; H, G^{\pm})
\notag \\
&\qquad
-\frac{m_{W}^{2}}{2}s_{\beta-\alpha}c_{\beta-\alpha}B_{0}(p_{1}^{2}; W^{\pm}, h)
+\frac{m_{W}^{2}}{2}s_{\beta-\alpha}c_{\beta-\alpha}B_{0}(p_{1}^{2}; W^{\pm}, H)
\Big], \\
&(16\pi^{2}){\Gamma^{2, {\rm 1PI}}_{H^{+}\gamma W^{-}}\red{(p_{1}^{2}, p_{2}^{2}, q^{2})}}_{B} \notag \\
&\quad=
\frac{2eg}{v}\Big[
-\lambda_{H^{+}H^{-}h}vc_{\beta-\alpha}(C_{12}+C_{23})(h, H^{\pm}, H^{\pm}) \notag \\
&\qquad
+\lambda_{H^{+}H^{-}H}vs_{\beta-\alpha}(C_{12}+C_{23})(H, H^{\pm}, H^{\pm}) \notag \\
&\qquad
-s_{\beta-\alpha}c_{\beta-\alpha}(m_{H^{\pm}}^{2}-m_{h}^{2})(C_{12}+C_{23})(h, G^{\pm}, G^{\pm}) \notag \\
&\qquad
+s_{\beta-\alpha}c_{\beta-\alpha}(m_{H^{\pm}}^{2}-m_{H}^{2})(C_{12}+C_{23})(H, G^{\pm}, G^{\pm}) \notag \\
&\qquad
-\frac{m_{W}^{2}}{2}s_{\beta-\alpha}c_{\beta-\alpha}(C_{23}-C_{12})(h, G^{\pm}, W^{\pm})
+\frac{m_{W}^{2}}{2}s_{\beta-\alpha}c_{\beta-\alpha}(C_{23}-C_{12})(H, G^{\pm}, W^{\pm}) \notag \\
&\qquad
+\frac{m_{W}^{2}}{2}s_{\beta-\alpha}c_{\beta-\alpha}(4C_{11}-3C_{12}-C_{23})(h, W^{\pm}, W^{\pm}) \notag \\
&\qquad
-\frac{m_{W}^{2}}{2}s_{\beta-\alpha}c_{\beta-\alpha}(4C_{11}-3C_{12}-C_{23})(H, W^{\pm}, W^{\pm}) 
\Big], \\
&(16\pi^{2}){\Gamma^{3, {\rm 1PI}}_{H^{+}\gamma W^{-}}\red{(p_{1}^{2}, p_{2}^{2}, q^{2})}}_{B}=0.
\end{align}}

\subsection{Vertex functions for $H^{-}ff',\, H^{-}W^{+}\phi$ and $H^{-}VW^{+}$}\label{sec:ApB3}
We here show the vertex functions with opposite electric sign as compared to those in Appendix~\ref{sec:vertex}.
Contributions from the 1PI diagrams for the $H^{-}\bar{d}u$ vertice are given by
\begin{align}
\Gamma_{H^{-}\bar{d}u}^{S}(p_{d}^{2}, p_{u}^{2}, q^{2}) &= +\Gamma_{H^{+}\bar{u}d}^{S}(p_{d}^{2}, p_{u}^{2}, q^{2}), \label{eq: H-ffp_S} \\
\Gamma_{H^{-}\bar{d}u}^{P}(p_{d}^{2}, p_{u}^{2}, q^{2}) &= -\Gamma_{H^{+}\bar{u}d}^{P}(p_{d}^{2}, p_{u}^{2}, q^{2}), \\
\Gamma_{H^{-}\bar{d}u}^{V_{1}}(p_{d}^{2}, p_{u}^{2}, q^{2}) &= -\Gamma_{H^{+}\bar{u}d}^{V_{2}}(p_{d}^{2}, p_{u}^{2}, q^{2}), \\
\Gamma_{H^{-}\bar{d}u}^{V_{2}}(p_{d}^{2}, p_{u}^{2}, q^{2}) &= -\Gamma_{H^{+}\bar{u}d}^{V_{1}}(p_{d}^{2}, p_{u}^{2}, q^{2}), \\
\Gamma_{H^{-}\bar{d}u}^{A_{1}}(p_{d}^{2}, p_{u}^{2}, q^{2}) &= -\Gamma_{H^{+}\bar{u}d}^{A_{2}}(p_{d}^{2}, p_{u}^{2}, q^{2}), \\
\Gamma_{H^{-}\bar{d}u}^{A_{2}}(p_{d}^{2}, p_{u}^{2}, q^{2}) &= -\Gamma_{H^{+}\bar{u}d}^{A_{1}}(p_{d}^{2}, p_{u}^{2}, q^{2}), \\
\Gamma_{H^{-}\bar{d}u}^{T}(p_{d}^{2}, p_{u}^{2}, q^{2}) &= +\Gamma_{H^{+}\bar{u}d}^{T}(p_{d}^{2}, p_{u}^{2}, q^{2}), \\
\Gamma_{H^{-}\bar{d}u}^{PT}(p_{d}^{2}, p_{u}^{2}, q^{2}) &= -\Gamma_{H^{+}\bar{u}d}^{PT}(p_{d}^{2}, p_{u}^{2}, q^{2}), \label{eq: H-ffp_PT}
\end{align}
where $p_{u}\, (p_{d})$ is a momentum of an up-type quark (a down-type quark).
We note that the moemntum arguments are $p_{1}=p_{d}$ and $p_{2}=p_{u}$ in the $H^{+}\bar{u}d$ vertex, while $p_{1}=p_{u}$ and $p_{2}=p_{d}$ in the $H^{-}\bar{d}u$ vertex.
From Eqs.~\eqref{eq: H-ffp_S}-\eqref{eq: H-ffp_PT}, one can obtain the results for the $H^{-}\bar{\ell}\nu$ vertex by replacing $u\to \nu$ and $d\to \ell$.

Contributions from the 1PI diagrams for the $H^{-}W^{+}\phi$ vertices $(\phi=h,\, H,\, A)$ are given by
\begin{align}
\Gamma_{H^{-}W^{+}\phi}(p_{1}^{2}, p_{2}^{2}, q^{2}) = 
\begin{dcases}
-\Gamma_{H^{+}W^{-}\phi}(p_{1}^{2}, p_{2}^{2}, q^{2}) \quad (\phi = h,\, H), \\
+\Gamma_{H^{+}W^{-}\phi}(p_{1}^{2}, p_{2}^{2}, q^{2}) \quad (\phi = A).
\end{dcases}
\end{align}

Contributions from the 1PI diagrams for the $H^{-}VW^{+}$ vertices $(V=Z,\, \gamma)$ are given by
\begin{align}
\Gamma_{H^{-}VW^{+}}^{1}(p_{1}^{2}, p_{2}^{2}, q^{2}) &= +\Gamma_{H^{+}VW^{-}}^{1}(p_{1}^{2}, p_{2}^{2}, q^{2}), \\
\Gamma_{H^{-}VW^{+}}^{2}(p_{1}^{2}, p_{2}^{2}, q^{2}) &= +\Gamma_{H^{+}VW^{-}}^{2}(p_{1}^{2}, p_{2}^{2}, q^{2}), \\
\Gamma_{H^{-}VW^{+}}^{3}(p_{1}^{2}, p_{2}^{2}, q^{2}) &= -\Gamma_{H^{+}VW^{-}}^{3}(p_{1}^{2}, p_{2}^{2}, q^{2}).
\end{align}

\section{Formulae for the real photon emissions}\label{sec:real}
We here give the decay rate of a massive particle with $p_{0}^{2}=m_{0}^{2}$ into two massive particles with $p_{1}^{2}=m_{1}^{2}$ and $p_{2}^{2}=m_{2}^{2}$ and a photon with $q^{2}=\mu^{2}$, where we introduce the small photon mass as an IR regulator.
We need to evaluate the following phase space integrals,
\begin{align}
I_{i_{1},\cdots,i_{n}}^{j_{1},\cdots j_{m}}(m_{0}, m_{1}, m_{2})
=
32\pi^{3}\int d\Phi_{3}\frac{(\pm 2q\vdot p_{j_{1}})\cdots(\pm 2q\vdot p_{j_{m}})}{(\pm 2q\vdot p_{i_{1}})\cdots(\pm 2q\vdot p_{i_{n}})}, \label{eq: brems_int}
\end{align}
where $j_{k}, i_{\ell}=0, 1, 2$ and the plus signs belong to $p_{1}, p_{2}$, the minus signs to $p_{0}$.
The analytic formulae for $I_{i_{1},\cdots,i_{n}}^{j_{1},\cdots j_{m}}(m_{0}, m_{1}, m_{2})$ are listed in Ref.~\cite{Denner:1991kt}.

The decay rate for $\Hpm\to f\bar{f'}\gamma$ can be written as
\begin{align}
&\Gamma(H^{\pm}\to f\bar{f'}\gamma) \notag \\
&\quad=\frac{e^{2}N_{c}^{f}}{(4\pi)^{3}m_{H^{\pm}}}
\Big\{(\abs{c_{L}}^{2}+\abs{c_{R}}^{2})\big[Q_{H^{\pm}}^{2}\Omega^{LL}_{00}+Q_{f'}^{2}\Omega^{LL}_{11}+Q_{f}^{2}\Omega^{LL}_{22}+Q_{H^{\pm}}Q_{f'}\Omega^{LL}_{01} \notag \\
&\qquad
+Q_{H^{\pm}}Q_{f}\Omega^{LL}_{02}+Q_{f'}Q_{f}\Omega^{LL}_{12}\big]
+(c_{L}c_{R}^{*}+c_{L}^{*}c_{R})\big[Q_{H^{\pm}}^{2}\Omega^{LR}_{00}+Q_{f'}^{2}\Omega^{LR}_{11}+Q_{f}^{2}\Omega^{LR}_{22} \notag \\
&\qquad
+Q_{H^{\pm}}Q_{f'}\Omega^{LR}_{01}+Q_{H^{ \pm}}Q_{f}\Omega^{LR}_{02}+Q_{f'}Q_{f}\Omega^{LR}_{12}\big]\Big\}
\end{align}
with $c_{L}=g_{S}^{H^{\pm}}-g_{P}^{H^{\pm}}$ and $c_{R}=g_{S}^{H^{\pm}}+g_{P}^{H^{\pm}}$.
We have also introduced $Q_{H^{\pm}}$ as the electric charge of the $H^{\pm}$.
The functions $\Omega^{LL}_{ij}$ and $\Omega^{LR}_{ij}$ are given by
\begin{align}
\Omega^{LL}_{00} &=
-4m_{H^{\pm}}^{2}I_{0}-4m_{H^{\pm}}^{2}(m_{H^{\pm}}^{2}-m_{f'}^{2}-m_{f}^{2})I_{00}, \\
\Omega^{LL}_{11} &=
-2I+2(m_{H^{\pm}}^{2}+m_{f'}^{2}-m_{f}^{2})I_{1}-4m_{f'}^{2}(m_{H^{\pm}}^{2}-m_{f'}^{2}-m_{f}^{2})I_{11}, \\
\Omega^{LL}_{22} &=
-2I+2(m_{H^{\pm}}^{2}-m_{f'}^{2}+m_{f}^{2})I_{2}-4m_{f}^{2}(m_{H^{\pm}}^{2}-m_{f'}^{2}-m_{f}^{2})I_{22}, \\
\Omega^{LL}_{01} &=
-2I_{1}^{0}-2(3m_{H^{\pm}}^{2}-m_{f'}^{2}-3m_{f}^{2})I_{1}+4(m_{f'}^{2}+m_{f}^{2})I_{0} \notag \\
&\quad
-4(m_{H^{\pm}}^{2}-m_{f'}^{2}-m_{f}^{2})(m_{H^{\pm}}^{2}+m_{f'}^{2}-m_{f}^{2})I_{01}, \\
\Omega^{LL}_{02} &=
2I_{2}^{0}+2(3m_{H^{\pm}}^{2}-3m_{f'}^{2}-m_{f}^{2})I_{2}-4(m_{f'}^{2}+m_{f}^{2})I_{0} \notag \\
&\quad
+4(m_{H^{\pm}}^{2}-m_{f'}^{2}-m_{f}^{2})(m_{H^{\pm}}^{2}-m_{f'}^{2}+m_{f}^{2})I_{02}, \\
\Omega^{LL}_{12} &=
8I+2I_{1}^{2}+2I_{2}^{1}-2(3m_{H^{\pm}}^{2}-m_{f'}^{2}-3m_{f}^{2})I_{1}-2(3m_{H^{\pm}}^{2}-3m_{f'}^{2}-m_{f}^{2})I_{2} \notag \\
&\quad
+4(m_{H^{\pm}}^{2}-m_{f'}^{2}-m_{f}^{2})^{2}I_{12}, \\
\Omega^{LR}_{00} &=
8m_{H^{\pm}}^{2}m_{f'}m_{f}I_{00}, \\
\Omega^{LR}_{11} &=
8m_{f'}^{3}m_{f}I_{11}, \\
\Omega^{LR}_{22} &=
8m_{f'}m_{f}^{3}I_{22}, \\
\Omega^{LR}_{01} &=
8m_{f'}m_{f}\qty[I_{0}+I_{1}+(m_{H^{\pm}}^{2}+m_{f'}^{2}-m_{f}^{2})I_{01}], \\
\Omega^{LR}_{02} &=
-8m_{f'}m_{f}\qty[I_{0}+I_{2}+(m_{H^{\pm}}^{2}-m_{f'}^{2}+m_{f}^{2})I_{02}], \\
\Omega^{LR}_{12} &=
8m_{f'}m_{f}\qty[I_{1}+I_{2}-(m_{H^{\pm}}^{2}-m_{f'}^{2}-m_{f}^{2})I_{12}].
\end{align}
Although one might think that our formulae are slightly different from those in Ref.~\cite{Goodsell:2017pdq}, our results coincide with those in Ref.~\cite{Goodsell:2017pdq}.
Summing all terms, the differences between our results and those in Ref.~\cite{Goodsell:2017pdq} vanish due to the charge conservation.

The partial decay width for $H^{\pm}\rightarrow W^{\pm}\phi\gamma$ is given by
\begin{align}
&\Gamma(H^{\pm}\to W^{\pm}\phi\gamma) \notag \\
&\quad=
-\frac{e^{2}\abs{g_{\phi W^{\mp}H^{\pm}}}^{2}m_{H^{\pm}}^{3}}{16\pi^{3}}\Bigg\{ \lambda\qty(\frac{m_{\phi}^{2}}{m_{H^{\pm}}^{2}}, \frac{m_{W}^{2}}{m_{H^{\pm}}^{2}})
\bigg[I_{22}+\frac{m_{H^{\pm}}^{2}}{m_{W}^{2}}I_{00}
+\left(1+\frac{m_{H^{\pm}}-m_{\phi}^{2}}{m_{W}^{2}}\right)I_{02} \notag \\
&\qquad
+\frac{1}{m_{W}^{2}}(I_{2}+I_{0})\bigg]
-\frac{2}{\mHp^{4}}I_{22}^{11}\Bigg\}.
\end{align}

